\documentclass[prx,twocolumn,superscriptaddress,longbibliography]{revtex4-2}
\usepackage{bm}
\usepackage{amsmath, amsfonts}
\usepackage{amssymb}
\usepackage{times}
\usepackage{graphicx}
\usepackage{xcolor}
\usepackage{mathtools}
\usepackage[colorlinks=true,allcolors=blue,bookmarks=false,final]{hyperref}
\usepackage{cleveref}
\usepackage{tikz}
\usetikzlibrary{tikzmark}
\usetikzlibrary{arrows.meta}

\DeclareMathOperator{\tr}{Tr}

\newcommand\varmp{\mathbin{\vcenter{\hbox{%
  \oalign{\hfil$\scriptstyle-$\hfil\cr
          \noalign{\kern-.5ex}
          $\scriptscriptstyle({+})$\cr}%
}}}}
\def\onecircle{\raisebox{.5pt}{\textcircled{\raisebox{-.9pt} {1}}}}
\def\twocircle{\raisebox{.5pt}{\textcircled{\raisebox{-.9pt} {2}}}}
\usepackage[normalem]{ulem}
\begin{document}

\title{Engineering Dynamical Sweet Spots to Protect Qubits from 1/$f$ Noise}
\author{Ziwen Huang}
\address{Department of Physics and Astronomy, Northwestern University, Evanston, Illinois 60208, USA}
\author{Pranav S. Mundada}
\address{Department of Electrical Engineering, Princeton University, Princeton, New Jersey 08544, USA}
\author{Andr\'as Gyenis}
\address{Department of Electrical Engineering, Princeton University, Princeton, New Jersey 08544, USA}
\author{David I. Schuster}
\address{ The James Franck Institute and Department of Physics, University of Chicago,  Chicago, Illinois 60637, USA}
\author{Andrew A. Houck}
\address{Department of Electrical Engineering, Princeton University, Princeton, New Jersey 08544, USA}
\author{Jens Koch}
\address{Department of Physics and Astronomy, Northwestern University, Evanston, Illinois 60208, USA}
\address{Northwestern–Fermilab Center for Applied Physics and Superconducting Technologies, Northwestern University, Evanston, Illinois 60208, USA}

\begin{abstract}
Protecting superconducting qubits from low-frequency noise is essential for advancing superconducting quantum computation.  Based on the application of a periodic drive field, we develop a protocol for engineering dynamical sweet spots which reduce the susceptibility of a qubit to low-frequency noise. Using the framework of Floquet theory, we prove rigorously that there are manifolds of dynamical sweet spots marked by extrema in the quasi-energy differences of the driven qubit.
In particular, for the example of fluxonium biased slightly away from half a flux quantum, we predict an enhancement of pure-dephasing by three orders of magnitude. Employing the Floquet eigenstates as the computational basis, we show that high-fidelity single- and two-qubit gates can be implemented while maintaining dynamical sweet-spot operation. We further confirm that qubit readout can be performed by adiabatically mapping the Floquet states back to the static qubit states, and subsequently applying standard measurement techniques. Our work provides an intuitive tool to encode quantum information in robust, time-dependent states, and may be extended to alternative architectures for quantum information processing.
\end{abstract}
\maketitle

\section{Introduction}
\begin{figure}[b]
\includegraphics[width = \columnwidth]{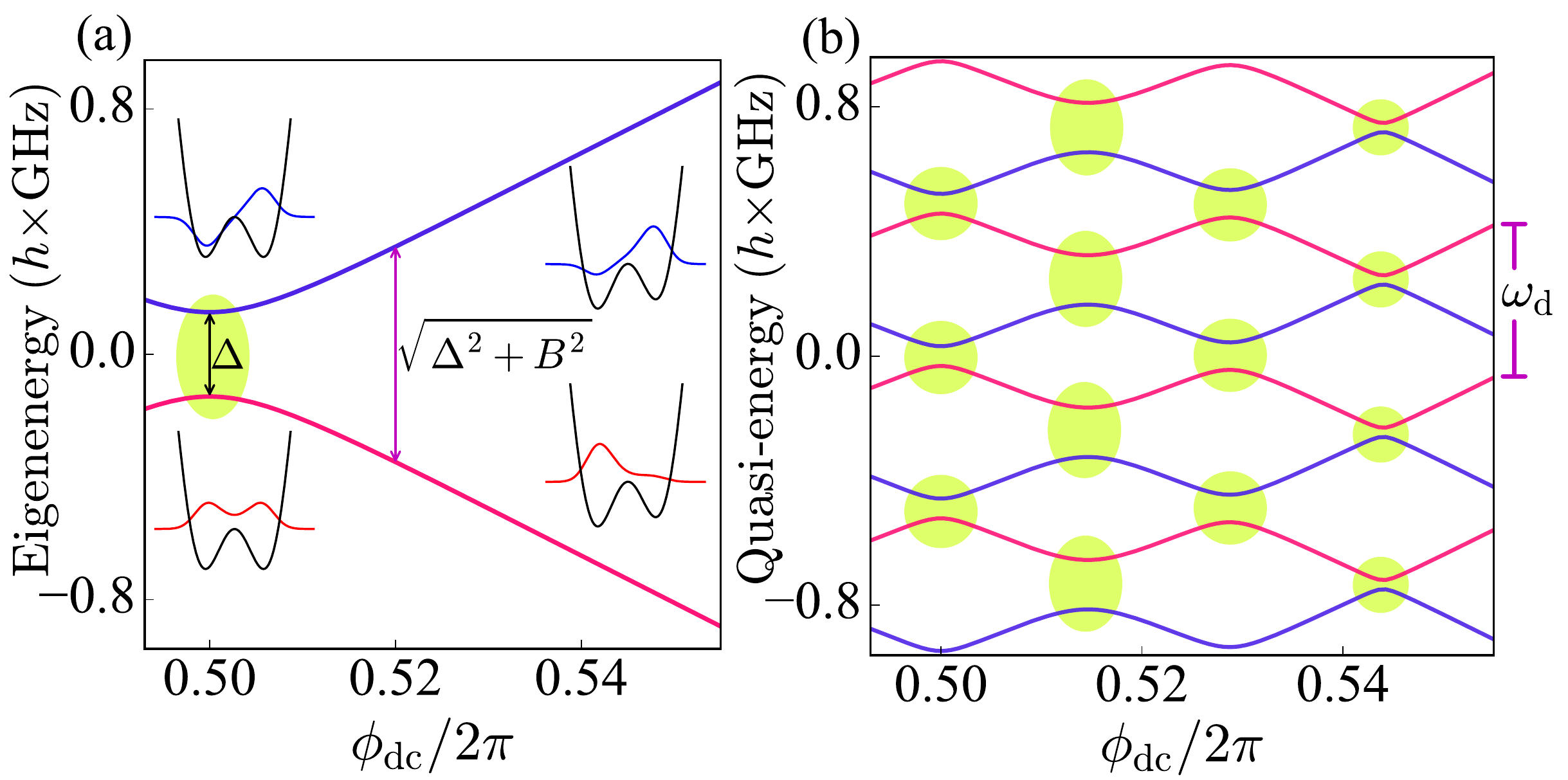}
\caption{(a) Static fluxonium spectrum as a function of magnetic flux. Insets show the qubit eigenfunctions at the sweet spot and slightly away from it ($\phi_{\mathrm{dc}}/2\pi=0.52$). The parameters used are: $E_\mathrm{C}/h=0.5\,\text{GHz}$, $E_\mathrm{J}/h=4.0\,\text{GHz}$ and $E_\mathrm{L}/h=1.3\,\text{GHz}$ for the capacitive, Josephson and inductive energy, respectively.
 (b) Quasi-energy spectrum of the driven qubit for flux $\phi_\mathrm{ac}/2\pi = 0.028$ and drive frequency $\omega_\mathrm{d}/2\pi = 490\,$MHz. The highlighted regions in both panels mark the flux sweet spots. The drive produces numerous dynamical sweet spots at different dc flux values, as opposed to only one in the static case.}
\label{Fig1}
\end{figure}

 Low-frequency noise has been a limiting factor for dephasing times of many solid-state based qubits \cite{Charge_qubit,Nakamura_charge_1/f,JSTsai_flux_1/f,JClarke_flux_qubit_1/f,Ithier_decoherence_analysis,Koch_transmon_theory,Fluxonium_hc,Schuster_heavy_fluxonium_control,Rigetti_flux_sweet_spot,IBM_frequency_tunable_transmon,Barends_xmon,Schuster_heavy_fluxonium,Manucharyan_heavy_fluxonium,Nanowire_fluxonium,Martinis_decoherence_bias_noise,Martinis_1/f_Crossover,Devoret_fluxonium_molecule,Oliver_flux_qubit_dd,Didier_dynamical_sweet_spot,Rigetti_ac_sweet_spot,Rigetti_ac_sweet_spot_exp,Rigetti_broad_band_noise,Coppersmith_two_qubit_sweet_spot,Sillanpaa_hybrid_circuit,Wunderlich_dressed_qubit_nature,Wunderlich_dressed_qubit_NJP,Friesen_charge_qubit_Floquet,Gossard_spin_qubit,Vandersypen_spin_qubit_echo,Yacoby_spin_qubit_CPMG,Sadeghpour_trapped_ion_1/f_theory,Haffner_ion_trap_heat_rate,Blatt_ion_trap_ef_noise,Matthew_bifluxon,Yu_critical_current,Clarke_critical_current}. Superconducting qubits especially suffer from 1/$f$ charge and flux noise \cite{Charge_qubit,Nakamura_charge_1/f,JSTsai_flux_1/f,JClarke_flux_qubit_1/f,Ithier_decoherence_analysis,Koch_transmon_theory,IBM_frequency_tunable_transmon,Barends_xmon,Schuster_heavy_fluxonium,Manucharyan_heavy_fluxonium,Fluxonium_hc,Nanowire_fluxonium,Schuster_heavy_fluxonium_control,Martinis_decoherence_bias_noise,Martinis_1/f_Crossover,Devoret_fluxonium_molecule,Oliver_flux_qubit_dd,Rigetti_flux_sweet_spot,Didier_dynamical_sweet_spot,Rigetti_ac_sweet_spot_exp,Rigetti_ac_sweet_spot,Rigetti_broad_band_noise}. A conventional way to improve dephasing times is to operate the qubit at so-called sweet spots \cite{Ithier_decoherence_analysis,Koch_transmon_theory,Fluxonium_hc,Schuster_heavy_fluxonium_control,Rigetti_flux_sweet_spot}. These sweet spots correspond to extrema of the qubit's transition frequency \cite{Ithier_decoherence_analysis}, see Fig.~\ref{Fig1}(a) for an example. Another established  method for improving dephasing times is dynamical-decoupling (DD) \cite{Oliver_flux_qubit_dd, Lidar_DD_2005,Lidar_transmon_dd, DasSarma_DD, Lloyd_DD_theory,Uhrig_DD}, which is well-known in the context of NMR echo sequences \cite{Hahn_echo,Carr_DD,Meiboom_DD}, and has been successfully applied to superconducting qubits \cite{Oliver_flux_qubit_dd,Lidar_transmon_dd}.

In this paper, we propose a qubit protection protocol based on \textit{dynamical} sweet spots \cite{Rigetti_ac_sweet_spot,Rigetti_ac_sweet_spot_exp,Rigetti_broad_band_noise,Didier_dynamical_sweet_spot,Coppersmith_two_qubit_sweet_spot,Sillanpaa_hybrid_circuit}. Inspired by static sweet-spot operation and dynamical decoupling, this protocol employs a periodic drive to mitigate the dephasing usually induced by $1/f$ noise. 
Utilizing Floquet theory, we show that dynamical sweet spots represent extrema in the qubit's quasi-energy difference, and thus generalize the concept of static sweet spots [Fig.~\ref{Fig1}(b)]. Notably, dynamical sweet spots are generally not isolated points, but rather form extended sweet-spot manifolds in parameter space. The multi-dimensional nature of dynamical sweet spots provides additional freedom to tune qubit properties such as the transition frequency while maintaining dynamical protection. We show that dynamical sweet-spot operation can simultaneously yield both long depolarization ($T_1$) and pure-dephasing times ($T_\phi$).

This protection scheme can also be interpreted as a continuous version of DD \cite{Knill_DD}. Here, the sequences of ultra-short pulses widely used in many DD experiments are replaced by a periodic drive on the qubit, which is much easier to realize experimentally. In addition to earlier explorations in this direction \cite{Ithier_decoherence_analysis,Coppersmith_two_qubit_sweet_spot,Pan_transmon_dd,Oliver_rotating_frame_relaxation,Hu_drive_spin_qubit,Smirnov_Raib_decoherence,Friesen_charge_qubit_Floquet,Sillanpaa_hybrid_circuit,Wunderlich_dressed_qubit_NJP,Wunderlich_dressed_qubit_nature,Rigetti_ac_sweet_spot,Rigetti_ac_sweet_spot_exp,Rigetti_broad_band_noise,Didier_dynamical_sweet_spot}, we here provide a systematic and general framework for locating dynamical sweet-spot manifolds in the control parameter space. This framework is general enough to cover a variety of qubit systems beyond the specific example discussed here, and can be adapted to different types of drives as well as noise environments. Indeed, some of the previously developed protection schemes \cite{Ithier_decoherence_analysis,Didier_dynamical_sweet_spot,Rigetti_ac_sweet_spot,Rigetti_ac_sweet_spot_exp,Rigetti_broad_band_noise,Pan_transmon_dd,Oliver_rotating_frame_relaxation,Coppersmith_two_qubit_sweet_spot,Hu_drive_spin_qubit,Smirnov_Raib_decoherence} based on qubit-frequency modulation or on-resonant Rabi drives, can be understood as special limits of the framework presented here (see Supplemental Material \cite{supplementary} for details).  The theoretical approach we develop allows us not only to predict the improvement of pure-dephasing times, but also to assess how dynamical depolarization times are affected by the driving. We further show that the protection scheme is compatible with concurrent single- and two-qubit gate operations, thus making it suitable for both quantum information storage and processing. In a companion  experimental paper \cite{Houck_Floquet_exp}, our theoretical results are demonstrated  to lead to a significant improvement in the dephasing time of a flux-modulated fluxonium qubit.

The paper is structured as follows. Throughout this paper, we consider the superconducting fluxonium qubit \cite{Fluxonium_Devoret,Devoret_fluxonium_quasiparticle,Devoret_fluxonium_molecule,Pop_granular_fluxonium,Manucharyan_heavy_fluxonium,Schuster_heavy_fluxonium,Fluxonium_hc,Schuster_heavy_fluxonium_control,Matthew_bifluxon,Nanowire_fluxonium} a platform for illustrating the dynamical-protection protocol. We begin in Section II, with a description of this qubit and discuss its static coherence times in the absence of external driving. In Section III, we employ Floquet theory to derive  expressions for the dynamical coherence times of the driven qubit. We then evaluate these expressions numerically in Section IV, and discuss the nature of the observed dynamical sweet spots associated with increases in coherence times. 
We illustrate how to perform gate operations and readout on such a driven qubit (Floquet qubit) in Section V, and further show how to implement a Floquet two-qubit gate in Section VI. Finally, we present our conclusions in Section VII.

\section{Two-level system subject to $\mathbf{1/f}$ noise}

For concreteness, we discuss the application of the protection scheme to the most recent genereation of fluxonium qubits \cite{Fluxonium_Devoret,Devoret_fluxonium_quasiparticle,Devoret_fluxonium_molecule,Pop_granular_fluxonium,Manucharyan_heavy_fluxonium,Schuster_heavy_fluxonium,Fluxonium_hc,Schuster_heavy_fluxonium_control,Matthew_bifluxon,Nanowire_fluxonium}, though the general theoretical framework is not limited to this choice. Fluxonium qubits biased close to half-integer flux exhibit attractive properties including increased coherence times as compared to other superconducting qubits \cite{Manucharyan_heavy_fluxonium,Schuster_heavy_fluxonium,Fluxonium_hc,Schuster_heavy_fluxonium_control,Houck_Floquet_exp,Devoret_fluxonium_quasiparticle}. When the external flux bias in the circuit loop is tuned to the sweet spot at half a flux quantum, both depolarization and dephasing times of the fluxonium circuit exceed 100$\,\mu$s and is strongly anharmonic  \cite{Fluxonium_hc, Schuster_heavy_fluxonium_control,Houck_Floquet_exp}. However, this sweet spot is point-like, and the qubit regains  sensitivity to 1/$f$ flux noise when the external flux is tuned slightly away from the half-integer point \cite{Schuster_heavy_fluxonium,Manucharyan_heavy_fluxonium,Fluxonium_hc,Schuster_heavy_fluxonium_control,Devoret_fluxonium_molecule,Nanowire_fluxonium}. This sensitivity leads to increased pure dephasing of the fluxonium qubit. We note that in multi-loop circuits with shared inductance, the local nature of 1/$f$ flux noise produces further constraints on the existence of static flux sweet spots \cite{Devoret_fluxonium_molecule}. It is thus desirable to find alternative means of  protection from 1/$f$ flux noise, in order to improve coherence times, and advance the promising direction of quantum-information processing with fluxonium qubits.

Our protection scheme is based on introducing a modulation of the external flux close to the static sweet spot, i.e.,\ $\phi_\text{ext}(t) = \phi_\mathrm{ac}\cos(\omega_\mathrm{d}t)+\phi_\mathrm{dc}$. Here, $\phi_\mathrm{ext}=2\pi \Phi_\text{ext}/\Phi_0$ denotes the reduced external flux, $\Phi_0$ is the flux quantum, and $\phi_\mathrm{ac}$, $\phi_\mathrm{dc}$ are its ac modulation amplitude and dc offset, respectively.
Upon truncation to two levels, the effective Hamiltonian of the driven fluxonium circuit is given by
\begin{align}
\hat{H}_\mathrm{q}(t) = \frac{\Delta}{2}\hat{\sigma}_x + \left(A\cos\omega_\mathrm{d}t+\frac{B}{2}\right)\hat{\sigma}_z,
\label{H_q_two_level}
\end{align}
see Appendix A for details. Here, $\Delta$ denotes the qubit splitting at $\phi_\mathrm{dc}=\pi$, and $A\propto\phi_\mathrm{ac}$ and $B\propto\delta \phi_\mathrm{dc} = \phi_\mathrm{dc}-\pi$ are the effective  drive amplitude and  dc bias away from the static sweet spot. Note that we also set $\hbar=1$ in this expression. The resulting eigenenergies of the static qubit ($\phi_\mathrm{ac}=0$) are plotted in Fig.~\ref{Fig1}(a) as a function of $\phi_\mathrm{dc}$. The full Hamiltonian including the qubit-bath coupling is given by $\hat{H} = \hat{H}_\mathrm{q}(t) + \hat{H}_\mathrm{B} + \hat{H}_\mathrm{int}$, where $H_\mathrm{B}$ and $H_\mathrm{int}$ denote the Hamiltonian of the bath and the qubit-bath interaction.  We consider two major noise sources that often limit fluxonium coherence times: 1/$f$ flux noise and dielectric loss \cite{Manucharyan_heavy_fluxonium,Fluxonium_hc,Nanowire_fluxonium, Martinis_decoherence_bias_noise,Schuster_heavy_fluxonium_control,Martinis_1/f_Crossover}. The corresponding interaction Hamiltonian thus takes the form $\hat{H}_\text{int}=(\hat{\eta}_\mathrm{f}+\hat{\eta}_\mathrm{d})\hat{\sigma}_z$, where $\hat{\eta}_\mathrm{f}$ and $\hat{\eta}_\mathrm{d}$ are the bath operators through which 1/$f$ flux noise and dielectric loss are induced. The noise spectra characterizing these channels are given by $S_\mathrm{f}(\omega)= \mathcal{A}^2_\mathrm{f} |\omega/2\pi|^{-1}$ and $S_\mathrm{d}(\omega)=\alpha(\omega, T)\mathcal{A}_\mathrm{d}(\omega/2\pi)^2$ \cite{footnote_one_over_f}. Here, $\alpha(\omega, T) = |\coth(\omega/2k_\mathrm{B}T)+1|/2$ is a thermal factor, $k_\mathrm{B}$ and $T$ denote the Boltzmann constant and temperature, and $\mathcal{A}_\mathrm{f}$ and $\mathcal{A}_\mathrm{d}$ denote the noise amplitudes.

As reference for our discussion of dynamical coherence times in Secs.\ III and IV, we first briefly review the static coherence times of the undriven qubit.  The decoherence rates depend on the matrix elements of the qubit operator coupling to the noise as well as  the noise spectra. For a non-singular noise spectrum $S(\omega)$, the rates for relaxation, excitation and pure dephasing are 
\begin{align}\label{static_depolarization}
    \gamma_\mp =&\,|\sigma_{z}^{ge}|^2\,S(\pm \Omega_{ge}),\\
    \gamma_\phi=&\,|\sigma_{z}^{ee} - \sigma_{z}^{gg}|^2\,S(0)/2.
    \label{static_dephasing}
\end{align}
As usual, these expressions are derived within Bloch-Redfield theory. Here, $\vert g\rangle$ and $\vert e\rangle$ denote the qubit ground and first excited state, $\Omega_{ge}=\sqrt{\Delta^2+B^2}$ the corresponding eigenenergy difference, and $\sigma_{z}^{ll'} \equiv \langle l\vert \hat{\sigma}_z\vert l'\rangle$ ($l,l' = g,e$) the relevant matrix elements. (Since these matrix elements will appear rather frequently, we choose to introduce this slightly more compact notation.) The quantity $|\sigma_{z}^{ee} - \sigma_{z}^{gg}|$ governing the pure-dephasing rate $\gamma_\phi$ turns out to be proportional to the flux dispersion of the eigenenergy difference $|\partial\Omega_{ge}/\partial\phi_\mathrm{dc}|$, in agreement with the well-known proportionality $\gamma_\phi\propto (\partial\Omega_{ge}/\partial\phi_\mathrm{dc})^2$ \cite{Ithier_decoherence_analysis,Koch_transmon_theory}. For the realistic noise spectrum  $S(\omega) = S_\mathrm{d}(\omega) + S_\mathrm{f}(\omega)$, however, there is a divergence at $\omega=0$ from the $1/f$ flux noise. In this case, our evaluation of dephasing times includes careful consideration of frequency cutoffs, see Refs.\ \cite{Ithier_decoherence_analysis, Groszkowski_Zero_pi_theory,Koch_transmon_theory, Rigetti_ac_sweet_spot}.

The resulting coherence times differ characteristically according to the flux bias. Away from the flux sweet spot, the qubit has wavefunctions with disjoint support [insets of Fig.~\ref{Fig1}(a)]. This leads to a suppression of the coefficient $|\sigma_{z}^{ge}|^2$ relevant for relaxation and excitation, and hence to a relatively long depolarization time of $T_1 = 770\,\mu\text{s}$ (see Table\ \ref{tabel1} caption for our specific choice of parameters). The pure-dephasing time of $T_{\phi} =0.88 \,\mu\text{s}$ is rather short, on the other hand, since the flux dispersion $\partial\Omega_{ge}/\partial\phi_\mathrm{dc}$ is significant away from the flux sweet spot. At the flux sweet spot, the situation changes:  disjointness of eigenfunctions is lost and depolarization times are correspondingly shorter, $T_1 = 360\,\mu$s. Since the flux dispersion $\partial\Omega_{ge}/\partial\phi_\mathrm{dc}$ vanishes at the sweet spot, the qubit is less sensitive to 1/$f$ noise, resulting in a pure-dephasing time exceeding $10\,$ms \cite{Fluxonium_hc, Devoret_two_Cooper_pair_qubit}, limited only by second-order contributions from 1/$f$ flux noise.  In realistic settings, the pure-dephasing times will be limited by other sources including photon shot noise, critical current noise, etc \cite{Rigetti_photon_shot,Schoelkopf_photon_shot_noise,Yu_critical_current,Clarke_critical_current}.

\section{Dynamical coherence times of the driven qubit}
The analysis of coherence times must be modified when including a periodic drive acting on the qubit. Based on an open-system Floquet theory \cite{Hanggi_Floquet_Markovian, Breuer_open_quantum_system}, the coherence times are most conveniently characterized in the basis formed by the qubit's Floquet states.
The quasi-energies $\epsilon_j$ and time-periodic Floquet states $\vert w_j(t)\rangle$ of the driven qubit, labeled by index $j$, are the counterparts of the ordinary eigenstates and eigenenergies in the undriven case \cite{Chu_flux_qubit_Floquet,Lupascu_flux_qubit_Floquet,Grifoni_dissipative_Floquet,Breuer_open_quantum_system}. They are obtained as solutions of the Floquet equation 
\begin{align}
    \left[\hat{H}_\mathrm{q}(t)-i \frac{\partial}{\partial t}\right]\vert w_j(t)\rangle = \epsilon_j\vert w_j(t)\rangle.
    \label{Floquet_eq}
\end{align}
In the absence of noise, the evolution operator 
$U_\mathrm{q}(t,0) = \sum_{j=0,1}\vert w_j(t)\rangle\langle w_j(0)\vert \exp(-i\epsilon_jt)$ governs the evolution of  the driven qubit. As a result, the population in each Floquet state remains invariant, while the relative phase accumulates at a rate given by the quasi-energy difference $\epsilon_{01}\equiv \epsilon_1-\epsilon_0$.

The matrix elements and noise frequencies relevant for the decoherence of the driven qubit crucially differ from the undriven case. By casting the expression for the decoherence rates into the form 
\begin{equation}\label{filtering}
    \gamma_\mu=\int^{\infty}_{-\infty}F_{\mu}(\omega)S(\omega)\mathrm{d}\omega,
\end{equation} these differences are conveniently captured as a change in the filter function $F_\mu(\omega)$  \cite{Martinis_decoherence_bias_noise, sidenote_pure_dephasing}. Here, $\mu=\mp, \phi$ denotes the different noise channels corresponding to relaxation, excitation and pure dephasing.

For the undriven qubit, $F_\mu(\omega)$ is strongly peaked at the filter frequencies $\omega=\pm\Omega_{ge}$ and $\omega=0$. The  integrated peak areas, referred to as weights, are given by the quantities $\vert\sigma_{z}^{ge}\vert^2$, $\vert\sigma_{z}^{eg}|^2$ and $|\sigma_{z}^{ee}-\sigma_{z}^{gg}|^2/2$ associated with the three noise channels. By contrast, for the driven qubit, $F_\mu(\omega)\sim \sum_k |g_{k\mu}|^2 \delta (\omega-\bar{\omega}_{k\mu} )$ develops additional sideband peaks, resulting in filter frequencies $\bar{\omega}_{k\mp}=\pm\epsilon_{01}+k\omega_\mathrm{d}$ and $\bar{\omega}_{k\phi} = k\omega_\mathrm{d}$ ($k\in\mathbb{Z}$). The corresponding weights are $|g_{k\mp}|^2$ and $2|g_{k\phi}|^2$, where
\begin{align}\label{gkplus}
    g_{k+} =& \frac{\omega_\mathrm{d}}{2\pi} \int_0^{2\pi/\omega_\mathrm{d}} dt\, e^{ik \omega_\mathrm{d} t}\, \tr \left(\sigma_z |w_0(t)\rangle\langle w_1(t)|\right),
\end{align}
and similar expressions hold for $g_{k,-}$ and $g_{k,\phi}$ (see Appendix B). Expressed in terms of these weights, the decoherence rates are given by
\begin{align}\label{rates}
\gamma_{\mp} =&\, \sum_{k\in\mathbb{Z}}|g_{k\mp}|^2S(k\omega_\mathrm{d}\pm\epsilon_{01}),\\
\gamma_{\phi} =&\, \mathcal{A}_f|2g_{0\phi}|\sqrt{|\ln \omega_\text{ir}t_\mathrm{m}|}+\sum_{k\neq 0}2|g_{k\phi}|^2S(k\omega_\mathrm{d}),
\label{gammaphi}
\end{align}
where the infrared cutoff $\omega_\text{ir}$ and a finite measurement time $t_\mathrm{m}$ are introduced to regularize the singular behavior of the $1/f$ noise spectrum (see Appendix C). We note that Eqs.~\eqref{rates} and \eqref{gammaphi} are based on the rotating-wave approximation described in Appendix B. Further, it is instructive to mention that the expressions for the dynamical rates [Eqs.\ \eqref{rates} and \eqref{gammaphi}] reduce to the rates  obtained for the static case when the drive is switched off $(A=0)$. To see this, note that the Floquet states are time-independent for $A=0$. As a result, the filter weights vanish for $k\not=0$ [see, for example, Eq.\ \eqref{gkplus}]. The remaining quantities to be identified are simply: $ \pm\Omega_{ge}\leftrightarrow \bar{\omega}_{0\mp}$, $0\leftrightarrow \bar{\omega}_{0\phi}$, and $\vert\sigma_{z}^{ge}\vert^2$, $\vert\sigma_{z}^{eg}|^2,\,|\sigma_{z}^{ee}-\sigma_{z}^{gg}|^2/2 \; \leftrightarrow \; |g_{0\mp}|^2$ and $2|g_{0\phi}|^2$.

\section{Dynamical sweet spots}

We numerically calculate the dynamical coherence times as a function of drive frequency and amplitude, for a flux bias fixed close to the half-integer point. Results of pure-dephasing times are presented in Fig.~\ref{Fig2} (a), and show broad regions where $T_\phi= \gamma_\phi^{-1}$ remains close to the value of the undriven qubit, but also exhibit well-defined maxima where pure-dephasing times exceed $1\,$ms. (This value is based on the noise sources included in our analysis, but may ultimately be limited by other noise channels.)  Fig.~\ref{Fig2}(b) shows the corresponding depolarization times $T_1=(\gamma_++\gamma_-)^{-1}$. While there are point-like dropouts of $T_1$ for certain drive parameters, the majority of the predicted $T_1$'s are well over $100\,\mu$s. Table \ref{tabel1} summarizes the coherence times for two example working points \onecircle{} and \twocircle{} aligned with local maxima of $T_\phi$. The pure-dephasing times for both points exceed $1\,$ms, much longer than those of the undriven qubit.  The depolarization times at those two points are around $500\,\mu$s, which are favorable compared to the static sweet-spot value. 

\begin{table}[h]
\caption{Calculated coherence times for four operating points. Without a drive and operated away from the sweet spot ($\delta\phi_\mathrm{dc}=0.02$), the qubit has the longest $T_1$ but the shortest $T_\phi$. At the sweet spot, this behavior reverses: the static $T_1$ reaches maximum values, but $T_\phi$ becomes relatively short. By comparison, Floquet operation at dynamical sweet spots yields $T_1$ and $T_\phi$ values that do not exceed the static maximal values, but are well above the minimal ones. [Underlying parameter choices: The noise amplitudes used are
$\mathcal{A}_\mathrm{d}=\pi^2\tan\delta_\mathrm{C}| \tilde{\varphi}_{ge}| ^2/E_\mathrm{C}$
and 
$\mathcal{A}_\mathrm{f} = 2\pi \delta_\mathrm{f}E_\mathrm{L} |\tilde{\varphi}_{ge}|$, where $\tilde{\varphi}_{ge}=\langle g|\hat{\varphi}\vert e\rangle$ is evaluated at $\phi_\mathrm{dc}/2\pi=0.5$, and $\hat{\varphi}$ is the phase operator. We assume the loss tangent $\tan\delta_\mathrm{C}=1.1 \times 10^{-6}$, flux-noise amplitude $\delta_\mathrm{f} = 1.8\times 10^{-6}$, and a temperature of $15\,$mK. The noise parameters used here are typical for recent fluxonium experiments, see e.g., Ref. \cite{Fluxonium_hc,Schuster_heavy_fluxonium_control}.]}
\begin{center}
\begin{ruledtabular}
\begin{tabular}{ccc}
Working points & $T_1$ ($\mu$s) & $T_\phi$ ($\mu$s)  \\\hline
Away from the static sweet spot & \tikzmark{a}\textbf{770} & \tikzmark{c}0.88\\ 
Dynamical sweet spot \onecircle{} &590 & 1200\\ 
Dynamical sweet spot \twocircle{} &490 &1750 \\ 
Static sweet spot & \tikzmark{b}360 & \tikzmark{d}$>\!\textbf{10}^\textbf{4}$ \\ 
\end{tabular}
\begin{tikzpicture}[overlay, remember picture]
    \draw [-{latex[length=2mm]}] ([xshift=-6pt]{pic cs:b})  to ([xshift=-6pt,yshift=2pt]{pic cs:a});
    \draw [-{latex[length=2mm]}] ([xshift=-8.5pt, yshift = 2pt]{pic cs:c})  to ([xshift=-6pt]{pic cs:d});
  \end{tikzpicture}
\end{ruledtabular}
\end{center}
\label{tabel1}
\end{table}

\subsection{Asymptotic behavior of sweet manifolds for weak and strong drive}
\begin{figure}
\includegraphics[width=\columnwidth]{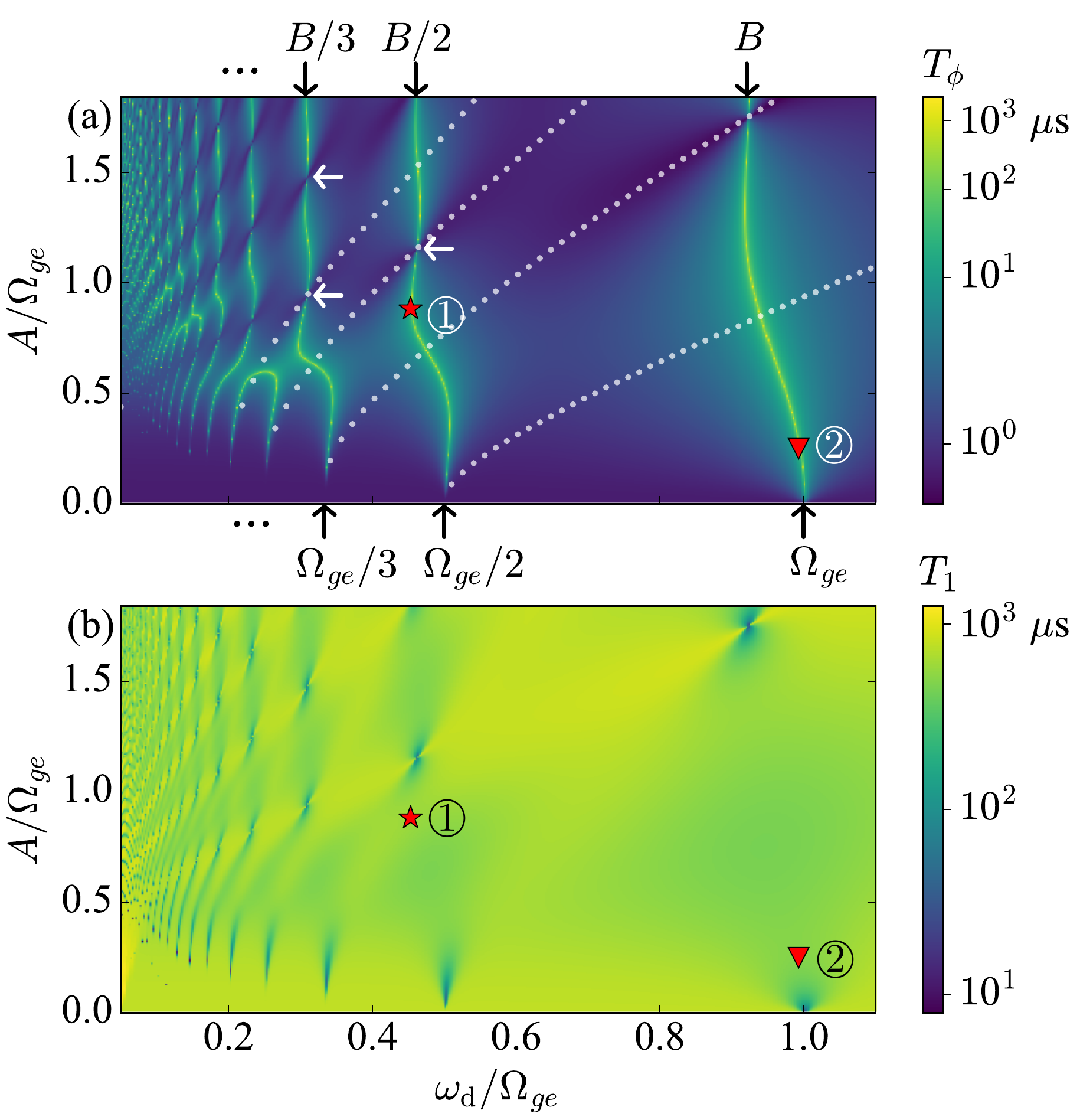}
\includegraphics[width=\columnwidth]{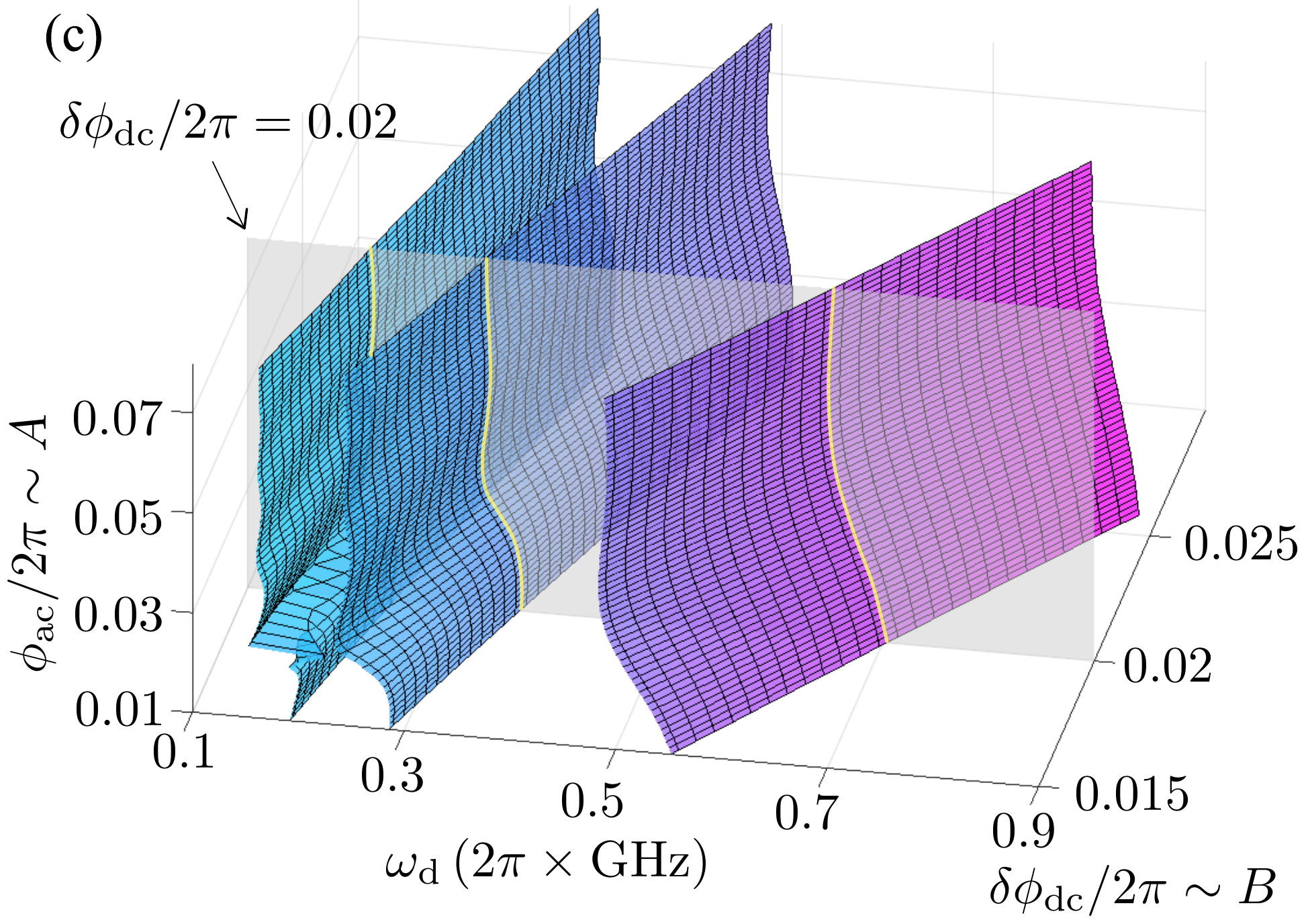}
\caption{(a) Dynamical pure-dephasing time $T_\phi$ (color-coded) as a function of drive frequency $\omega_\mathrm{d}$ (horizontal axis) and drive amplitude $A$ (vertical axis). Results are calculated via Eq.~(\ref{gammaphi}) for flux  $\delta\phi_\mathrm{dc}/2\pi=0.02$. The curves visible by their bright-yellow coloring are the dynamical sweet spots  characterized by large $T_\phi$. In the weak ($A\ll \Omega_{ge}$) and strong drive limit ($A\gtrsim\Omega_{ge}$) these curves asymptotically line up with $\omega_\mathrm{d} = \Omega_{ge}/m$ and $\omega_\mathrm{d} = B/m$ (black arrows). The curves formed by the dynamical sweet spots are interrupted by cuts marked by  white arrows. The overlaid white dotted curves depict the ac dynamical sweet spots corresponding to $\partial\epsilon_{01}/\partial\phi_\mathrm{ac}=0$. (b) Depolarization time $T_1$ calculated by Eq.~\eqref{rates}. The majority of the obtained $T_1$ values are at the order of 500$\,\mu$s, except for point-like dropouts shown by the dark-blue coloring. 
The red star and triangle specify the dynamical sweet spots \onecircle{} and \twocircle{}. The noise parameters used for the calculation are given in the caption of Table \ref{tabel1}.
(c) Sweet-spot manifolds embedded in the 3d parameter space, with axes corresponding to $\omega_\mathrm{d}$, $\delta\phi_\mathrm{dc}$ and $\phi_\mathrm{ac}$. The semi-transparent plane given by $\delta\phi_\mathrm{dc}/2\pi=0.02$, intersects the manifolds and thus yields the sweet-spot curves shown in (a) as cross sections (bright-yellow coloring).}
\label{Fig2}
\end{figure}

The regions where $T_\phi$ becomes maximal, form curves in the plane spanned by the drive frequency and amplitude, with distinct behavior in the two regimes of weak driving, $A\ll \Omega_{ge}$ [bottom of Fig.\ \ref{Fig2}(a)],  and strong driving, $A\gtrsim\Omega_{ge}$ [top of Fig.\ \ref{Fig2}(a)]. These curves are the cross-sections of the sweet-spot manifolds at a fixed dc flux value $\delta\phi_\text{dc}$, see Fig.~\ref{Fig2}(c). The curves of maximal pure-dephasing times show simple asymptotic behavior in these two limits, where they approach fixed-frequency intercepts in the $A$--$\omega_\text{d}$ plane. In the strong-drive limit, these curves are interrupted by cuts (see white arrows) where the width of the peak in $T_\phi(\omega_\text{d})|_{A=\text{const}}$ goes to zero. No such cuts are present in the weak-drive regime; rather, the peak width gradually decreases as drive amplitude $A$ is lowered.

This behavior of pure-dephasing times of the driven qubit can be explained and approximated analytically using Floquet theory.
Away from dynamical sweet manifolds, $T_\phi$ is limited by contributions from the (regularized) pole of the $1/f$ spectrum, see the first term on the right-hand side of Eq.\ \eqref{gammaphi}. Thus, $\gamma_\phi\propto |g_{0\phi}|$ which in turn can be shown to be proportional to $\partial \epsilon_{01}/\partial \phi_\text{dc}$ (see derivation in Appendix D), i.e., the dynamical  flux-noise sensitivity given by the flux dispersion of the Floquet quasi-energy difference. We emphasize that this result is analogous to the more familiar case of the undriven qubit, where the pure-dephasing rate is proportional to the static flux dispersion $\partial \Omega_{ge}/\partial \phi_\text{dc}$, with quasi-energies replaced by eigenenergies.

Pure-dephasing times are maximal whenever
\begin{equation}\label{eq:g0phi}
 \frac{\partial\epsilon_{01}}{\partial\phi_\mathrm{dc}} \propto |g_{0\phi}| =0,
\end{equation}
which generically occurs at avoided crossings in the extended quasi-energy spectrum [Fig.~\ref{Fig1}(b)]. The latter, analogous to the extended Brillouin zone in spatially periodic systems, consists of  the extended set of quasi-energies  $\epsilon_{j,n}=\epsilon_j+n\omega_\mathrm{d}$ ($n\in\mathbb{Z}$) \cite{Breuer_open_quantum_system, Chu_flux_qubit_Floquet,Petta_double_quantum_well_Floquet}. This extended spectrum shows numerous avoided level crossings, and hence a multitude of regions of maximal $T_\phi$. These operation points are called \textit{dynamical sweet spots}; see Refs.~\cite{Didier_dynamical_sweet_spot,Coppersmith_two_qubit_sweet_spot,Sillanpaa_hybrid_circuit} for previous studies of this concept. Here, we specifically use this term to refer to the working points where the derivative of $\epsilon_{01}$ with respect to the noise parameter vanishes. 

As shown in Fig.~\ref{Fig2}(a), these spots form a set of curves with maximal $T_\phi$ in the $A$-$\omega_\text{d}$ plane. Once we  account for the additional perpendicular axis representing $B$, we find that each curve is the cross-section of a continuous surface of sweet spots, 
which we refer to as a sweet-spot manifold. The locations of sweet spots can be predicted in the limits of weak and strong drive,  by treating either the drive $A \cos \omega_\mathrm{d} t \,\hat{\sigma}_z$ or the transverse qubit Hamiltonian $\Delta\, \hat{\sigma}_x/2$ perturbatively.

\emph{Weak-drive limit.---} For $A\ll \Omega_{ge}$, the unperturbed quasi-energies are the static eigenenergies up to the addition of integer multiples of the drive frequency, $\epsilon_{\pm,n}=\pm\Omega_{ge}/2+n\omega_\mathrm{d}\,(n \in\mathbb{Z})$. Two levels exhibit a crossing, $\epsilon_{+,n}=\epsilon_{-,n'}$, whenever the qubit frequency is an integer multiple of the drive frequency, $\Omega_{ge}=m \omega_\mathrm{d}$ where $m=n'-n\in\mathbb{N}$. The perturbation lifts these degeneracies and generates avoided crossings. As a result, the sweet spots observed towards the bottom of Fig.\ \ref{Fig2}(a) asymptotically take the form of vertical lines at drive frequencies set by  $\omega_\mathrm{d}=\Omega_{ge}/m$. 
The width of maxima in $T_\phi(\omega_\mathrm{d})$ is significant for the issue of parameter deviations: the wider the maximum, the larger is the robustness of the coherence-time increase with respect to small detunings from the dynamical sweet spot. This width is proportional to the gap size of the avoided crossing and given by  $\Delta_{m} \approx A^m|\sin\theta\cos^{m-1}\theta|/\omega_\mathrm{d}^{m-1}(m-1)!$ in the weak-drive limit, where $\theta = \tan^{-1} (\Delta/B)$   (see derivation in Appendix D). For decreasing drive strength $A$, the width narrows with $\sim A^m$ consistent with the behavior observed in Fig.~\ref{Fig2}(a).

\emph{Strong-drive limit.---} For $A\gtrsim\Omega_{ge}$, the unperturbed quasi-energies are given by $\epsilon_{\pm,n} = \pm B/2 +n \omega_\mathrm{d}$ ($n\in\mathbb{Z}$) and  cross whenever $B=m\omega_\mathrm{d}$ ($m\in\mathbb{N}$) \cite{Chu_flux_qubit_Floquet,Nori_strong_drive_Floquet,Grifoni_dissipative_Floquet,Paraoanu_frequency_modulation_review}. The perturbation $\Delta\hat{\sigma}_x/2$ generically opens up gaps. The resulting sweet spots asymptotically line up with vertical intercepts $\omega_\mathrm{d}=B/m$, as shown in Fig.~\ref{Fig2}(a). The proportionality between the width of the maximal $T_\phi$ and the gap size also holds in this limit, with the latter given by $\Delta_{m} \approx \Delta| J_m(2A/\omega_\mathrm{d})|$ (see derivation in Appendix D). Whenever $\Delta_{m}=0$, i.e., $2A/\omega_\mathrm{d}$ is one of the roots of the Bessel function $J_m$, the width goes to zero and the sweet-spot curve is interrupted with a cut.  

The dropouts of $T_1$ visible in Fig.~\ref{Fig2}(b) are similarly related to the vanishing gap size of the avoided crossings. If the gap opening of the avoided crossing, i.e., the quasi-energy difference of the qubit at the dynamical sweet spot becomes smaller, the terms $|g_{0\mp}|^2S(\pm \epsilon_{01})$ in Eq.~(\ref{rates}) rapidly increase in magnitude as the regularized divergence of $S(\omega)$ is sampled. In other words, the low-frequency 1/$f$ noise significantly suppresses the dynamical $T_1$ whenever $\epsilon_{01}$ vanishes. Therefore, the low-$T_1$ features observed in Fig.~\ref{Fig2}(b) match the locations of strong narrowing of the maximal $T_\phi$ regions in Fig.~\ref{Fig2}(a) ($\Delta_m\to0$), including the discussed cuts in the strong-drive limit, as well as the gradual narrowing in the weak-drive limit. In our example, the widths of $T_\phi$ peaks surrounding the sweet-spot manifolds are generally sufficiently wide and, hence, gap sizes sufficiently large, such that 1/$f$ flux noise does not limit the dynamical $T_1$.

Driving the qubit, as discussed above, efficiently decouples it from the low-frequency dc flux noise. Recent experimental evidence points to the relevance of additional noise in the ac drive amplitude \cite{Rigetti_ac_sweet_spot, Rigetti_ac_sweet_spot_exp,Rigetti_broad_band_noise,Didier_dynamical_sweet_spot,Houck_Floquet_exp}. While the magnitude and  power spectrum of this noise are not well established, it is useful to note that there exist simultaneous sweet spots for the dc and ac flux amplitude, $\partial\epsilon_{01}/\partial\phi_\mathrm{dc} = \partial\epsilon_{01}/\partial \phi_\mathrm{ac} =0$. These doubly-sweet spots correspond to intersection points of the white dotted curves  ($\partial\epsilon_{01}/\partial \phi_\mathrm{ac} =0$) and the underlying dc sweet-spot curves obtained for $\partial\epsilon_{01}/\partial \phi_\mathrm{dc} =0$ [see Fig.~\ref{Fig2}(a)] \cite{sweet_spot_stability}. Depending on the magnitude of this ac noise, we expect such doubly-sweet spots to form the optimal working points.

\begin{figure}
\includegraphics[width = 8.6cm]{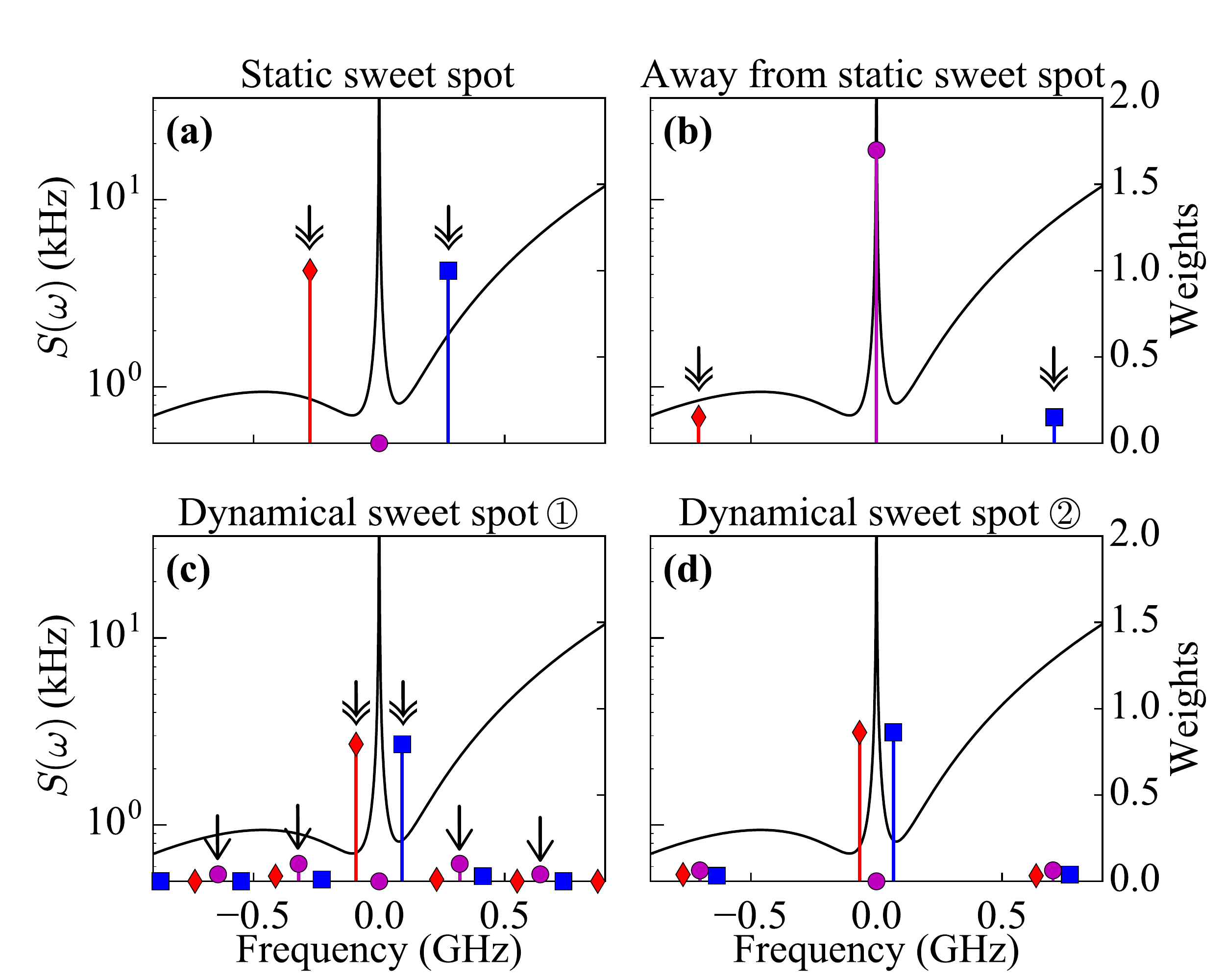}
\caption{Noise spectra and filter weights centered at the corresponding filter frequencies. The panels refer to four different working points:  (a) the static sweet spot, (b) static operation away from the sweet spot ($\phi_\mathrm{dc}/2\pi=0.52$), (c)-(d)dynamical sweet-spot operation at the working points \onecircle{} and  \twocircle{}. The symbols represent relaxation (blue squares), excitation (red diamonds), and pure dephasing (purple dots). The noise spectrum is plotted concurrently in (a)-(d). The positions of filter frequencies and the associated filter weights determine which components of the noise spectrum  contribute significantly to the rates $\gamma_\pm$ and $\gamma_\phi$ [see Eqs.~\eqref{rates} and \eqref{gammaphi}]. (See the main text for the discussion of filter frequencies marked by arrows.)}

\label{Fig3}
\end{figure}

\subsection{Interpretation of coherence times in terms of filter functions}

We observe that, although the obtained dynamical $T_1$ and $T_\phi$ times in the sweet manifolds do not exceed the maximal values at the two static working points (see Table \ref{tabel1}), they are well above the corresponding static minimal values. To understand this behavior, it is instructive to interpret the decoherence rates in terms of the sampling of the noise spectrum by the filter function [Eq.\ \eqref{filtering}]. For that purpose, Fig.~\ref{Fig3} shows the noise spectrum $S(\omega)$ along with information characterizing the filter function $F_\mu(\omega)$ in terms of the relevant filter frequencies and weights. 
The noise spectrum (black curve) is peaked at $\omega=0$ due to the $1/f$ flux noise; away from that peak, dielectric loss dominates. For each filter frequency, the value of the corresponding filter weight is shown and marked by symbols distinguishing between depolarization and pure-dephasing channels. While there are  only three filter frequencies in the static case, the dynamical case in principle produces an infinite number of filter frequencies $\bar{\omega}_{k\mu}$. The appearance of additional filter frequencies corresponds to the sampling of the noise spectral density at sideband frequencies, a point previously discussed for weakly driven systems in  Refs.~\cite{Hu_drive_spin_qubit,Oliver_rotating_frame_relaxation,Smirnov_Raib_decoherence}.

We first interpret the behavior of pure-dephasing times. The weight related to filter frequency $\bar{\omega}_{0\phi}$ is suppressed to zero for both static and dynamical sweet spots [see Fig.~\ref{Fig3}(a),(c),(d)], but is large for the working point away from sweet spot [see Fig.~\ref{Fig2}(b)]. This weight reflects the qubit's sensitivity to $1/f$ flux noise. Therefore, the $T_\phi$ times at the sweet spots (both static and dynamical) are significantly longer than the one at the non-sweet spot. Compared with $T_\phi$ at the static sweet spot, the dynamical sweet spots exhibit somewhat  lower values of $T_\phi$. This is related to the small but nonzero pure-dephasing weights at filter frequencies $\bar{\omega}_{k\phi}\neq 0$,  absent for static sweet spots. Figure \ref{Fig3}(c) illustrates this for the working point \onecircle{},  where the relevant weights resulting in the dynamical $T_\phi\approx 1\,$ms are marked by single-headed arrows. (The same reasoning applies to the other working point \twocircle{}.)

The behavior of depolarization times $T_1$ at and away from sweet spots is reversed relative to that of $T_\phi$. Specifically, $T_1$ is longest at the static \emph{non-sweet spot}, where disjoint support of wave functions leads to the strongly suppressed weights marked by double-headed arrows in Fig.~\ref{Fig3}(b). By contrast, depolarization weights for \emph{sweet spots} [both static and dynamical, Figs.\ \ref{Fig3}(a),(c)] are not subject to this suppression and produce correspondingly lower $T_1$. [The $T_1$ trend obtained from the analysis of weight suppression is partially offset by the fact that  $S(\omega)$ is filtered at different frequencies in the sweet-spot vs.\ non-sweet-spot case.] Next, the comparison  shows that the static depolarization time at the sweet spot is smaller than the dynamical $T_1$. The reason for this can be traced to the difference in filter frequencies and corresponding magnitudes of the noise power spectrum, see Fig.\ \ref{Fig3}(a) vs.\ (c). In the static case, the filter frequencies for depolarization are $\pm \Omega_{ge}$, and $S(\pm \Omega_{ge})$ is relatively large compared to the dynamical case in \ref{Fig3}(c) where the dominant contributions arise from $S(\bar{\omega}_{0\pm})$. Indeed, these latter contributions closely match the minima of the noise power spectrum -- a situation which can be established simply by tuning the drive parameters.

 Inspection of Tab.\ \ref{tabel1} reveals a trend of $T_1$ and $T_\phi$ being anti-correlated: larger $T_1$ tend to coincide with with smaller $T_\phi$ and vice versa. This trend originates from the conservation of the cumulative filter weight, 
 \begin{equation}\label{conserve}
 (W_+ + W_-) + W_\phi = 2,
 \end{equation}
 where $W_{\pm} = \sum_k |g_{k\pm}|^2$ governs depolarization and $W_{\phi} = \sum_k 2|g_{k\phi}|^2$ pure dephasing. (A proof of this conservation law is given in Appendix C.) Increases in depolarization weights thus go along with decreases in the pure-dephasing weight, creating a tendency for trade-off between depolarization and dephasing which is exact only in the special case of white noise. This conservation rule is analogous to the sum rule that emerges in the context of dynamical decoupling \cite{DasSarma_DD,Oliver_flux_qubit_dd}. It is crucial that the conservation rule applies to filter weights rather than the rates. This enables one to manipulate the distribution of weights and filter frequencies to our advantage, putting the largest weights at or near minima in the noise spectrum.
 
For simplicity, our discussion has been based exclusively on a two-level approximation of the fluxonium qubit. In general, the presence of higher qubit levels can induce leakage to states outside the computational subspace. This concern is less significant for qubits with relatively large anharmonicity like the fluxonium circuit considered here. Through numerical calculations including higher levels we have confirmed that this leads to quantitative changes of the dynamical decoherence rates above, but does not affect the results reported above qualitatively.

\section{Gates and Readout of a single Floquet qubit}

The above results suggest that use of the driven Floquet states as computational qubit states can be advantageous due to the long coherence times reached at the dynamical sweet spots. We refer to this dynamically protected qubit as the \textit{Floquet qubit}, which belongs to the broader class of \textit{dressed-state qubits}. A host of previous work has studied gate operations on such dynamically encoded qubits \cite{Paraoanu_dressed_qubit,Wunderlich_dressed_qubit_nature,Wunderlich_dressed_qubit_NJP,Wineland_dressed_qubit,Plenio_dressed_qubit_gates}. Here, we specifically discuss how to maintain dynamical-sweet-spot operation while performing gates in order to maximize protection from $1/f$ noise. In the following, we show that Floquet qubits can easily be integrated into gate and readout protocols necessary for quantum-information processing.

\begin{figure}[h!]
\includegraphics[width = \columnwidth]{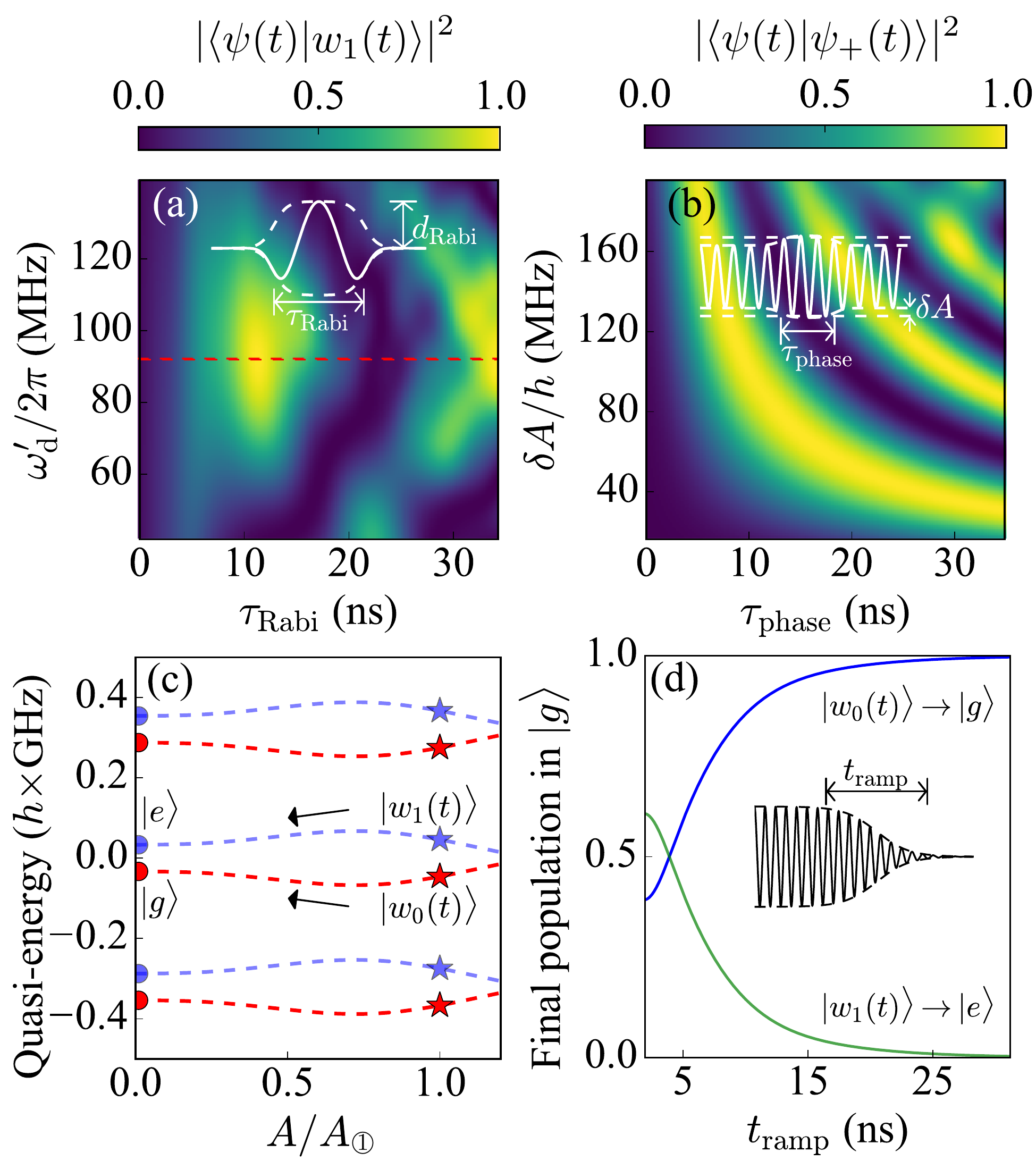}
\caption{Concurrent gates for the Floquet qubit and adiabatic mapping protocol for readout. (a) Adding a secondary pulse (inset) to the Floquet drive induces Rabi oscillations, which are sufficient for implementing $X$ gate operations. The plot shows the final population in Floquet eigenstate $\vert w_1(t)\rangle$ as a function of pulse duration $\tau_\mathrm{Rabi}$ and carrier frequency $\omega_\mathrm{d}'$, for the initial state $\vert w_0(t)\rangle$. Full Rabi oscillations are observed when the secondary drive frequency matches the  quasi-energy difference $\epsilon_{01}$ (dashed line) (b) Phase gates can be realized by a temporary increase in the Floquet drive strength (inset). The change in drive strength modulates the  quasi-energy and thus enables phase gates such as $S\,(\pi/2)$ and $T\,(\pi/4)$. The plot shows the final population in $\vert \psi_+(t)\rangle$, as a function of the pulse duration $\tau_\mathrm{phase}$  and the drive amplitude variation $\delta A$, with the qubit initialized in $\vert \psi_-(t)\rangle$ (see main text for definition for $\vert \psi_\pm(t)\rangle$). (c) shows the quasi-energy spectrum as a function of $A$ (from 0 to $A_{\scriptsize{\onecircle{}}}$), with Floquet drive frequency fixed at $\omega_{\mathrm{d}\scriptsize{\onecircle{}}}$. ($A_{\scriptsize{\onecircle{}}}$ and $\omega_{\mathrm{d}\scriptsize{\onecircle{}}}$ are the drive parameters at working point \onecircle{}.) Red and blue star symbols mark the two Floquet states at point \onecircle{}, whereas dots of the same color represent the states $\vert g\rangle$ and $\vert e\rangle$ of the undriven fluxonium. An adiabatic mapping from Floquet states to static qubit eigenstates can by realized with a sufficiently slow switch-off of the drive from $A_{\scriptsize\onecircle{}}$ to 0, given the nonzero gap between the quasi-energies. (d) Simulation of the adiabatic mapping achieved by continuously switching off the drive (ramp-down in inset). The final population in $\vert g\rangle$ is plotted as a function of the ramp time $t_\mathrm{ramp}$, with the qubit initiated in $\vert w_0(t)\rangle$ (blue) or $\vert w_1(t)\rangle$ (green). The results confirm the feasibility of an adiabatic mapping with high fidelity, thus enabling readout of the Floquet states.}
\label{Fig4}
\end{figure}

 \subsection{Gate operations}
 We show that we can realize direct single-qubit gates on the Floquet qubit.
 For example, $X$ and $\sqrt{X}$ gates can be realized by inducing Rabi oscillations among Floquet eigenstates. This is accomplished by applying an additional pulse with carrier frequency $\omega_\mathrm{d}'\approx \epsilon_{01}$, duration $\tau_\mathrm{Rabi}$, and  maximal amplitude $d_\mathrm{Rabi}$, see inset of Fig.~\ref{Fig4}(a). We verify the presence of Rabi oscillations numerically by simulating the time evolution for the working point \onecircle{}. For a fixed initial state $\vert w_0(t)\rangle$, the final population of $\vert w_1(t)\rangle$ shows oscillatory behavior as a function of $\tau_\mathrm{Rabi}$ and $\omega'_\mathrm{d}$, see Fig.~\ref{Fig4}(a). Full Rabi cycles only occur when $\omega'_\mathrm{d}$ matches $\epsilon_{01}$. Computation of the gate fidelities for the examples of $X$ and $\sqrt{X}$ gates yields a value of $99.99\%$ in both cases.
 
Single-qubit phase gates can be implemented by modulating the quasi-energy difference $\epsilon_{01}$ through a temporary increase $\delta A$ of the drive amplitude  [see inset of Fig.~\ref{Fig4}(b)]. This modifies the dynamical phase acquired over the gate duration $\tau_\text{phase}$, enabling $S$ and $T$ gates, for example. 
For numerical verification, we initialize the qubit in one of the Floquet superposition states $\vert \psi_\pm(t)\rangle = [\vert w_0(t)\rangle\pm \vert w_1(t)\rangle\mathrm{e}^{-i\epsilon_{01}t}]/\sqrt{2}$ (equator of the Bloch sphere) and monitor the population in the orthogonal state as a function of $\tau_\text{phase}$ and $\delta A$. The observed oscillations [Fig.~\ref{Fig4}(b)] in this population indicates that the computational states accumulate a relative phase as expected. The computed fidelity for $S$ ($\pi/2$) and $T$ ($\pi/4$) gates realized both exceed $99.99\%$.

\subsection{Readout}
Floquet states can be adiabatically mapped \cite{GuerinFloquet,Lupascu_flux_qubit_Floquet} to the eigenstates of the driven qubit by slowly ramping down the flux modulation, provided that gaps in the quasi-energy spectrum are sufficiently large.  For \onecircle{}, Fig.~\ref{Fig4}(c) shows that quasi-energy gaps do not close as $A$ is decreased to 0, thus enabling the adiabatic state transfer. We verify this mapping numerically by simulating the closed-system evolution with either of the driven Floquet qubit eigenstates $\vert w_{0(1)}(t)\rangle$ as the initial state and a smooth ramp-down of duration $t_\text{ramp}$. Fig.~\ref{Fig4}(d) shows the calculated population in the undriven qubit eigenstates  $\vert g(e)\rangle$ as a function of time. The resulting state-transfer fidelity is high for ramp times of the order of tens of ns, ($99.6\%$ for $t_\mathrm{ramp}= 30\,$ns). Conventional dispersive readout techniques, applicable to fluxonium qubits \citep{Manucharyan_heavy_fluxonium,Schuster_heavy_fluxonium,Nanowire_fluxonium,Fluxonium_hc,Schuster_heavy_fluxonium_control}, can then be employed subsequently in order to infer the original dynamical state. 

In future work, it may be interesting to explore alternative readout protocols similar to the one presented in \cite{Schuster_heavy_fluxonium_control}.  In an extension of that scheme, a higher fluxonium level that produces a large dispersive shift on the readout resonator would be excited conditionally, based on the occupied computational Floquet state.

\section{Floquet two-qubit gates}
The fact that dynamical sweet spots form entire manifolds in the control-parameter space provides sufficient flexibility to perform two-qubit gates among Floquet qubits without ever giving up the dynamical protection. Thanks to the one-to-one relation between quasi-energies and Floquet states on one hand, and ordinary eigenenergies and eigenstates on the other hand, it is possible to transfer existing protocols for two-qubit gates to the case of Floquet qubits. In the following, we present a protocol for implementing a $\sqrt{i\textrm{SWAP}}$ gate between two Floquet qubits, again based on flux-modulated fluxonium qubits. Related protocols for implementing two-qubit gates with dynamical protection have been discussed for slightly different systems involving either near-adiabatic parametric modulation of the qubit frequency \cite{Didier_dynamical_sweet_spot,Rigetti_ac_sweet_spot,Rigetti_ac_sweet_spot_exp,Coppersmith_two_qubit_sweet_spot} or requiring a tunable coupler between qubits \cite{Friesen_all_sweet_two_q_gate,Oliver_tunable_coupling}. The two-qubit gate proposed here is designed for the protected Floquet regime discussed above. It is compatible with direct driving of the qubit and circumvents the need for tunable coupling, thus providing a relatively simple scheme for future experimental realization.




\subsection{Analytical description}

A simple method of implementing $\sqrt{i\textrm{SWAP}}$ gates, for example among two transmon qubits, consists of bringing the pair of weakly coupled qubits into resonance for a certain gate duration. For two Floquet qubits, we show that $\sqrt{i\textrm{SWAP}}$ gates can realized in a similar manner by tuning the quasi-energy differences into and out of resonance. An important advantage of the Floquet two-qubit gate is the ability to keep both  qubits within the dynamical sweet manifolds for the complete duration of the gate, thus reducing the error due to the qubits' coupling to 1/$f$ noise. 

We establish this Floquet-gate protocol for a composite system of two coupled fluxonium qubits, each of which is flux-modulated, described by
\begin{align}
\hat{H}_\mathrm{LR}(t) = \hat{H}_\mathrm{L}(t) + \hat{H}_\mathrm{R}(t) + \hat{H}_\mathrm{int}.
\end{align}
Here, $\hat{H}_\mathrm{L}(t)$ and $\hat{H}_\mathrm{R}(t)$ denote the Hamiltonians of the two periodically driven fluxonium qubits, and $\hat{H}_\mathrm{int}$ is the time-independent coupling between them.
The flux-modulation frequencies associated with the two qubits are given by $\omega^\mathrm{L}_\mathrm{d}$ and $\omega^\mathrm{R}_\mathrm{d}$, respectively. As appropriate for a fluxonium with large anharmonicity, we may simplify the description by truncating the Hilbert space to a two-level subspace. We propose to induce the necessary qubit-qubit interaction $H_\mathrm{int}$ via a mutual inductance between the two fluxonium loops. In this case, the coupling term takes the form $\hat{H}_\mathrm{int} = J\hat{\sigma}^\mathrm{L}_{z}\hat{\sigma}^\mathrm{R}_{z}$, with $J$ denoting the coupling strength. For later convenience, we introduce the abbreviation $\hat{H}_0(t) = \hat{H}_\mathrm{L}(t)+\hat{H}_\mathrm{R}(t)$ for the bare qubit Hamiltonian.

When the two Floquet qubits are in resonance, i.e., their quasi-energies are degenerate, then the static coupling induces excitation swapping between the Floquet states (rather than between bare qubit eigenstates). To describe this process, we move to the interaction picture using the time-dependent unitary $\hat{U}_0(t) =\, \mathcal{T}\exp[-i\int_{0}^t\hat{H}_0(t')dt'] = \hat{U}^\mathrm{L}_\mathrm{q}(t)\otimes \hat{U}^\mathrm{R}_\mathrm{q}(t)$. Here, $\hat{U}^\mathrm{L(R)}_\mathrm{q}(t) = \sum_{j=0,1}\vert w^\mathrm{L(R)}_j(t)\rangle\langle w^\mathrm{L(R)}_j(0)\vert\exp[-i\epsilon^\mathrm{L(R)}_jt]$, and $\vert w_j^\mathrm{L(R)}(t)\rangle$ and $\epsilon^\mathrm{L(R)}_j$ denote the $j$-th Floquet state and corresponding quasi-energy of the left (right) qubit. In this interaction picture, the Hamiltonian is given by
\begin{align}\label{full_H_tilde}
\tilde{H}_\mathrm{LR}(t) =& J\, \hat{U}^{\dagger}_0(t) \hat{\sigma}^\mathrm{L}_{z}\hat{\sigma}^\mathrm{R}_{z} \hat{U}_0(t)\nonumber\\
=& J\!\sum_{k,k'\in\mathbb{Z}}\sum_{\mu,\mu'=\pm,\phi}\! g^\mathrm{L}_{k\mu} g^\mathrm{R}_{k'\mu'} \hat{c}^\mathrm{L}_{\mu}\hat{c}^\mathrm{R}_{\mu'}\nonumber\\
&\times\exp\!\left[-i(\bar{\omega}^\mathrm{L}_{k\mu}\!+\!\bar{\omega}^\mathrm{R}_{k'\mu'})t\right]\!,
\end{align}
where $\bar{\omega}^\mathrm{L(R)}_{k\mu}$ and $g^\mathrm{L(R)}_{k\mu}$ are the filter frequencies and the Fourier coefficients of the $\hat{\sigma}^\mathrm{L(R)}_z$'s matrix elements in the Floquet basis, associated with the left (right) qubit, respectively. The operators $\hat{c}^\mathrm{L(R)}_\mu$ denote the Pauli matrices defined in the Floquet basis (see Appendix B for details).

Following the conventional strategy, we perform a $\sqrt{i\textrm{SWAP}}$ gate by bringing the Floquet qubits into resonance  ($\epsilon^\mathrm{L}_{01}=\epsilon^\mathrm{R}_{01}$) through an suitable change of the drive parameters. After rotating-wave approximation, the effective Hamiltonian at the degeneracy point reduces to 
\begin{align}
    \tilde{H}'=J\,g^\mathrm{L}_{0+}g^\mathrm{R}_{0-}\hat{c}^\mathrm{L}_+\hat{c}^\mathrm{R}_-
    +\mathrm{h.c.},
\end{align}
which is the flip-flop interaction necessary for the $\sqrt{i\textrm{SWAP}}$ gate. We note that the term proportional to $g^\mathrm{L}_{0\phi}g^\mathrm{R}_{0\phi}\hat{c}^\mathrm{L}_\phi\hat{c}^\mathrm{R}_\phi$ corresponds to an unwanted $Z\!Z$ interaction between Floquet qubits. This term exactly vanishes as soon as at least one of the qubits is at a  dynamical sweet spot where $g_{0\phi}^\mathrm{L,R}=0$ [Eq.\ \eqref{eq:g0phi}].



Based on the full interaction Hamiltonian \eqref{full_H_tilde}, we next verify numerically that this simple strategy indeed yields high-fidelity two-qubit gates.

\begin{figure}
    \centering
    \includegraphics[width = \columnwidth]{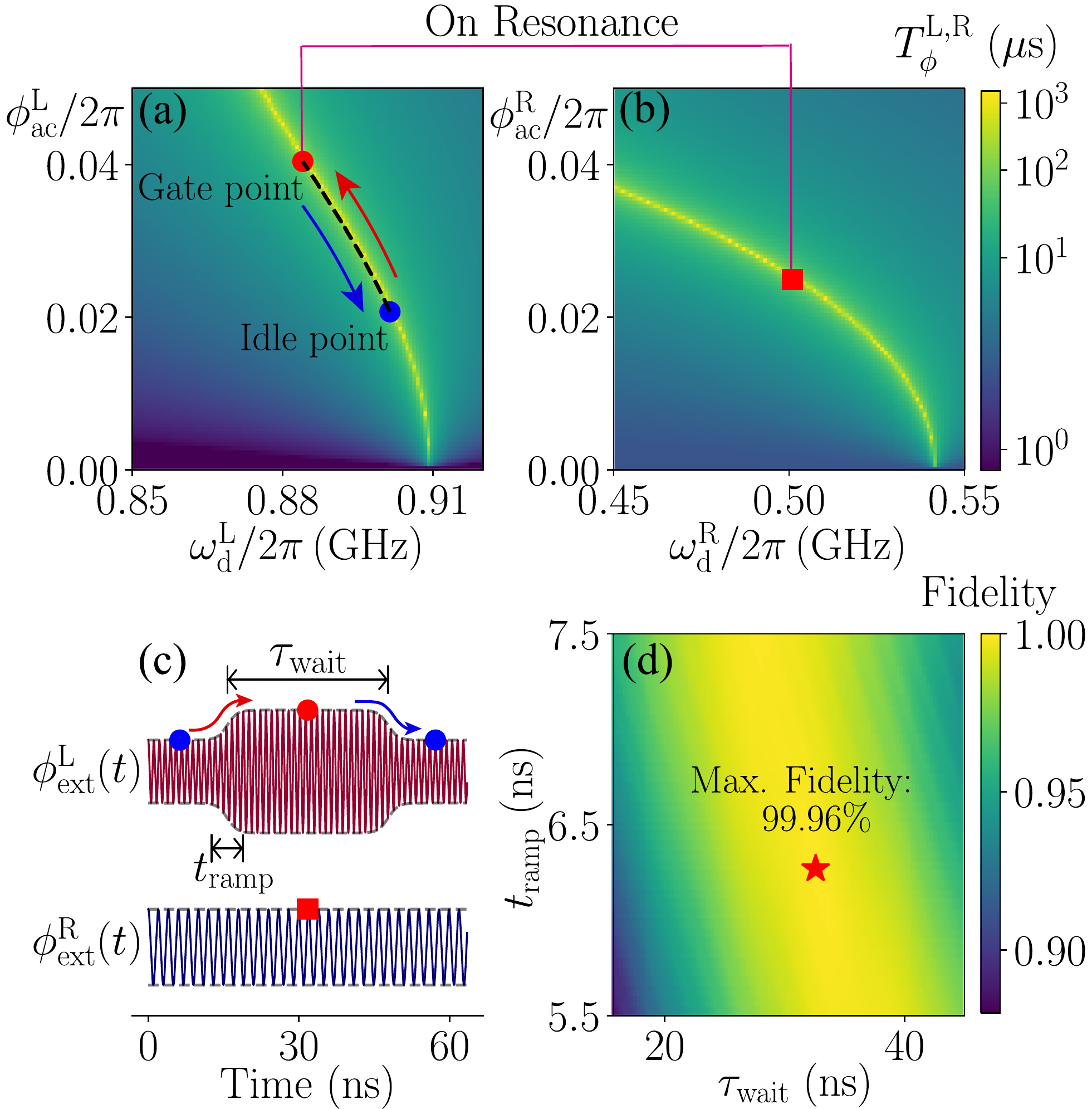}
    \caption{Simulation of a Floquet-$\sqrt{i\textrm{SWAP}}$ gate for two inductively coupled fluxonium qubits. (a) and (b) show the dynamical sweet manifolds corresponding to the two driven fluxonium qubits, at given dc flux biases. The red and blue dots in (a) indicate the gate and idle points for the left fluxonium, which is tuned along the path marked by the black-dashed curve. The red square in (b) represents the operating point for the right fluxonium qubit. (c) shows the drive pulses realizing this gate operation with $\tau_\mathrm{wait}$ and $t_\mathrm{ramp}$ denoting gate duration and ramp time, respectively. (d) depicts the calculated gate fidelity as a function of ramp time and gate duration, where the red star marks the position of maximal gate fidelity. (Parameters are as follows: left fluxonium -- $E_\mathrm{C}^\mathrm{L}/h = 1.2\,$GHz, $E_\mathrm{J}^\mathrm{L}/h = 6.0\,$GHz; right fluxonium --  $E_\mathrm{L}^\mathrm{L}/h = 0.95\,$ GHz, while $E_\mathrm{C}^\mathrm{R}/h = 1.0\,$GHz, $E_\mathrm{J}^\mathrm{R}/h = 4.1\,$GHz, and  $E_\mathrm{L}^\mathrm{R}/h = 0.7\,$ GHz. The interaction strength is set by $J/h = 4.8\,$MHz. The dc fluxes are fixed to $\phi^{\mathrm{L(R)}}_\mathrm{dc}/2\pi = 0.529(0.520)$, and noise parameters are the same as in Table \ref{tabel1}.)}
    \label{fig5}
\end{figure}

\subsection{Numerical simulation}
To construct our $\sqrt{i\textrm{SWAP}}$ gate, we first identify appropriate drive parameters for sweet-spot operation and for bringing the qubits into and out of resonance.
Fig.~\ref{fig5}(a) and (b) show the relevant sweet-spot manifolds for the two fluxonium qubits.
Within these manifolds, the quasi-energy difference $\epsilon^\mathrm{L,R}_{01}$ varies continuously, making it possible to establish degeneracy of the two Floquet qubit quasi-energies, $\epsilon^\mathrm{L}_{01} = \epsilon^\mathrm{R}_{01}$. In the example we selected, the right qubit is maintained at a fixed dynamical sweet spot [Fig.~\ref{fig5}(b), red square] while the left qubit can be tuned within its sweet-spot manifold from an idle point (blue dot) into resonance at the gate point (red dot) and back [Fig.~\ref{fig5}(a)]. 

The detailed pulse shapes of the drives enacting the gate are shown in Fig.~\ref{fig5}(c). For the left fluxonium qubit, amplitude and frequency of the flux modulation are adjusted in a way to smoothly tune the qubit from its idle point to the gate point (within the ramp time $t_\mathrm{ramp}$). Pulse shaping  allows one to choose a path (black-dashed curve) that keeps the Floquet qubit within the sweet manifold [Fig.~\ref{fig5}(a)]. After leaving the qubit at the gate point for a suitable waiting time $\tau_\mathrm{wait}$, the drive parameters are tuned back to the idle point. We calculate the $\sqrt{i\textrm{SWAP}}$-gate fidelity by an open-system simulation of this composite system (again taking into account of $1/f$ flux noise and dielectric loss). The results in Fig.~\ref{fig5}(d) show a broad region of gate parameters $\tau_\mathrm{wait}$ and $t_\mathrm{ramp}$ with high gate fidelities up to $99.96\%$. (The discussion of the effect of stray two-qubit interactions at the idle point is beyond the scope of this paper, but see Refs.~\cite{Martinis_tunable_coupling,Oliver_tunable_coupling,Yu_two_qubit_gates,Houck_two_qubit_gates,Houck_tunable_coupling} for mitigation strategies.)

\color{black}

\section{Conclusions}
 Operation of superconducting qubits at static sweet spots is a well-established means to reducing 1/$f$ noise sensitivity. However, one limitation is the abrupt symmetry-induced change in the nature of wavefunctions at the sweet spot, which can negatively impact depolarization times at the sweet spot. We have presented a protocol for engineering dynamical sweet spots which partially overcome this limitation. In contrast to static sweet-spot operation, the Floquet scheme can yield long dynamical $T_\phi$ and $T_1$ simultaneously.  The possibility to directly perform both single- and two-qubit gate operations as well as readout on Floquet qubits makes them promising for both quantum information storage and processing. A companion experimental work has implemented this proposed protocol using a flux-modulated fluxonium qubit \cite{Houck_Floquet_exp}, and a 40-fold improvement in dephasing time due to dynamical sweet-spot operation is reported.
 
Although the example we have demonstrated only makes use of the simplest single-tone drives, it is possible that non-sinusoidal or multichromatic  drives could further expand the sweet-spot manifolds and yield even higher qubit coherence times. Future work may explore building networks of larger numbers of Floquet qubits, which could be particularly beneficial for quantum information processing thanks to enhanced dynamical coherence times and tunability.


\begin{acknowledgments} The work was supported by the Army Research Office under Grant No.\ W911NF-19-1-0016. We thank Xinyuan You, Daniel Weiss and Brian Baker for helpful discussion. 
\end{acknowledgments}

\appendix
\section{Effective model for fluxonium, and coupling to noise sources}

The Hamiltonian describing a flux-modulated fluxonium is given by \cite{You_flux_allocation}
\begin{align}
H_\mathrm{q,full}(t) = 4E_\mathrm{C} \hat{n}^2 + \frac{1}{2}E_\mathrm{L} [\hat{\varphi}+\phi_\mathrm{ext}(t)]^2 - E_\mathrm{J} \cos\hat{\varphi},
\label{H_full}
\end{align}
where $E_\mathrm{C}$, $E_\mathrm{L}$ and $E_\mathrm{J}$ represent the capacitive, inductive and Josephson energies of the fluxonium qubit, and $\phi_\mathrm{ext}(t) = \phi_\mathrm{dc}+\phi_\mathrm{ac}\cos(\omega_\mathrm{d}t)$. We use $\hat{\varphi}$ and $\hat{n}$ to denote the flux and conjugate charge operator of the qubit, respectively.

The static eigenenergies $\Omega_l$ and corresponding eigenstates $\vert l\rangle$ ($l=g,e,f, \cdots$) are obtained by diagonalizing $H_\mathrm{q,full}$, and depend on the dc flux component $\phi_\text{dc}$. We will refer to the specific solutions at the static sweet spot $\phi_\mathrm{dc}=\pi$ by $ \widetilde{\Omega}_l$ and $\big\vert \Tilde{l}\big\rangle$. These eigenstates, expressed in the phase basis, have alternating parities (for example, $\vert \tilde{g}\rangle$ and $\vert \tilde{e}\rangle$ have even and odd parities respectively). 

To avoid leakage into higher fluxonium states under flux modulation, we choose fluxonium parameters resulting in a large anharmonicity at half-integer flux, $\widetilde{\Omega}_f-\widetilde{\Omega}_e\gg \widetilde{\Omega}_e-\widetilde{\Omega}_g$. If we limit the external flux $\phi_\mathrm{ext}(t)$ to values in the vicinity of $\phi_\mathrm{dc}=\pi$, and avoid resonance with the $e-f$ transition, $\omega_\mathrm{d}\ll \widetilde{\Omega}_f-\widetilde{\Omega}_e$,
then Eq.~(\ref{H_full}) can be approximated by the effective two-level Hamiltonian (\ref{H_q_two_level}). In that Hamiltonian, $\Delta = \widetilde{\Omega}_e-\widetilde{\Omega}_g$, $A=E_\mathrm{L}\phi_\mathrm{ac}\tilde{\varphi}_{ge}$, $B=2E_\mathrm{L}(\phi_\mathrm{dc}-\pi)\tilde{\varphi}_{ge}$; here, $\tilde{\varphi}_{ge} =\vert \langle \tilde{g}\vert \hat{\varphi}\vert \tilde{e}\rangle\vert$. Different from the usual convention, we define the Pauli matrices as
\begin{align}
\label{sigma_xz}
    \sigma_x = \vert \tilde{e}\rangle \langle \tilde{e}\vert - \vert \tilde{g}\rangle \langle \tilde{g}\vert, \quad \sigma_z = \vert \tilde{g}\rangle \langle \tilde{e}\vert + \vert \tilde{e}\rangle \langle \tilde{g}\vert,
\end{align}
which is a common choice in the context of flux qubits \cite{Chu_flux_qubit_Floquet,Lupascu_flux_qubit_Floquet}.

Given this effective model, it is important to revisit the question of how the fluxonium qubit couples to the limiting environment degrees of freedom.
In Section II, it is posited that the noise sources of interest couple to the qubit through its $\sigma_z$ operator which can be motivated as follows. The fluxonium's interaction with the $1/f$ flux noise source can be modeled as mutual inductance between the fluxonium's inductor and the bath, hence the coupling to the noise is via the qubit operator $\hat{\varphi}$. Experimental results are further consistent with dielectric noise coupling to the qubit's phase operator \cite{Fluxonium_hc,Manucharyan_heavy_fluxonium, Nanowire_fluxonium, Schuster_heavy_fluxonium_control,Martinis_1/f_Crossover}. Note that operator $\hat{\varphi}$ only couples states with different parities. Therefore, based on Eq.~\eqref{sigma_xz}, it is projected to $\sigma_z$ in the two-level subspace, which results in the $H_\mathrm{int}$ used in our model.

\section{Floquet master equation}
This appendix sketches the derivation of the Floquet master equation \cite{Hanggi_Floquet_Markovian, Breuer_open_quantum_system, Grifoni_dissipative_Floquet} which we use in the subsequent appendix to calculate the dynamical decoherence rates. The full Hamiltonian is given by $H(t) = H_\mathrm{q}(t) +  H_\mathrm{B} + H_\mathrm{int}$ with time-periodic qubit Hamiltonian, and time-independent bath and interaction Hamiltonian. The latter is taken to be of the form $H_\mathrm{int} = \hat{\sigma}\hat{\eta}$, where $\hat{\sigma}$ and $\hat{\eta}$ are qubit and bath operators, respectively.

We start from the Redfield equation of the driven qubit
\begin{align}
    \frac{\mathrm{d}\tilde{\rho}_\mathrm{q}(t)}{\mathrm{d}t}\!=\!-\!\int_0^t d\tau \text{Tr}_\mathrm{B}\! \left[\tilde{H}_\mathrm{int}(t), \left[\tilde{H}_\mathrm{int}(t-\tau), \tilde{\rho}_\mathrm{q}(t)\!\otimes\!\tilde{\rho}_\mathrm{B}   \right]\right]\!,
    \label{Redfield}
\end{align}
which describes the evolution of the qubit density matrix $\tilde{\rho}_\mathrm{q}$ (in the interaction picture). Here, $\mathrm{Tr}_\mathrm{B}$ denotes a partial trace on the bath degrees of freedom, and $\tilde{\rho}_\mathrm{B}$ is the density matrix of the bath in the interaction picture, which is assumed to stay in thermal equilibrium. The term $\tilde{H}_\mathrm{int}(t) = U_0^\dagger(t) H_\mathrm{int} U_0(t)$ is the qubit-bath coupling expressed in the interaction picture, where $U_0(t)=U_\mathrm{q}(t)U_\mathrm{B}(t)$, $U_\mathrm{q}(t) = \sum_{j=0,1}\vert w_j(t)\rangle\langle w_j(0)\vert \exp(-i\epsilon_jt)$, and $U_\mathrm{B}(t)=\exp(-iH_\mathrm{B}t)$. The interaction term can be further expressed as $\tilde{H}_\mathrm{int}(t)=\tilde{\sigma}(t)\tilde{\eta}(t)$, where $\tilde{\sigma}(t)= U^\dagger_\mathrm{q}(t)\hat{\sigma}U_\mathrm{q}(t)$ and $\tilde{\eta}(t)=U^\dagger_\mathrm{B}(t)\hat{\eta}U_\mathrm{B}(t)$.

Eq.~\eqref{Redfield} is an integro-differential equation, and not convenient for reading off decoherence rates. To derive expressions for $\gamma_{\mu}$ ($\mu=\pm,\phi$), we first simplify this equation by employing the rotating-wave approximation. In order to identify the fast-rotating terms, we decompose  $\tilde{\sigma}(t)$ into different frequency components, 
\begin{align}\label{decompose}
    \tilde{\sigma}(t) 
    =\,\sum_{k\in\mathbb{Z},\mu=\pm,\phi}g_{k\mu}\hat{c}_\mu(0)\exp(-i\bar{\omega}_{k\mu}t).
\end{align}
Here, we define the Floquet counterparts of the Pauli matrices by
\begin{align}
    \hat{c}_+(t) =&\, \vert w_1(t)\rangle \langle w_0(t)\vert,\nonumber\\
    \hat{c}_-(t) =&\, \vert w_0(t)\rangle \langle w_1(t)\vert,\nonumber\\
    \hat{c}_\phi (t)=&\, \vert w_1(t)\rangle \langle w_1(t)\vert - \vert w_0(t)\rangle \langle w_0(t)\vert. \label{cdef}
\end{align}
The frequencies $\bar{\omega}_{k,\mu}$ appearing in Eq.\ \eqref{decompose}
are the filter frequencies defined in the Section III, namely $\bar{\omega}_{k\pm} = \mp \epsilon_{01}+k\omega_\mathrm{d}$ and $\bar{\omega}_{k\phi} =k\omega_\mathrm{d}$.
Furthermore, the Fourier-transformed coupling matrix elements are given by
\begin{align}
    g_{k,\pm} =& \frac{\omega_\mathrm{d}}{2\pi} \int_0^{2\pi/\omega_\mathrm{d}} \mathrm{d}t\, e^{ik \omega_\mathrm{d} t}\, \tr_\mathrm{q} \left[\hat{\sigma} \hat{c}_\mp(t)\right],\nonumber\\
    g_{k,\phi} =& \frac{\omega_\mathrm{d}}{4\pi} \int_0^{2\pi/\omega_\mathrm{d}} \mathrm{d}t\, e^{ik \omega_\mathrm{d} t}\, \tr_\mathrm{q} \left[\hat{\sigma} \hat{c}_\phi(t)\right],
    \label{gk's}
\end{align}
where $\tr_\mathrm{q}$ is the partial trace over the qubit degrees of freedom.

The qubit interaction operator $\tilde{\sigma}(t)$, expanded in this way, is substituted into Eq.~\eqref{Redfield} resulting in a sum terms each involving the coefficient $\exp[-i(\bar{\omega}_{k'\mu'}-\bar{\omega}_{k\mu})t]$. Under certain conditions, every term  with $\bar{\omega}_{k'\mu'}-\bar{\omega}_{k\mu}\neq 0$ can be treated as fast-rotating and be neglected. This strategy is appropriate if both the minimal quasi-energy difference and the drive frequency $\omega_\mathrm{d}$ are much larger than the inverse of the relevant time scale, i.e., the coherence time. With this Eq.~\eqref{Redfield} is cast into the simplified form
\begin{align}
\label{Redfield_RWA}
    \frac{\mathrm{d}\tilde{\rho}_\mathrm{q}(t)}{\mathrm{d}t} = \sum_{\mu=\pm,\phi} \zeta_{\mu}\left[\int_{-\infty}^{\infty}{\mathrm{d}\omega} F_\mu (\omega,t)S(\omega)\right]\mathbb{D}[\hat{c}_\mu]\tilde{\rho}_\mathrm{q}(t),
\end{align}
where \begin{align}
    F_\mu(\omega, t) = \zeta^{-1}_\mu\sum_{k} \pi^{-1}t\,\textrm{sinc}[(\omega-\bar{\omega}_{k\mu})t]|g_{k\mu}|^2
    \label{Filter_functions}
\end{align} 
denotes the filter functions from Section III, $\mathbb{D}[L]\tilde{\rho}_\mathrm{q} = L\tilde{\rho}_\mathrm{q}L^\dagger - (L^\dagger L \tilde{\rho}_\mathrm{q} 
+ \tilde{\rho}_\mathrm{q}L^\dagger L)/2$ is the usual damping superoperator, and $S(\omega)=\int_{-\infty}^{\infty}\mathrm{d}t\,e^{i\omega t}\tr_\mathrm{B}[\tilde{\eta}(t)\tilde{\eta}(0)\tilde{\rho}_\mathrm{B}]$
  is the noise spectrum. We have further introduced the abbreviations $\zeta_\pm \equiv 1$ and $\zeta_\phi\equiv1/2$, and used  $\hat{c}_\mu(0) \to \hat{c}_\mu$. [We note that terms contributing to the Lamb shift have been omitted in Eq.~\eqref{Redfield_RWA}.]

The simplified Redfield equation \eqref{Redfield_RWA} is reminiscent of the Lindblad form, and includes three distinct terms $\mu=\pm$ and $\mu=\phi$ that describe relaxation, excitation and pure dephasing of the Floquet qubit. However, in place of fixed rates associated with the individual jump terms, Eq.\ \eqref{Redfield_RWA} still involves time-dependent rate coefficients given by    
\begin{equation}\label{Kintegral}
K_\mu(t) = \int_{-\infty}^{\infty}\mathrm{d}\omega F_\mu(\omega, t)S(\omega).
\end{equation}
We discuss in the subsequent appendix how to evaluate these rate coefficients for concrete choices of the  noise spectrum $S(\omega)$.

\section{Evaluation of  decoherence rates}
This appendix discusses the evaluation of the decoherence rate coefficients associated with the simplified Redfield equation \eqref{Redfield_RWA}, focusing on the specific noise spectrum $S(\omega)$ adopted in the main text. To simplify the integral $K_\mu(t)$, we first inspect the structure of the filter  functions $F_\mu(\omega, t)$ defined in Eq.~\eqref{Filter_functions}. These functions are peaked at the filter frequencies $\bar{\omega}_{k,\mu}$, with the peak width given by $2\pi t^{-1}$. We distinguish two separate scenarios: (1) the case of noise spectra that can be approximated as constant within each peak width, and (2) the case of noise spectra, such as $1/f$ spectra, where this approximation does not hold for all peaks.

\emph{Case (1).}---If the spectrum $S(\omega)$ is sufficiently flat within each peak-width frequency range, we can approximate $(t/\pi)\,\mathrm{sinc}(\omega t)\approx \delta(\omega)$ in Eq.~\eqref{Filter_functions}, and arrive at the Markovian Floquet master equation 
\begin{align}
    \frac{\mathrm{d}\tilde{\rho}_\mathrm{q}(t)}{\mathrm{d}t} = \sum_{k\in\mathbb{Z},\mu=\pm,\phi}|g_{k\mu}|^2S(\bar{\omega}_{k\mu})\mathbb{D}[\hat{c}_\mu]\tilde{\rho}_\mathrm{q}(t).
    \label{Lindblad_master}
\end{align}
This form allows one to directly read off the resulting rates which are given by  $\gamma_{\pm} = \sum_{k}|g_{k\mu}|^2S(\bar{\omega}_{k\pm})$ and $\gamma_\phi = \sum_{k}2|g_{k\phi}|^2S(\bar{\omega}_{k\phi})$.

\emph{Case (2).}---On the other hand, the noise spectrum $S(\omega)$ at depolarization filter frequencies $\bar{\omega}_{k\pm}$ is considered flat, therefore the resulting expressions of $\gamma_\pm$ is the same as shown in Case (1). Finally, we arrive at the results shown in Eqs.~\eqref{rates} and \eqref{gammaphi} in the main text.

Whenever the noise spectrum varies significantly across one filter-function peak width, the above approximation fails. This is, in particular, the case for $1/f$ noise near $\omega=0$ where $S(\omega)$ is purely dominated by the contribution $S_\mathrm{f}(\omega)=\mathcal{A}_f^2|\omega/2\pi|^{-1}$. For filter frequencies away from $\omega=0$, we continue treating $S(\omega)$ as sufficiently flat. Zero and nonzero filter frequencies hence play distinct roles. For depolarization, relevant filter frequencies $\bar{\omega}_{k\pm}$ are non-zero and the discussion of Case (1) carries over, yielding Eq.\ \eqref{rates}  for the depolarization rates.

The appearance of a zero filter frequency for dephasing  motivates us to separate the integral $K_\phi(t)$ [Eq.\ \eqref{Kintegral}] into a low-frequency and a high-frequency part. We focus on the low-frequency part first, which is given by
\begin{align}
    I(t) =2|g_{0\phi}|^2\int_{-\pi/t}^{\pi/t}\mathrm{d}\omega\, \frac{t}{\pi}\,\mathrm{sinc}(\omega t) S_\mathrm{f}(\omega),
\end{align}
where the integration range is set by the peak width $2\pi/t$. To regularize the logarithmic divergence of this integral, we employ infrared cutoffs $\pm\omega_\mathrm{ir}$ \cite{Ithier_decoherence_analysis, Devoret_fluxonium_molecule, Rigetti_ac_sweet_spot, Groszkowski_Zero_pi_theory}. The cutoff is of the order of $1\,$Hz \cite{Devoret_fluxonium_molecule}, much smaller than the inverse of the measurement time. In this case, the integral can be approximated by
\begin{align}
\label{integral_low_frequency}
    I(t) \approx 8\mathcal{A}^2_\mathrm{f} t|\ln \omega_\mathrm{ir}t||g_{0\phi}|^2.
\end{align}

For the integral over the remaining high-frequency range, the $\delta$-function approximation we made in Case (1) is again valid. After combining the low and high-frequency contributions, the approximated $K_\phi(t)$ is a time-dependent function, given by
\begin{align}
    K_\phi(t)\approx I(t) + 2\sum_{k\neq 0} |g_{k\phi}|^2S(\bar{\omega}_{k\phi}).
    \label{K_phi_t}
\end{align}
According to this, $K_\phi(t)$ is reminiscent of a time-dependent rate for pure dephasing that  grows linearly in time (up to logarithmic corrections). Consequently, the off-diagonal elements of the density matrix do not follow an exponential decay. Instead, the decay is given by 
\begin{align}
\label{Off_d_decay}
    \tilde{\rho}_{\mathrm{q},01}\sim\, \exp\Big[&-4\mathcal{A}^2_\mathrm{f}|g_{0\phi}|^2t^2|\ln\omega_\mathrm{ir}t|\nonumber\\
   &-\sum_{k\neq 0} 2|g_{k\phi}|^2S(\bar{\omega}_{k\phi})t\Bigg],
\end{align}
which is a product of a Gaussian (again, up to logrithmic corrections) and pure exponential. (Note that to estimate the pure-dephasing rate, the contribution of depolarization to the decay of $\tilde{\rho}_{\mathrm{q},01}$ is excluded in the expression above.) 
Based on the $1/e$ decay time, we obtain 

\begin{align}
   \gamma_\phi \simeq \mathcal{A}_\mathrm{f}|2g_{0\phi}|\sqrt{|\ln\omega_\mathrm{ir}t_\mathrm{m}|} + \sum_{k\neq 0} 2|g_{k\phi}|^2S(\bar{\omega}_{k\phi}),
\end{align}
as a simple approximation bounding the pure-dephasing rate from above. Here, $t_\mathrm{m}$ is the characteristic measurement time; a representative value of the factor $\sqrt{|\ln\omega_\mathrm{ir}t_\mathrm{m}|}$ found in a recent experiment \cite{Devoret_fluxonium_molecule} is close to 4.

As discussed in Section IV.B, there exists an interrelation constraining the depolarization and pure-dephasing rates. This constraint originates from the conservation rule \eqref{conserve} for the filter weights which we prove in the following. Without loss of generality, we take the qubit coupling operator $\hat{\sigma}$ in $H_\mathrm{int}=\hat{\sigma}\hat{\eta}$ to be traceless with eigenvalues $\pm1$. (Any trace contribution renormalizes the bath Hamiltonian, and the scale factor rendering the eigenvalues $\pm1$ can be absorbed into $\hat{\eta}$.)
Employing the decomposition of the identity in terms of the Floquet states, $\openone = \sum_{j={0,1}}\vert w_j(t)\rangle\langle w_j(t)\vert$, and making use of Eqs.\ \eqref{cdef} and \eqref{gk's}, we find
\begin{align}
\label{conservation_proof}
    \tr_\mathrm{q}(\hat{\sigma}^2)
      =&\, \tr_\mathrm{q}(\hat{\sigma}\openone\hat{\sigma}\openone) = \sum_{j,j'=0,1}\vert \langle w_j(t)\vert \hat{\sigma}\vert w_{j'}(t)\rangle\vert^2 \nonumber\\
      =&\,\frac{1}{2}\big\vert\tr_\mathrm{q}[\hat{\sigma}\hat{c}_z(t)] \big\vert^2+ \sum_{\mu=\pm}\big\vert\tr_\mathrm{q}[\hat{\sigma}\hat{c}_\pm(t)] \big\vert^2\nonumber\\
      =&\, \sum_{\mu=\pm,\phi}\zeta^{-1}_\mu\Bigg\vert\sum_{k\in\mathbb{Z}} g_{k\mu}\,e^{-ik\omega_\mathrm{d}t}  \Bigg\vert^2=2.
\end{align}
Time averaging this expression over one drive period $2\pi/\omega_\mathrm{d}$ finally yields the claimed conservation rule
\begin{align}
\label{conservation}
     W_+ + W_- + W_\phi = \sum_{k\in\mathbb{Z}}\left(|g_{k+}|^2+ |g_{k-}|^2 + 2|g_{k\phi}|^2\right) = 2.
\end{align}
We further note that Eq.~\eqref{conservation} also imposes a constraint on the filter functions, namely
\begin{align}
    \sum_{\mu=\pm,\phi}\int_{-\infty}^{\infty}\mathrm{d}\omega F_\mu (\omega, t) = 2.
\end{align}

\section{Analytical approach for solving Floquet equations}
In this appendix, we first introduce a framework useful for solving the Floquet equation, and later employ this framework to derive several results discussed in Sections III and IV.

Solutions $\vert w_j(t)\rangle$ of the Floquet equation \eqref{Floquet_eq} are required to be time-periodic in $2\pi/\omega_\mathrm{d}$. Each such  wavefunction can be considered an element in the vector space $\mathcal{F}$ of $2\pi/\omega_\mathrm{d}$-periodic functions of the type $f:\, \mathbb{R}\to \mathbb{C}^2$. We choose the basis vectors of $\mathcal{F}$ to be $\vert f_{\sigma,k}(t)\rangle=\vert \sigma\rangle \exp(-ik\omega_\mathrm{d}t)$, where $\vert \sigma= z^{\pm}\rangle$ are the eigenvectors of the operator $\hat{\sigma}_z$ and $k\in\mathbb{Z}$. In this basis, the Floquet state $\vert w_j(t)\rangle$ has the decomposition
\begin{align}
\label{wjFourier}
    \vert w_j(t)\rangle =&\, \sum_{k\in\mathbb{Z}} \sum_{\sigma = z^{\pm}}u_{j\sigma k}\vert \sigma\rangle  e^{-ik\omega_\mathrm{d}t},
\end{align}
which is the Fourier expansion of $\vert w_j(t)\rangle$ with $u_{j\sigma k}$ as Fourier coefficients. It is useful to define an inner product for elements of $\mathcal{F}$ via the time average of their product over one drive period. Based on this definition, the basis $\{\vert f_{\sigma,k}(t)\rangle\}$ is orthonormal, since
\begin{align}
\label{innerprod}
 \frac{\omega_\mathrm{d}}{2\pi}\int_0^{2\pi/\omega_\mathrm{d}}\mathrm{d}t\, \langle f_{\sigma,k}(t)\vert f_{\sigma',k'}(t)\rangle =\delta_{\sigma\sigma'}\delta_{kk'}.
\end{align}
The decomposition \eqref{wjFourier} maps the periodic function $\vert w_j(t)\rangle\in\mathcal{F}$ to a vector $\vec{\mathbf{u}}_j\in \mathcal{V}=\mathbb{C}^2\otimes \mathbb{C}^\infty$. Here, the basis vectors $\vert f_{\sigma,k}(t)\rangle$ of $\mathcal{F}$ are mapped to the canonical unit vectors $(\vec{\mathbf{u}})_{\sigma',k'} = \delta_{\sigma\sigma'}\delta_{kk'}$ which we also denote by $|\sigma,k\rangle$. Following this basis change, the Floquet state Eq.~\eqref{wjFourier} is now represented as a vector in $\mathcal{V}$,
\begin{align}
    \vert \bar{w}_j\rangle = \sum_{\sigma=z^\pm}\sum_{k\in\mathbb{Z}}u_{j\sigma k}\vert \sigma, k\rangle.
\end{align}

Applying the basis change to the Floquet equation, one finds that it converts to an ordinary eigenvalue problem. To carry out this step, we consider the two operators $H_\mathrm{q}(t)$ and $-i\partial/\partial t$ acting on $\vert w_j(t)\rangle$ on the left-hand side of Eq.~\eqref{Floquet_eq}. Both of them map basis functions $\vert f_{\sigma, k}(t)\rangle$ to other time-periodic functions in $\mathcal{F}$, and hence correspond to matrices acting on elements in $\mathcal{V}$. Specifically, we have
\begin{align}
\label{operation_basis}
    H_\mathrm{q}(t) \vert f_{\sigma, k}(t)\rangle =& \sum_{\sigma,\sigma'=z^{\pm}} \sum_{k'\in\mathbb{Z}}h_{\sigma'\sigma k'}\vert f_{\sigma',k+k'}(t)\rangle,\nonumber\\
    -i\frac{\partial}{\partial t}\vert f_{\sigma, k}(t)\rangle =& -k\omega_\mathrm{d}\vert f_{\sigma, k}(t)\rangle,
\end{align}
where 
\begin{align*}
    h_{\sigma'\sigma k'} = \frac{\omega_\mathrm{d}}{2\pi}\int_0^{2\pi/\omega_\mathrm{d}}\mathrm{d}t\, e^{ik'\omega_\mathrm{d}t}\langle \sigma' \vert H_\mathrm{q}(t)\vert \sigma\rangle.
\end{align*}
Using Eq.~\eqref{operation_basis}, we can easily express the matrices representing $H_\mathrm{q}(t)$ and $-i\partial/\partial t$ as
\begin{align}
\label{Hbarq}
    \bar{H}_\mathrm{q} =& \sum_{\sigma,\sigma'=z^{\pm}} \sum_{k,k'\in\mathbb{Z}}h_{\sigma'\sigma'k'} \vert \sigma',k+k'\rangle\langle \sigma,k\vert,\\
    \bar{\Lambda} =& -\sum_{\sigma=z^\pm} \sum_{k\in\mathbb{Z}}k\omega_{\mathrm{d}}\vert \sigma,k\rangle \langle \sigma,k\vert.\nonumber
\end{align}
With this the Floquet equation takes on the form
\begin{align}
\label{eigenvalue_eq}
    \bar{H}\vert\bar{w}_j\rangle =\epsilon_j \vert \bar{w}_j\rangle,
\end{align}
where $\bar{H}=\bar{H}_\mathrm{q}+\bar{\Lambda}$.

Solving this eigenvalue equation yields an infinite number of eigenvectors and corresponding eigenvalues (quasi-energies). The structure of this equation is such that any given eigenpair $\vert\bar{w}_{j}\rangle$, $\epsilon_j$ generates an infinite set of solutions defined via
\begin{align}
    \vert \bar{w}_{j,n}\rangle =&\, \sum_{\sigma=z^\pm}\sum_{k\in\mathbb{Z}} u_{j\sigma,k} \vert \sigma, k-n\rangle,\nonumber\\
    \epsilon_{j,n}=&\,\epsilon_{j}+n\omega_\mathrm{d} \,\, (n\in\mathbb{Z}).
\end{align}
Reverting back to the function space $\mathcal{F}$, the above states have the form $\vert w_{j,n}(t)\rangle = \vert w_j(t)\rangle \exp(-in\omega_\mathrm{d}t)$. Accordingly, at the level of the  underlying Hilbert space of quantum states, only two of these states ($j=0,1$) are linearly independent. 

In the following, we employ this Floquet framework to the specific Hamiltonian \eqref{H_q_two_level}. For this analysis, it is useful to provide explicit expressions for the transformed $\bar{H}_\mathrm{q}$ from $H_\mathrm{q}(t)$. 
$H_\mathrm{q}(t)$ involves three distinct operations: $\hat{\sigma}_x$, $\hat{\sigma}_z$, and $\hat{\sigma}_z\cos\omega_\mathrm{d}t$ which are all valid linear operators on the function space $\mathcal{F}$. Applying again the basis transformation that led from Eq.~\eqref{operation_basis} to Eq.~\eqref{Hbarq}, these operators are transformed to the following matrices in the $\vert \sigma, k\rangle$ basis:
\begin{align}
    \bar{\sigma}_x =&\, \sum_{k\in\mathbb{Z}}\vert z^+,k\rangle \langle z^-,k\vert+\vert z^-,k\rangle \langle z^+,k\vert,\nonumber\\
    \bar{\sigma}_z =&\, \sum_{k\in\mathbb{Z}}\vert z^+,k\rangle \langle z^+,k\vert-\vert z^-,k\rangle \langle z^-,k\vert, \label{barsigma} \\
    \bar{\sigma}_{z,\mathrm{d}}\! =&\, \frac{1}{2}\!\sum_{k'=\pm 1}\sum_{k\in\mathbb{Z}}\vert z^+,k+k'\rangle \langle z^+,k\vert-\vert z^-,k+k'\rangle \langle z^-,k\vert.\nonumber
\end{align}
The resulting $\bar{H}_\mathrm{q}$ can then be compactly written as
\begin{align}\label{Hq}
    \bar{H}_\mathrm{q} = \frac{\Delta}{2} \bar{\sigma}_x + \left(A\bar{\sigma}_{z,\mathrm{d}}+\frac{B}{2}\bar{\sigma}_z\right).
\end{align}

\subsection{Relating $\partial\epsilon_{01}/\partial B$ to $g_{k\mu}$}
Here, we establish the relation between the derivative $\partial\epsilon_{01}/\partial B\sim\partial \epsilon_{01}/\partial\phi_\mathrm{dc}$ and the coefficients $g_{k\mu}$. We consider a small perturbation affecting the Floquet Hamiltonian \eqref{Hq} of the type $\bar{H}_\mathrm{q}\to \bar{H}_\mathrm{q} +  \delta B\, \bar{\sigma}_z/2$.  The first-order correction to the quasi-energy difference $\epsilon_{01}$ is given by
\begin{align}
    \delta \epsilon_{01}^{(1)}
    = \frac{\delta B}{2}\left(\langle \bar{w}_1\vert \bar{\sigma}_z \vert \bar{w}_1 \rangle - \langle \bar{w}_0\vert \bar{\sigma}_z \vert \bar{w}_0 \rangle \right).
\end{align}
Making use of the definition of $\bar\sigma_z$ in Eq.\ \eqref{barsigma} and the inner product, we find
\begin{align}
\langle \bar{w}_j\vert \bar{\sigma}_z \vert \bar{w}_{j'} \rangle = \frac{\omega_\mathrm{d}}{2\pi}\int_{0}^{2\pi/\omega_\mathrm{d}} \mathrm{d}t\, \langle w_j(t)\vert \hat{\sigma}_z\vert w_{j'}(t)\rangle,
\label{inner_product}
\end{align}
and thus arrive at the identity
\begin{align}
        \delta \epsilon_{01}^{(1)}=&\, \frac{\delta B}{2}\times \frac{\omega_\mathrm{d}}{2\pi}\int_{0}^{2\pi/\omega_\mathrm{d}} \mathrm{d}t\, \tr_\mathrm{q}[\hat{\sigma}_z\hat{c}_z(t)]=  \delta B \,g_{0\phi},
\end{align}
where the last step uses the definition of $g_{0\phi}$ from Eq.\ \eqref{gk's}. We thus conclude that $\partial \epsilon_{01}/\partial B = g_{0\phi}$.

\subsection{Avoided crossings in the strong-drive limit}
In this and the following subsections, we employ perturbation theory to estimate the gap sizes of avoided crossings, in the strong-drive ($A\gtrsim \Omega_{ge}$) and weak-drive ($A\ll \Omega_{ge}$) limit. In the strong-drive limit, we treat the first term $\bar{V}=\Delta\bar{\sigma}_x/2$ in \eqref{Hq} perturbatively while $\bar{H}_0 = \bar{H}-\bar{V}$ acts as the unperturbed Hamiltonian. The exact eigenstates and eigenvalues of $\bar{H}_0$ are  \cite{Chu_flux_qubit_Floquet, Lupascu_flux_qubit_Floquet}
\begin{align}\label{Floquet_Fourier}
    &\big\vert \bar{w}^{(0)}_{\pm,n}\big\rangle = \sum_{k\in\mathbb{Z}} J_k\left(\mp\frac{A}{\omega_\mathrm{d}}\right)\vert z^{\pm},k-n\rangle,\\
    &\epsilon_{\pm,n}=\pm B/2 + n\omega_\mathrm{d}.\nonumber
\end{align}
Here, we have chosen to adjuster notation according to $j=0,1\to \pm$ which helps keep expressions in the following more compact, but should not be confused with the notation $z^\pm$.

Whenever the drive frequency matches $\omega_\mathrm{d} = B/m$ ($m\in\mathbb{N}$), one finds that the unperturbed quasi-energies $\epsilon_{+,n}$ become $\epsilon_{-,n+m}$ degenerate. This degeneracy is lifted when including corrections of first order in $\Delta$. Perturbation theory yields
\begin{align}
    \Delta^{(1)}_m =&\, 2\big\vert\big\langle \bar{w}^{(0)}_{+,0}\big\vert \bar{V} \big\vert \bar{w}^{(0)}_{-,m}\rangle\big\vert 
    = \Delta\big\vert\big\langle \bar{w}^{(0)}_{+,0}\big\vert \bar{\sigma}_x \big\vert \bar{w}^{(0)}_{-,m}\rangle\big\vert\nonumber.
\end{align}
To proceed, we convert the Floquet states back into the time domain via
\begin{align}
\label{Floquet_time}
    \big\vert w^{(0)}_{\pm, n}(t)\big\rangle = \exp\left(\mp i\frac{A}{\omega_\mathrm{d}}\sin\omega_\mathrm{d}t+in\omega_\mathrm{d}t\right)\vert z^{\pm}\rangle.
\end{align}
(Note that Eqs.~\eqref{Floquet_Fourier} and \eqref{Floquet_time} are related through the Jacobi-Anger expansion.) This enables the evaluation of the leading-order gap size:
\begin{align}
\label{strong_gap_size}
     \Delta^{(1)}_m =&\, \Delta\left|\frac{\omega_\mathrm{d}}{2\pi}\int^{2\pi/\omega_\mathrm{d}}_{0} \mathrm{d}t\, \big\langle w^{(0)}_{+,0}(t)\big\vert \hat{\sigma}_x\big\vert w^{(0)}_{-,m}(t)\big\rangle\right|\nonumber\\
    =&\, \Delta\left|J_m\left(\frac{2A}{\omega_\mathrm{d}}\right)\right|.
\end{align}

\subsection{Avoided crossings in the weak-drive limit}
In the weak-drive limit ($A\ll \Omega_{ge}$), we instead treat the drive-related term $\bar{V}=A\,\bar{\sigma}_{z,\mathrm{d}}$ perturbatively. The unperturbed eigenvalues and eigenstates of $\bar{H}_0 = \bar{H}-\bar{V}$ are given by
\begin{align}
    \big\vert \bar{w}^{(0)}_{\pm,n}\big\rangle &= \cos\frac{\theta}{2}\,\vert z^{\pm},-n\rangle \pm \sin\frac{\theta}{2}\,\vert z^{\mp},-n\rangle,\\
    \epsilon_{\pm,n}&=\pm \Omega_{ge}/2 + n\omega_\mathrm{d}.
\end{align}
These are closely related to the eigenstates and eigenvalues of the undriven qubit. Here, we employ the definitions  $\theta = \tan^{-1}(\Delta/B)$, and $\Omega_{ge}=\sqrt{\Delta^2+B^2}$. 

Whenever the drive frequency obeys $\omega_\mathrm{d}=\Omega_{ge}/m$ ($m\in\mathbb{N}$), the quasi-energies $\epsilon_{+,n} = \epsilon_{-,n+m}$ become degenerate. Again, this degeneracy is lifted by the perturbation $\bar{V}$. For $m=1$, the calculation resembles the one for the strong-drive limit and results in a leading-order gap size of
\begin{align}
    \Delta^{(1)}_{m=1} =&\, 2\big\vert\big\langle \bar{w}^{(0)}_{+,0}\big\vert \bar{V} \big\vert \bar{w}^{(0)}_{-,1}\rangle\big\vert
    = A|\sin\theta|.
\end{align}
The calculation of the gap sizes for $m>1$ requires higher-order degenerate perturbation theory, which we perform using Brillouin-Wigner expansion. This approach converts Eq.~\eqref{eigenvalue_eq} into a reduced equation that only involves the degenerate eigenvector pair $\vert \bar{w}^{(0)}_{+,0}\rangle$ and $\vert \bar{w}^{(0)}_{-,m}\rangle$. 

To facilitate the derivation of the reduced equation, we define the projection operators
\begin{align*}
\bar{P} = \big\vert \bar{w}^{(0)}_{+,0}\big\rangle \big\langle \bar{w}^{(0)}_{+,0}\big\vert+ \big\vert \bar{w}^{(0)}_{-,m}\big\rangle \big\langle \bar{w}^{(0)}_{-,m}\big\vert
\end{align*}
and $\bar{Q} = \bar{\openone}-\bar{P}$, which project vectors in $\mathcal{V}$ onto the degenerate subspace, and onto the subspace orthogonal to it, respectively. Here, $\bar{\openone}$ is the identity operator on  $\mathcal{V}$. According to Brillouin-Wigner theory, the two exact eigenvectors $\vert \bar{w}_j\rangle$ with quasi-energy $\epsilon_j$ obey the equation
\begin{align}
\label{H_deg_eigen}
    \bar{H}_\mathrm{deg} \vert \bar{w}_j\rangle = \epsilon_j\vert \bar{w}_j\rangle,
\end{align}
where
\begin{align}
\bar{H}_{\mathrm{deg}} = \bar{P}(\bar{V}+\bar{V}\bar{T}\bar{V}+ \bar{V}\bar{T}\bar{V}\bar{T}\bar{V}+\cdots )\bar{P},
\label{H_deg}
\end{align}
and
\begin{align}
    \bar{T} = \frac{\bar{Q}}{\epsilon_j-\bar{H}_0}.
\end{align}
Note that despite its appearance, Eq.~\eqref{H_deg_eigen} is not an ordinary eigenvalue problem, since both sides contain the eigenvalue $\epsilon_j$. It is possible to find a solution for the eigenvalues iteratively. To avoid excessive notation, we focus on the $j=0$ eigenvalue and omit unnecessary subscripts in the following. In the first iteration, we insert the unperturbed quasi-energy $\epsilon^{(0)}=\epsilon_{+,0}$ into the left-hand side of Eq.~\eqref{H_deg_eigen}, and solve for $\epsilon^{(1)}$ on the right-hand side. Using the new quasi-energy approximation, we then repeat these steps to include higher-order corrections. With this procedure, we find that, to leading order in $A$, the gap size is given by
\begin{align}
    \Delta_m \approx  |\sin\theta\cos^{m-1}\theta| \frac{A^m}{(m-1)!\,\omega^{m-1}_\mathrm{d}}.
\end{align}

\subsection{Gap size and the width of $T_\phi$ peaks surrounding sweet-spot manifolds}
 In this subsection, we establish the relation between the gap size $\Delta_m$ and the width of the $T_\phi$ peaks along the drive-frequency axis surrounding sweet-spot manifolds.  We derive this relation only for the strong-drive limit; the derivation for the weak-drive limit is analogous.
 
 Generically, the pure-dephasing rate of a Floquet qubit is likely to be dominated by the $1/f$ noise contributions away from sweet spots. In our case, that noise correspond to flux noise which limits the system, whenever the derivative of the quasi-energy difference with respect to flux is nonzero, $\partial\epsilon_{01}/\partial B\neq 0$.  Under these conditions, Eq.~\eqref{gammaphi} implies that  $T_\phi$ is inversely proportional to  $|\partial\epsilon_{01}/\partial B|$. Therefore, to find the drive-frequency width of the $T_\phi$ peaks, it is useful to first explore how $|\partial\epsilon_{01}/\partial B|$ depends on $\omega_\mathrm{d}$.

 For a dynamical sweet spot in the strong-drive limit, $A_0\gg \Omega_{ge}$,  the drive parameters satisfy $\omega_{\mathrm{d},m}=B_0/m$. At the sweet spot, the quasi-energy derivative vanishes, $\partial\epsilon_{01}/\partial B=0$. Let us consider values $B$ and $\omega_\mathrm{d}$ in the vicinity of the sweet-spot point given by $B_0$ and $\omega_{\mathrm{d},m}$. Using  Eq.\ \eqref{Floquet_Fourier}, we see that the Hamiltonian in the relevant subspace is
 \begin{align}
 \bar{H}&=\bar{H}_0+\bar{V} \\\nonumber
 &= \epsilon_{+,0}|\bar{w}_{+,0}^{(0)}\rangle\langle\bar{w}_{+,0}^{(0)}|+\epsilon_{-,m}|\bar{w}_{-,m}^{(0)}\rangle\langle\bar{w}_{-,m}^{(0)}|+\Delta\bar{\sigma}_x/2,
 \end{align}
which results in the quasi-energy difference 
\begin{align}
     \epsilon_{01}\approx \sqrt{\Delta_m^2 
     +(B-m \omega_\mathrm{d})^2}.
\end{align}
The derivative of $\epsilon_{01}$ with respect to $B$ is thus
\begin{align}
     \frac{\partial \epsilon_{01}}{\partial B}  \approx \frac{ B-m\omega_\mathrm{d}}{\epsilon_{01}}.
\end{align}
Since we are interested in the width of the sweet manifold along the $\omega_\mathrm{d}$-axis, we set $B=B_0$, and consider variations of $\omega_\mathrm{d}$ around $\omega_{\mathrm{d},m}$. As a function of $\omega_\mathrm{d}$, the derivative $|\partial\epsilon_{01}/\partial B|$ takes on its minimum value of zero at $\omega_\mathrm{d} = \omega_{\mathrm{d},m}$. Away from this sweet spot, $|\partial\epsilon_{01}/\partial B|$ has an upper bound of 1, which is reached asymptotically in the limit $m|\omega_\mathrm{d}-\omega_{\mathrm{d},m}| \gg \Delta_m$. Based on this, we can use the full width at half minimum (FWHm) of $|\partial \epsilon_{01}/\partial B|$ as an estimate of the peak width of $T_\phi$. The condition $|\partial\epsilon_{01}/\partial B|=1/2$ for reaching the half-minimum value, results in the equation 
\begin{align}\label{FWHm}
    \frac{m|\omega_\mathrm{d}-\omega_{\mathrm{d},m}|}{\sqrt{\Delta^2_m+m^2(\omega_\mathrm{d}-\omega_{\mathrm{d},m})^2}} = \frac{1}{2}.
\end{align}
The corresponding two solutions $\omega_\text{d}^{(1,2)}$ yield the FWHm $|\omega_\text{d}^{(2)}-\omega_\text{d}^{(1)}|$. Due the dependence of $\Delta_m$ on $\omega_\text{d}$ involving a Bessel function [Eq.\ \eqref{strong_gap_size}], the above equation \eqref{FWHm} is transcendental. We can obtain analytical approximations as follows. We rewrite Eq. \eqref{FWHm} in the form  $\sqrt{3} m|\omega_\mathrm{d}-\omega_{\mathrm{d},m}| = \Delta_m$,  and expanding the latter in  $\omega_\text{d}$ around $\omega_{\text{d},m}$. The result of this is another transcendental equation, in which the problematic Bessel function term can, however, be neglected if $\partial \Delta_m/\partial\omega_\mathrm{d}\ll \sqrt{3}m$ holds. We have verified the validity of this inequality for our parameters numerically, and this way finally obtain the approximate FWHm
\begin{equation}
\Delta \omega_\text{FWHm} =2\Delta_{m,0}/\sqrt{3}m,
\end{equation}
where $\Delta_{m,0}= \Delta|J_m(2A_0/\omega_{\mathrm{d},m})|$.

\bibliography{mybib_1}

\begin{thebibliography}{80}%
\makeatletter
\providecommand \@ifxundefined [1]{%
 \@ifx{#1\undefined}
}%
\providecommand \@ifnum [1]{%
 \ifnum #1\expandafter \@firstoftwo
 \else \expandafter \@secondoftwo
 \fi
}%
\providecommand \@ifx [1]{%
 \ifx #1\expandafter \@firstoftwo
 \else \expandafter \@secondoftwo
 \fi
}%
\providecommand \natexlab [1]{#1}%
\providecommand \enquote  [1]{``#1''}%
\providecommand \bibnamefont  [1]{#1}%
\providecommand \bibfnamefont [1]{#1}%
\providecommand \citenamefont [1]{#1}%
\providecommand \href@noop [0]{\@secondoftwo}%
\providecommand \href [0]{\begingroup \@sanitize@url \@href}%
\providecommand \@href[1]{\@@startlink{#1}\@@href}%
\providecommand \@@href[1]{\endgroup#1\@@endlink}%
\providecommand \@sanitize@url [0]{\catcode `\\12\catcode `\$12\catcode
  `\&12\catcode `\#12\catcode `\^12\catcode `\_12\catcode `\%12\relax}%
\providecommand \@@startlink[1]{}%
\providecommand \@@endlink[0]{}%
\providecommand \url  [0]{\begingroup\@sanitize@url \@url }%
\providecommand \@url [1]{\endgroup\@href {#1}{\urlprefix }}%
\providecommand \urlprefix  [0]{URL }%
\providecommand \Eprint [0]{\href }%
\providecommand \doibase [0]{https://doi.org/}%
\providecommand \selectlanguage [0]{\@gobble}%
\providecommand \bibinfo  [0]{\@secondoftwo}%
\providecommand \bibfield  [0]{\@secondoftwo}%
\providecommand \translation [1]{[#1]}%
\providecommand \BibitemOpen [0]{}%
\providecommand \bibitemStop [0]{}%
\providecommand \bibitemNoStop [0]{.\EOS\space}%
\providecommand \EOS [0]{\spacefactor3000\relax}%
\providecommand \BibitemShut  [1]{\csname bibitem#1\endcsname}%
\let\auto@bib@innerbib\@empty
\bibitem [{\citenamefont {van~der Wal}\ \emph {et~al.}(2000)\citenamefont
  {van~der Wal}, \citenamefont {ter Haar}, \citenamefont {Wilhelm},
  \citenamefont {Schouten}, \citenamefont {Harmans}, \citenamefont {Orlando},
  \citenamefont {Lloyd},\ and\ \citenamefont {Mooij}}]{Charge_qubit}%
  \BibitemOpen
  \bibfield  {author} {\bibinfo {author} {\bibfnamefont {C.~H.}\ \bibnamefont
  {van~der Wal}}, \bibinfo {author} {\bibfnamefont {A.~C.~J.}\ \bibnamefont
  {ter Haar}}, \bibinfo {author} {\bibfnamefont {F.~K.}\ \bibnamefont
  {Wilhelm}}, \bibinfo {author} {\bibfnamefont {R.~N.}\ \bibnamefont
  {Schouten}}, \bibinfo {author} {\bibfnamefont {C.~J. P.~M.}\ \bibnamefont
  {Harmans}}, \bibinfo {author} {\bibfnamefont {T.~P.}\ \bibnamefont
  {Orlando}}, \bibinfo {author} {\bibfnamefont {S.}~\bibnamefont {Lloyd}},\
  and\ \bibinfo {author} {\bibfnamefont {J.~E.}\ \bibnamefont {Mooij}},\
  }\bibfield  {title} {\bibinfo {title} {\textit{Quantum Superposition of
  Macroscopic Persistent-Current States}},\ }\href
  {https://doi.org/10.1126/science.290.5492.773} {\bibfield  {journal}
  {\bibinfo  {journal} {Science}\ }\textbf {\bibinfo {volume} {290}},\ \bibinfo
  {pages} {773} (\bibinfo {year} {2000})}\BibitemShut {NoStop}%
\bibitem [{\citenamefont {Nakamura}\ \emph {et~al.}(2002)\citenamefont
  {Nakamura}, \citenamefont {Pashkin}, \citenamefont {Yamamoto},\ and\
  \citenamefont {Tsai}}]{Nakamura_charge_1/f}%
  \BibitemOpen
  \bibfield  {author} {\bibinfo {author} {\bibfnamefont {Y.}~\bibnamefont
  {Nakamura}}, \bibinfo {author} {\bibfnamefont {Y.~A.}\ \bibnamefont
  {Pashkin}}, \bibinfo {author} {\bibfnamefont {T.}~\bibnamefont {Yamamoto}},\
  and\ \bibinfo {author} {\bibfnamefont {J.~S.}\ \bibnamefont {Tsai}},\
  }\bibfield  {title} {\bibinfo {title} {\textit{Charge Echo in a Cooper-Pair
  Box}},\ }\href {https://doi.org/10.1103/PhysRevLett.88.047901} {\bibfield
  {journal} {\bibinfo  {journal} {Phys. Rev. Lett.}\ }\textbf {\bibinfo
  {volume} {88}},\ \bibinfo {pages} {047901} (\bibinfo {year}
  {2002})}\BibitemShut {NoStop}%
\bibitem [{\citenamefont {Yoshihara}\ \emph {et~al.}(2006)\citenamefont
  {Yoshihara}, \citenamefont {Harrabi}, \citenamefont {Niskanen}, \citenamefont
  {Nakamura},\ and\ \citenamefont {Tsai}}]{JSTsai_flux_1/f}%
  \BibitemOpen
  \bibfield  {author} {\bibinfo {author} {\bibfnamefont {F.}~\bibnamefont
  {Yoshihara}}, \bibinfo {author} {\bibfnamefont {K.}~\bibnamefont {Harrabi}},
  \bibinfo {author} {\bibfnamefont {A.~O.}\ \bibnamefont {Niskanen}}, \bibinfo
  {author} {\bibfnamefont {Y.}~\bibnamefont {Nakamura}},\ and\ \bibinfo
  {author} {\bibfnamefont {J.~S.}\ \bibnamefont {Tsai}},\ }\bibfield  {title}
  {\bibinfo {title} {\textit{Decoherence of Flux Qubits Due to $1/f$ Flux
  Noise}},\ }\href {https://doi.org/10.1103/PhysRevLett.97.167001} {\bibfield
  {journal} {\bibinfo  {journal} {Phys. Rev. Lett.}\ }\textbf {\bibinfo
  {volume} {97}},\ \bibinfo {pages} {167001} (\bibinfo {year}
  {2006})}\BibitemShut {NoStop}%
\bibitem [{\citenamefont {Anton}\ \emph {et~al.}(2012)\citenamefont {Anton},
  \citenamefont {M\"uller}, \citenamefont {Birenbaum}, \citenamefont
  {O'Kelley}, \citenamefont {Fefferman}, \citenamefont {Golubev}, \citenamefont
  {Hilton}, \citenamefont {Cho}, \citenamefont {Irwin}, \citenamefont
  {Wellstood}, \citenamefont {Sch\"on}, \citenamefont {Shnirman},\ and\
  \citenamefont {Clarke}}]{JClarke_flux_qubit_1/f}%
  \BibitemOpen
  \bibfield  {author} {\bibinfo {author} {\bibfnamefont {S.~M.}\ \bibnamefont
  {Anton}}, \bibinfo {author} {\bibfnamefont {C.}~\bibnamefont {M\"uller}},
  \bibinfo {author} {\bibfnamefont {J.~S.}\ \bibnamefont {Birenbaum}}, \bibinfo
  {author} {\bibfnamefont {S.~R.}\ \bibnamefont {O'Kelley}}, \bibinfo {author}
  {\bibfnamefont {A.~D.}\ \bibnamefont {Fefferman}}, \bibinfo {author}
  {\bibfnamefont {D.~S.}\ \bibnamefont {Golubev}}, \bibinfo {author}
  {\bibfnamefont {G.~C.}\ \bibnamefont {Hilton}}, \bibinfo {author}
  {\bibfnamefont {H.-M.}\ \bibnamefont {Cho}}, \bibinfo {author} {\bibfnamefont
  {K.~D.}\ \bibnamefont {Irwin}}, \bibinfo {author} {\bibfnamefont {F.~C.}\
  \bibnamefont {Wellstood}}, \bibinfo {author} {\bibfnamefont {G.}~\bibnamefont
  {Sch\"on}}, \bibinfo {author} {\bibfnamefont {A.}~\bibnamefont {Shnirman}},\
  and\ \bibinfo {author} {\bibfnamefont {J.}~\bibnamefont {Clarke}},\
  }\bibfield  {title} {\bibinfo {title} {\textit{Pure Dephasing in Flux Qubits
  Due to Flux Noise with Spectral Density Scaling as}
  $1/{f}^{\ensuremath{\alpha}}$},\ }\href
  {https://doi.org/10.1103/PhysRevB.85.224505} {\bibfield  {journal} {\bibinfo
  {journal} {Phys. Rev. B}\ }\textbf {\bibinfo {volume} {85}},\ \bibinfo
  {pages} {224505} (\bibinfo {year} {2012})}\BibitemShut {NoStop}%
\bibitem [{\citenamefont {Ithier}\ \emph {et~al.}(2005)\citenamefont {Ithier},
  \citenamefont {Collin}, \citenamefont {Joyez}, \citenamefont {Meeson},
  \citenamefont {Vion}, \citenamefont {Esteve}, \citenamefont {Chiarello},
  \citenamefont {Shnirman}, \citenamefont {Makhlin}, \citenamefont {Schriefl},\
  and\ \citenamefont {Sch\"on}}]{Ithier_decoherence_analysis}%
  \BibitemOpen
  \bibfield  {author} {\bibinfo {author} {\bibfnamefont {G.}~\bibnamefont
  {Ithier}}, \bibinfo {author} {\bibfnamefont {E.}~\bibnamefont {Collin}},
  \bibinfo {author} {\bibfnamefont {P.}~\bibnamefont {Joyez}}, \bibinfo
  {author} {\bibfnamefont {P.~J.}\ \bibnamefont {Meeson}}, \bibinfo {author}
  {\bibfnamefont {D.}~\bibnamefont {Vion}}, \bibinfo {author} {\bibfnamefont
  {D.}~\bibnamefont {Esteve}}, \bibinfo {author} {\bibfnamefont
  {F.}~\bibnamefont {Chiarello}}, \bibinfo {author} {\bibfnamefont
  {A.}~\bibnamefont {Shnirman}}, \bibinfo {author} {\bibfnamefont
  {Y.}~\bibnamefont {Makhlin}}, \bibinfo {author} {\bibfnamefont
  {J.}~\bibnamefont {Schriefl}},\ and\ \bibinfo {author} {\bibfnamefont
  {G.}~\bibnamefont {Sch\"on}},\ }\bibfield  {title} {\bibinfo {title}
  {\textit{Decoherence in a Superconducting Quantum Bit Circuit}},\ }\href
  {https://doi.org/10.1103/PhysRevB.72.134519} {\bibfield  {journal} {\bibinfo
  {journal} {Phys. Rev. B}\ }\textbf {\bibinfo {volume} {72}},\ \bibinfo
  {pages} {134519} (\bibinfo {year} {2005})}\BibitemShut {NoStop}%
\bibitem [{\citenamefont {Koch}\ \emph {et~al.}(2007)\citenamefont {Koch},
  \citenamefont {Yu}, \citenamefont {Gambetta}, \citenamefont {Houck},
  \citenamefont {Schuster}, \citenamefont {Majer}, \citenamefont {Blais},
  \citenamefont {Devoret}, \citenamefont {Girvin},\ and\ \citenamefont
  {Schoelkopf}}]{Koch_transmon_theory}%
  \BibitemOpen
  \bibfield  {author} {\bibinfo {author} {\bibfnamefont {J.}~\bibnamefont
  {Koch}}, \bibinfo {author} {\bibfnamefont {T.~M.}\ \bibnamefont {Yu}},
  \bibinfo {author} {\bibfnamefont {J.}~\bibnamefont {Gambetta}}, \bibinfo
  {author} {\bibfnamefont {A.~A.}\ \bibnamefont {Houck}}, \bibinfo {author}
  {\bibfnamefont {D.~I.}\ \bibnamefont {Schuster}}, \bibinfo {author}
  {\bibfnamefont {J.}~\bibnamefont {Majer}}, \bibinfo {author} {\bibfnamefont
  {A.}~\bibnamefont {Blais}}, \bibinfo {author} {\bibfnamefont {M.~H.}\
  \bibnamefont {Devoret}}, \bibinfo {author} {\bibfnamefont {S.~M.}\
  \bibnamefont {Girvin}},\ and\ \bibinfo {author} {\bibfnamefont {R.~J.}\
  \bibnamefont {Schoelkopf}},\ }\bibfield  {title} {\bibinfo {title}
  {\textit{Charge-Insensitive Qubit Design Derived from the Cooper Pair Box}},\
  }\href {https://doi.org/10.1103/PhysRevA.76.042319} {\bibfield  {journal}
  {\bibinfo  {journal} {Phys. Rev. A}\ }\textbf {\bibinfo {volume} {76}},\
  \bibinfo {pages} {042319} (\bibinfo {year} {2007})}\BibitemShut {NoStop}%
\bibitem [{\citenamefont {Nguyen}\ \emph {et~al.}(2019)\citenamefont {Nguyen},
  \citenamefont {Lin}, \citenamefont {Somoroff}, \citenamefont {Mencia},
  \citenamefont {Grabon},\ and\ \citenamefont {Manucharyan}}]{Fluxonium_hc}%
  \BibitemOpen
  \bibfield  {author} {\bibinfo {author} {\bibfnamefont {L.~B.}\ \bibnamefont
  {Nguyen}}, \bibinfo {author} {\bibfnamefont {Y.-H.}\ \bibnamefont {Lin}},
  \bibinfo {author} {\bibfnamefont {A.}~\bibnamefont {Somoroff}}, \bibinfo
  {author} {\bibfnamefont {R.}~\bibnamefont {Mencia}}, \bibinfo {author}
  {\bibfnamefont {N.}~\bibnamefont {Grabon}},\ and\ \bibinfo {author}
  {\bibfnamefont {V.~E.}\ \bibnamefont {Manucharyan}},\ }\bibfield  {title}
  {\bibinfo {title} {\textit{High-Coherence Fluxonium Qubit}},\ }\href
  {https://doi.org/10.1103/PhysRevX.9.041041} {\bibfield  {journal} {\bibinfo
  {journal} {Phys. Rev. X}\ }\textbf {\bibinfo {volume} {9}},\ \bibinfo {pages}
  {041041} (\bibinfo {year} {2019})}\BibitemShut {NoStop}%
\bibitem [{\citenamefont {Zhang}\ \emph {et~al.}()\citenamefont {Zhang},
  \citenamefont {Chakram}, \citenamefont {Roy}, \citenamefont {Earnest},
  \citenamefont {Lu}, \citenamefont {Huang}, \citenamefont {Weiss},
  \citenamefont {Koch},\ and\ \citenamefont
  {Schuster}}]{Schuster_heavy_fluxonium_control}%
  \BibitemOpen
  \bibfield  {author} {\bibinfo {author} {\bibfnamefont {H.}~\bibnamefont
  {Zhang}}, \bibinfo {author} {\bibfnamefont {S.}~\bibnamefont {Chakram}},
  \bibinfo {author} {\bibfnamefont {T.}~\bibnamefont {Roy}}, \bibinfo {author}
  {\bibfnamefont {N.}~\bibnamefont {Earnest}}, \bibinfo {author} {\bibfnamefont
  {Y.}~\bibnamefont {Lu}}, \bibinfo {author} {\bibfnamefont {Z.}~\bibnamefont
  {Huang}}, \bibinfo {author} {\bibfnamefont {D.}~\bibnamefont {Weiss}},
  \bibinfo {author} {\bibfnamefont {J.}~\bibnamefont {Koch}},\ and\ \bibinfo
  {author} {\bibfnamefont {D.~I.}\ \bibnamefont {Schuster}},\ }\bibfield
  {title} {\bibinfo {title} {\textit{Universal Fast Flux Control of a Coherent,
  Low-Frequency Qubit}},\ }\Eprint {https://arxiv.org/abs/arXiv:2002.10653}
  {arXiv:2002.10653} \BibitemShut {NoStop}%
\bibitem [{\citenamefont {Sete}\ \emph {et~al.}(2017)\citenamefont {Sete},
  \citenamefont {Reagor}, \citenamefont {Didier},\ and\ \citenamefont
  {Rigetti}}]{Rigetti_flux_sweet_spot}%
  \BibitemOpen
  \bibfield  {author} {\bibinfo {author} {\bibfnamefont {E.~A.}\ \bibnamefont
  {Sete}}, \bibinfo {author} {\bibfnamefont {M.~J.}\ \bibnamefont {Reagor}},
  \bibinfo {author} {\bibfnamefont {N.}~\bibnamefont {Didier}},\ and\ \bibinfo
  {author} {\bibfnamefont {C.~T.}\ \bibnamefont {Rigetti}},\ }\bibfield
  {title} {\bibinfo {title} {\textit{Charge- and Flux-Insensitive Tunable
  Superconducting Qubit}},\ }\href
  {https://doi.org/10.1103/PhysRevApplied.8.024004} {\bibfield  {journal}
  {\bibinfo  {journal} {Phys. Rev. Applied}\ }\textbf {\bibinfo {volume} {8}},\
  \bibinfo {pages} {024004} (\bibinfo {year} {2017})}\BibitemShut {NoStop}%
\bibitem [{\citenamefont {Hutchings}\ \emph {et~al.}(2017)\citenamefont
  {Hutchings}, \citenamefont {Hertzberg}, \citenamefont {Liu}, \citenamefont
  {Bronn}, \citenamefont {Keefe}, \citenamefont {Brink}, \citenamefont {Chow},\
  and\ \citenamefont {Plourde}}]{IBM_frequency_tunable_transmon}%
  \BibitemOpen
  \bibfield  {author} {\bibinfo {author} {\bibfnamefont {M.~D.}\ \bibnamefont
  {Hutchings}}, \bibinfo {author} {\bibfnamefont {J.~B.}\ \bibnamefont
  {Hertzberg}}, \bibinfo {author} {\bibfnamefont {Y.}~\bibnamefont {Liu}},
  \bibinfo {author} {\bibfnamefont {N.~T.}\ \bibnamefont {Bronn}}, \bibinfo
  {author} {\bibfnamefont {G.~A.}\ \bibnamefont {Keefe}}, \bibinfo {author}
  {\bibfnamefont {M.}~\bibnamefont {Brink}}, \bibinfo {author} {\bibfnamefont
  {J.~M.}\ \bibnamefont {Chow}},\ and\ \bibinfo {author} {\bibfnamefont
  {B.~L.~T.}\ \bibnamefont {Plourde}},\ }\bibfield  {title} {\bibinfo {title}
  {\textit{Tunable Superconducting Qubits with Flux-Independent Coherence}},\
  }\href {https://doi.org/10.1103/PhysRevApplied.8.044003} {\bibfield
  {journal} {\bibinfo  {journal} {Phys. Rev. Applied}\ }\textbf {\bibinfo
  {volume} {8}},\ \bibinfo {pages} {044003} (\bibinfo {year}
  {2017})}\BibitemShut {NoStop}%
\bibitem [{\citenamefont {Barends}\ \emph {et~al.}(2013)\citenamefont
  {Barends}, \citenamefont {Kelly}, \citenamefont {Megrant}, \citenamefont
  {Sank}, \citenamefont {Jeffrey}, \citenamefont {Chen}, \citenamefont {Yin},
  \citenamefont {Chiaro}, \citenamefont {Mutus}, \citenamefont {Neill},
  \citenamefont {O'Malley}, \citenamefont {Roushan}, \citenamefont {Wenner},
  \citenamefont {White}, \citenamefont {Cleland},\ and\ \citenamefont
  {Martinis}}]{Barends_xmon}%
  \BibitemOpen
  \bibfield  {author} {\bibinfo {author} {\bibfnamefont {R.}~\bibnamefont
  {Barends}}, \bibinfo {author} {\bibfnamefont {J.}~\bibnamefont {Kelly}},
  \bibinfo {author} {\bibfnamefont {A.}~\bibnamefont {Megrant}}, \bibinfo
  {author} {\bibfnamefont {D.}~\bibnamefont {Sank}}, \bibinfo {author}
  {\bibfnamefont {E.}~\bibnamefont {Jeffrey}}, \bibinfo {author} {\bibfnamefont
  {Y.}~\bibnamefont {Chen}}, \bibinfo {author} {\bibfnamefont {Y.}~\bibnamefont
  {Yin}}, \bibinfo {author} {\bibfnamefont {B.}~\bibnamefont {Chiaro}},
  \bibinfo {author} {\bibfnamefont {J.}~\bibnamefont {Mutus}}, \bibinfo
  {author} {\bibfnamefont {C.}~\bibnamefont {Neill}}, \bibinfo {author}
  {\bibfnamefont {P.}~\bibnamefont {O'Malley}}, \bibinfo {author}
  {\bibfnamefont {P.}~\bibnamefont {Roushan}}, \bibinfo {author} {\bibfnamefont
  {J.}~\bibnamefont {Wenner}}, \bibinfo {author} {\bibfnamefont {T.~C.}\
  \bibnamefont {White}}, \bibinfo {author} {\bibfnamefont {A.~N.}\ \bibnamefont
  {Cleland}},\ and\ \bibinfo {author} {\bibfnamefont {J.~M.}\ \bibnamefont
  {Martinis}},\ }\bibfield  {title} {\bibinfo {title} {\textit{Coherent
  Josephson Qubit Suitable for Scalable Quantum Integrated Circuits}},\ }\href
  {https://doi.org/10.1103/PhysRevLett.111.080502} {\bibfield  {journal}
  {\bibinfo  {journal} {Phys. Rev. Lett.}\ }\textbf {\bibinfo {volume} {111}},\
  \bibinfo {pages} {080502} (\bibinfo {year} {2013})}\BibitemShut {NoStop}%
\bibitem [{\citenamefont {Earnest}\ \emph {et~al.}(2018)\citenamefont
  {Earnest}, \citenamefont {Chakram}, \citenamefont {Lu}, \citenamefont
  {Irons}, \citenamefont {Naik}, \citenamefont {Leung}, \citenamefont {Ocola},
  \citenamefont {Czaplewski}, \citenamefont {Baker}, \citenamefont {Lawrence},
  \citenamefont {Koch},\ and\ \citenamefont
  {Schuster}}]{Schuster_heavy_fluxonium}%
  \BibitemOpen
  \bibfield  {author} {\bibinfo {author} {\bibfnamefont {N.}~\bibnamefont
  {Earnest}}, \bibinfo {author} {\bibfnamefont {S.}~\bibnamefont {Chakram}},
  \bibinfo {author} {\bibfnamefont {Y.}~\bibnamefont {Lu}}, \bibinfo {author}
  {\bibfnamefont {N.}~\bibnamefont {Irons}}, \bibinfo {author} {\bibfnamefont
  {R.~K.}\ \bibnamefont {Naik}}, \bibinfo {author} {\bibfnamefont
  {N.}~\bibnamefont {Leung}}, \bibinfo {author} {\bibfnamefont
  {L.}~\bibnamefont {Ocola}}, \bibinfo {author} {\bibfnamefont {D.~A.}\
  \bibnamefont {Czaplewski}}, \bibinfo {author} {\bibfnamefont
  {B.}~\bibnamefont {Baker}}, \bibinfo {author} {\bibfnamefont
  {J.}~\bibnamefont {Lawrence}}, \bibinfo {author} {\bibfnamefont
  {J.}~\bibnamefont {Koch}},\ and\ \bibinfo {author} {\bibfnamefont {D.~I.}\
  \bibnamefont {Schuster}},\ }\bibfield  {title} {\bibinfo {title}
  {\textit{Realization of a $\mathrm{\ensuremath{\Lambda}}$ System with
  Metastable States of a Capacitively Shunted Fluxonium}},\ }\href
  {https://doi.org/10.1103/PhysRevLett.120.150504} {\bibfield  {journal}
  {\bibinfo  {journal} {Phys. Rev. Lett.}\ }\textbf {\bibinfo {volume} {120}},\
  \bibinfo {pages} {150504} (\bibinfo {year} {2018})}\BibitemShut {NoStop}%
\bibitem [{\citenamefont {Lin}\ \emph {et~al.}(2018)\citenamefont {Lin},
  \citenamefont {Nguyen}, \citenamefont {Grabon}, \citenamefont {San~Miguel},
  \citenamefont {Pankratova},\ and\ \citenamefont
  {Manucharyan}}]{Manucharyan_heavy_fluxonium}%
  \BibitemOpen
  \bibfield  {author} {\bibinfo {author} {\bibfnamefont {Y.-H.}\ \bibnamefont
  {Lin}}, \bibinfo {author} {\bibfnamefont {L.~B.}\ \bibnamefont {Nguyen}},
  \bibinfo {author} {\bibfnamefont {N.}~\bibnamefont {Grabon}}, \bibinfo
  {author} {\bibfnamefont {J.}~\bibnamefont {San~Miguel}}, \bibinfo {author}
  {\bibfnamefont {N.}~\bibnamefont {Pankratova}},\ and\ \bibinfo {author}
  {\bibfnamefont {V.~E.}\ \bibnamefont {Manucharyan}},\ }\bibfield  {title}
  {\bibinfo {title} {\textit{Demonstration of Protection of a Superconducting
  Qubit from Energy Decay}},\ }\href
  {https://doi.org/10.1103/PhysRevLett.120.150503} {\bibfield  {journal}
  {\bibinfo  {journal} {Phys. Rev. Lett.}\ }\textbf {\bibinfo {volume} {120}},\
  \bibinfo {pages} {150503} (\bibinfo {year} {2018})}\BibitemShut {NoStop}%
\bibitem [{\citenamefont {Hazard}\ \emph {et~al.}(2019)\citenamefont {Hazard},
  \citenamefont {Gyenis}, \citenamefont {Di~Paolo}, \citenamefont {Asfaw},
  \citenamefont {Lyon}, \citenamefont {Blais},\ and\ \citenamefont
  {Houck}}]{Nanowire_fluxonium}%
  \BibitemOpen
  \bibfield  {author} {\bibinfo {author} {\bibfnamefont {T.~M.}\ \bibnamefont
  {Hazard}}, \bibinfo {author} {\bibfnamefont {A.}~\bibnamefont {Gyenis}},
  \bibinfo {author} {\bibfnamefont {A.}~\bibnamefont {Di~Paolo}}, \bibinfo
  {author} {\bibfnamefont {A.~T.}\ \bibnamefont {Asfaw}}, \bibinfo {author}
  {\bibfnamefont {S.~A.}\ \bibnamefont {Lyon}}, \bibinfo {author}
  {\bibfnamefont {A.}~\bibnamefont {Blais}},\ and\ \bibinfo {author}
  {\bibfnamefont {A.~A.}\ \bibnamefont {Houck}},\ }\bibfield  {title} {\bibinfo
  {title} {\textit{Nanowire Superinductance Fluxonium Qubit}},\ }\href
  {https://doi.org/10.1103/PhysRevLett.122.010504} {\bibfield  {journal}
  {\bibinfo  {journal} {Phys. Rev. Lett.}\ }\textbf {\bibinfo {volume} {122}},\
  \bibinfo {pages} {010504} (\bibinfo {year} {2019})}\BibitemShut {NoStop}%
\bibitem [{\citenamefont {Martinis}\ \emph {et~al.}(2003)\citenamefont
  {Martinis}, \citenamefont {Nam}, \citenamefont {Aumentado}, \citenamefont
  {Lang},\ and\ \citenamefont {Urbina}}]{Martinis_decoherence_bias_noise}%
  \BibitemOpen
  \bibfield  {author} {\bibinfo {author} {\bibfnamefont {J.~M.}\ \bibnamefont
  {Martinis}}, \bibinfo {author} {\bibfnamefont {S.}~\bibnamefont {Nam}},
  \bibinfo {author} {\bibfnamefont {J.}~\bibnamefont {Aumentado}}, \bibinfo
  {author} {\bibfnamefont {K.~M.}\ \bibnamefont {Lang}},\ and\ \bibinfo
  {author} {\bibfnamefont {C.}~\bibnamefont {Urbina}},\ }\bibfield  {title}
  {\bibinfo {title} {\textit{Decoherence of a Superconducting Qubit Due to Bias
  Noise}},\ }\href {https://doi.org/10.1103/PhysRevB.67.094510} {\bibfield
  {journal} {\bibinfo  {journal} {Phys. Rev. B}\ }\textbf {\bibinfo {volume}
  {67}},\ \bibinfo {pages} {094510} (\bibinfo {year} {2003})}\BibitemShut
  {NoStop}%
\bibitem [{\citenamefont {Quintana}\ \emph {et~al.}(2017)\citenamefont
  {Quintana}, \citenamefont {Chen}, \citenamefont {Sank}, \citenamefont
  {Petukhov}, \citenamefont {White}, \citenamefont {Kafri}, \citenamefont
  {Chiaro}, \citenamefont {Megrant}, \citenamefont {Barends}, \citenamefont
  {Campbell}, \citenamefont {Chen}, \citenamefont {Dunsworth}, \citenamefont
  {Fowler}, \citenamefont {Graff}, \citenamefont {Jeffrey}, \citenamefont
  {Kelly}, \citenamefont {Lucero}, \citenamefont {Mutus}, \citenamefont
  {Neeley}, \citenamefont {Neill}, \citenamefont {O'Malley}, \citenamefont
  {Roushan}, \citenamefont {Shabani}, \citenamefont {Smelyanskiy},
  \citenamefont {Vainsencher}, \citenamefont {Wenner}, \citenamefont {Neven},\
  and\ \citenamefont {Martinis}}]{Martinis_1/f_Crossover}%
  \BibitemOpen
  \bibfield  {author} {\bibinfo {author} {\bibfnamefont {C.~M.}\ \bibnamefont
  {Quintana}}, \bibinfo {author} {\bibfnamefont {Y.}~\bibnamefont {Chen}},
  \bibinfo {author} {\bibfnamefont {D.}~\bibnamefont {Sank}}, \bibinfo {author}
  {\bibfnamefont {A.~G.}\ \bibnamefont {Petukhov}}, \bibinfo {author}
  {\bibfnamefont {T.~C.}\ \bibnamefont {White}}, \bibinfo {author}
  {\bibfnamefont {D.}~\bibnamefont {Kafri}}, \bibinfo {author} {\bibfnamefont
  {B.}~\bibnamefont {Chiaro}}, \bibinfo {author} {\bibfnamefont
  {A.}~\bibnamefont {Megrant}}, \bibinfo {author} {\bibfnamefont
  {R.}~\bibnamefont {Barends}}, \bibinfo {author} {\bibfnamefont
  {B.}~\bibnamefont {Campbell}}, \bibinfo {author} {\bibfnamefont
  {Z.}~\bibnamefont {Chen}}, \bibinfo {author} {\bibfnamefont {A.}~\bibnamefont
  {Dunsworth}}, \bibinfo {author} {\bibfnamefont {A.~G.}\ \bibnamefont
  {Fowler}}, \bibinfo {author} {\bibfnamefont {R.}~\bibnamefont {Graff}},
  \bibinfo {author} {\bibfnamefont {E.}~\bibnamefont {Jeffrey}}, \bibinfo
  {author} {\bibfnamefont {J.}~\bibnamefont {Kelly}}, \bibinfo {author}
  {\bibfnamefont {E.}~\bibnamefont {Lucero}}, \bibinfo {author} {\bibfnamefont
  {J.~Y.}\ \bibnamefont {Mutus}}, \bibinfo {author} {\bibfnamefont
  {M.}~\bibnamefont {Neeley}}, \bibinfo {author} {\bibfnamefont
  {C.}~\bibnamefont {Neill}}, \bibinfo {author} {\bibfnamefont {P.~J.~J.}\
  \bibnamefont {O'Malley}}, \bibinfo {author} {\bibfnamefont {P.}~\bibnamefont
  {Roushan}}, \bibinfo {author} {\bibfnamefont {A.}~\bibnamefont {Shabani}},
  \bibinfo {author} {\bibfnamefont {V.~N.}\ \bibnamefont {Smelyanskiy}},
  \bibinfo {author} {\bibfnamefont {A.}~\bibnamefont {Vainsencher}}, \bibinfo
  {author} {\bibfnamefont {J.}~\bibnamefont {Wenner}}, \bibinfo {author}
  {\bibfnamefont {H.}~\bibnamefont {Neven}},\ and\ \bibinfo {author}
  {\bibfnamefont {J.~M.}\ \bibnamefont {Martinis}},\ }\bibfield  {title}
  {\bibinfo {title} {\textit{Observation of Classical-Quantum Crossover of
  $1/f$ Flux Noise and Its Paramagnetic Temperature Dependence}},\ }\href
  {https://doi.org/10.1103/PhysRevLett.118.057702} {\bibfield  {journal}
  {\bibinfo  {journal} {Phys. Rev. Lett.}\ }\textbf {\bibinfo {volume} {118}},\
  \bibinfo {pages} {057702} (\bibinfo {year} {2017})}\BibitemShut {NoStop}%
\bibitem [{\citenamefont {Kou}\ \emph {et~al.}(2017)\citenamefont {Kou},
  \citenamefont {Smith}, \citenamefont {Vool}, \citenamefont {Brierley},
  \citenamefont {Meier}, \citenamefont {Frunzio}, \citenamefont {Girvin},
  \citenamefont {Glazman},\ and\ \citenamefont
  {Devoret}}]{Devoret_fluxonium_molecule}%
  \BibitemOpen
  \bibfield  {author} {\bibinfo {author} {\bibfnamefont {A.}~\bibnamefont
  {Kou}}, \bibinfo {author} {\bibfnamefont {W.~C.}\ \bibnamefont {Smith}},
  \bibinfo {author} {\bibfnamefont {U.}~\bibnamefont {Vool}}, \bibinfo {author}
  {\bibfnamefont {R.~T.}\ \bibnamefont {Brierley}}, \bibinfo {author}
  {\bibfnamefont {H.}~\bibnamefont {Meier}}, \bibinfo {author} {\bibfnamefont
  {L.}~\bibnamefont {Frunzio}}, \bibinfo {author} {\bibfnamefont {S.~M.}\
  \bibnamefont {Girvin}}, \bibinfo {author} {\bibfnamefont {L.~I.}\
  \bibnamefont {Glazman}},\ and\ \bibinfo {author} {\bibfnamefont {M.~H.}\
  \bibnamefont {Devoret}},\ }\bibfield  {title} {\bibinfo {title}
  {\textit{Fluxonium-Based Artificial Molecule with a Tunable Magnetic
  Moment}},\ }\href {https://doi.org/10.1103/PhysRevX.7.031037} {\bibfield
  {journal} {\bibinfo  {journal} {Phys. Rev. X}\ }\textbf {\bibinfo {volume}
  {7}},\ \bibinfo {pages} {031037} (\bibinfo {year} {2017})}\BibitemShut
  {NoStop}%
\bibitem [{\citenamefont {Bylander}\ \emph {et~al.}(2011)\citenamefont
  {Bylander}, \citenamefont {Gustavsson}, \citenamefont {Yan}, \citenamefont
  {Yoshihara}, \citenamefont {Harrabi}, \citenamefont {Fitch}, \citenamefont
  {Cory}, \citenamefont {Nakamura}, \citenamefont {Tsai},\ and\ \citenamefont
  {Oliver}}]{Oliver_flux_qubit_dd}%
  \BibitemOpen
  \bibfield  {author} {\bibinfo {author} {\bibfnamefont {J.}~\bibnamefont
  {Bylander}}, \bibinfo {author} {\bibfnamefont {S.}~\bibnamefont
  {Gustavsson}}, \bibinfo {author} {\bibfnamefont {F.}~\bibnamefont {Yan}},
  \bibinfo {author} {\bibfnamefont {F.}~\bibnamefont {Yoshihara}}, \bibinfo
  {author} {\bibfnamefont {K.}~\bibnamefont {Harrabi}}, \bibinfo {author}
  {\bibfnamefont {G.}~\bibnamefont {Fitch}}, \bibinfo {author} {\bibfnamefont
  {D.~G.}\ \bibnamefont {Cory}}, \bibinfo {author} {\bibfnamefont
  {Y.}~\bibnamefont {Nakamura}}, \bibinfo {author} {\bibfnamefont {J.-S.}\
  \bibnamefont {Tsai}},\ and\ \bibinfo {author} {\bibfnamefont {W.~D.}\
  \bibnamefont {Oliver}},\ }\bibfield  {title} {\bibinfo {title} {\textit{Noise
  Spectroscopy Through Dynamical Decoupling with a Superconducting Flux
  Qubit}},\ }\href {https://doi.org/10.1038/nphys1994} {\bibfield  {journal}
  {\bibinfo  {journal} {Nat. Phys.}\ }\textbf {\bibinfo {volume} {7}},\
  \bibinfo {pages} {565} (\bibinfo {year} {2011})}\BibitemShut {NoStop}%
\bibitem [{\citenamefont {Didier}()}]{Didier_dynamical_sweet_spot}%
  \BibitemOpen
  \bibfield  {author} {\bibinfo {author} {\bibfnamefont {N.}~\bibnamefont
  {Didier}},\ }\bibfield  {title} {\bibinfo {title} {\textit{Flux Control of
  Superconducting Qubits at Dynamical Sweet Spots}},\ }\Eprint
  {https://arxiv.org/abs/arXiv:1912.09416} {arXiv:1912.09416} \BibitemShut
  {NoStop}%
\bibitem [{\citenamefont {Didier}\ \emph {et~al.}(2019)\citenamefont {Didier},
  \citenamefont {Sete}, \citenamefont {Combes},\ and\ \citenamefont
  {da~Silva}}]{Rigetti_ac_sweet_spot}%
  \BibitemOpen
  \bibfield  {author} {\bibinfo {author} {\bibfnamefont {N.}~\bibnamefont
  {Didier}}, \bibinfo {author} {\bibfnamefont {E.~A.}\ \bibnamefont {Sete}},
  \bibinfo {author} {\bibfnamefont {J.}~\bibnamefont {Combes}},\ and\ \bibinfo
  {author} {\bibfnamefont {M.~P.}\ \bibnamefont {da~Silva}},\ }\bibfield
  {title} {\bibinfo {title} {\textit{ac Flux Sweet Spots in Parametrically
  Modulated Superconducting Qubits}},\ }\href
  {https://doi.org/10.1103/PhysRevApplied.12.054015} {\bibfield  {journal}
  {\bibinfo  {journal} {Phys. Rev. Applied}\ }\textbf {\bibinfo {volume}
  {12}},\ \bibinfo {pages} {054015} (\bibinfo {year} {2019})}\BibitemShut
  {NoStop}%
\bibitem [{\citenamefont {Hong}\ \emph {et~al.}(2020)\citenamefont {Hong},
  \citenamefont {Papageorge}, \citenamefont {Sivarajah}, \citenamefont
  {Crossman}, \citenamefont {Didier}, \citenamefont {Polloreno}, \citenamefont
  {Sete}, \citenamefont {Turkowski}, \citenamefont {da~Silva},\ and\
  \citenamefont {Johnson}}]{Rigetti_ac_sweet_spot_exp}%
  \BibitemOpen
  \bibfield  {author} {\bibinfo {author} {\bibfnamefont {S.~S.}\ \bibnamefont
  {Hong}}, \bibinfo {author} {\bibfnamefont {A.~T.}\ \bibnamefont
  {Papageorge}}, \bibinfo {author} {\bibfnamefont {P.}~\bibnamefont
  {Sivarajah}}, \bibinfo {author} {\bibfnamefont {G.}~\bibnamefont {Crossman}},
  \bibinfo {author} {\bibfnamefont {N.}~\bibnamefont {Didier}}, \bibinfo
  {author} {\bibfnamefont {A.~M.}\ \bibnamefont {Polloreno}}, \bibinfo {author}
  {\bibfnamefont {E.~A.}\ \bibnamefont {Sete}}, \bibinfo {author}
  {\bibfnamefont {S.~W.}\ \bibnamefont {Turkowski}}, \bibinfo {author}
  {\bibfnamefont {M.~P.}\ \bibnamefont {da~Silva}},\ and\ \bibinfo {author}
  {\bibfnamefont {B.~R.}\ \bibnamefont {Johnson}},\ }\bibfield  {title}
  {\bibinfo {title} {\textit{Demonstration of a Parametrically Activated
  Entangling Gate Protected from Flux Noise}},\ }\href
  {https://doi.org/10.1103/PhysRevA.101.012302} {\bibfield  {journal} {\bibinfo
   {journal} {Phys. Rev. A}\ }\textbf {\bibinfo {volume} {101}},\ \bibinfo
  {pages} {012302} (\bibinfo {year} {2020})}\BibitemShut {NoStop}%
\bibitem [{\citenamefont {Fried}\ \emph {et~al.}()\citenamefont {Fried},
  \citenamefont {Sivarajah}, \citenamefont {Didier}, \citenamefont {Sete},
  \citenamefont {da~Silva}, \citenamefont {Johnson},\ and\ \citenamefont
  {Ryan}}]{Rigetti_broad_band_noise}%
  \BibitemOpen
  \bibfield  {author} {\bibinfo {author} {\bibfnamefont {E.~S.}\ \bibnamefont
  {Fried}}, \bibinfo {author} {\bibfnamefont {P.}~\bibnamefont {Sivarajah}},
  \bibinfo {author} {\bibfnamefont {N.}~\bibnamefont {Didier}}, \bibinfo
  {author} {\bibfnamefont {E.~A.}\ \bibnamefont {Sete}}, \bibinfo {author}
  {\bibfnamefont {M.~P.}\ \bibnamefont {da~Silva}}, \bibinfo {author}
  {\bibfnamefont {B.~R.}\ \bibnamefont {Johnson}},\ and\ \bibinfo {author}
  {\bibfnamefont {C.~A.}\ \bibnamefont {Ryan}},\ }\bibfield  {title} {\bibinfo
  {title} {\textit{Assessing the Influence of Broadband Instrumentation Noise
  on Parametrically Modulated Superconducting Qubits}},\ }\Eprint
  {https://arxiv.org/abs/arXiv:1908.11370} {arXiv:1908.11370} \BibitemShut
  {NoStop}%
\bibitem [{\citenamefont {Frees}\ \emph {et~al.}(2019)\citenamefont {Frees},
  \citenamefont {Mehl}, \citenamefont {Gamble}, \citenamefont {Friesen},\ and\
  \citenamefont {Coppersmith}}]{Coppersmith_two_qubit_sweet_spot}%
  \BibitemOpen
  \bibfield  {author} {\bibinfo {author} {\bibfnamefont {A.}~\bibnamefont
  {Frees}}, \bibinfo {author} {\bibfnamefont {S.}~\bibnamefont {Mehl}},
  \bibinfo {author} {\bibfnamefont {J.~K.}\ \bibnamefont {Gamble}}, \bibinfo
  {author} {\bibfnamefont {M.}~\bibnamefont {Friesen}},\ and\ \bibinfo {author}
  {\bibfnamefont {S.~N.}\ \bibnamefont {Coppersmith}},\ }\bibfield  {title}
  {\bibinfo {title} {\textit{Adiabatic Two-Qubit Gates in Capacitively Coupled
  Quantum Dot Hybrid Qubits}},\ }\href
  {https://doi.org/10.1038/s41534-019-0190-7} {\bibfield  {journal} {\bibinfo
  {journal} {npj Quantum Inform.}\ }\textbf {\bibinfo {volume} {5}},\ \bibinfo
  {pages} {73} (\bibinfo {year} {2019})}\BibitemShut {NoStop}%
\bibitem [{\citenamefont {Pirkkalainen}\ \emph {et~al.}(2013)\citenamefont
  {Pirkkalainen}, \citenamefont {Cho}, \citenamefont {Li}, \citenamefont
  {Paraoanu}, \citenamefont {Hakonen},\ and\ \citenamefont
  {Sillanp{\"a}{\"a}}}]{Sillanpaa_hybrid_circuit}%
  \BibitemOpen
  \bibfield  {author} {\bibinfo {author} {\bibfnamefont {J.~M.}\ \bibnamefont
  {Pirkkalainen}}, \bibinfo {author} {\bibfnamefont {S.~U.}\ \bibnamefont
  {Cho}}, \bibinfo {author} {\bibfnamefont {J.}~\bibnamefont {Li}}, \bibinfo
  {author} {\bibfnamefont {G.~S.}\ \bibnamefont {Paraoanu}}, \bibinfo {author}
  {\bibfnamefont {P.~J.}\ \bibnamefont {Hakonen}},\ and\ \bibinfo {author}
  {\bibfnamefont {M.~A.}\ \bibnamefont {Sillanp{\"a}{\"a}}},\ }\bibfield
  {title} {\bibinfo {title} {\textit{Hybrid Circuit Cavity Quantum
  Electrodynamics with a Micromechanical Resonator}},\ }\href
  {https://doi.org/10.1038/nature11821} {\bibfield  {journal} {\bibinfo
  {journal} {Nature}\ }\textbf {\bibinfo {volume} {494}},\ \bibinfo {pages}
  {211} (\bibinfo {year} {2013})}\BibitemShut {NoStop}%
\bibitem [{\citenamefont {Timoney}\ \emph {et~al.}(2011)\citenamefont
  {Timoney}, \citenamefont {Baumgart}, \citenamefont {Johanning}, \citenamefont
  {Var\'on}, \citenamefont {Plenio}, \citenamefont {Retzker},\ and\
  \citenamefont {Wunderlich}}]{Wunderlich_dressed_qubit_nature}%
  \BibitemOpen
  \bibfield  {author} {\bibinfo {author} {\bibfnamefont {N.}~\bibnamefont
  {Timoney}}, \bibinfo {author} {\bibfnamefont {I.}~\bibnamefont {Baumgart}},
  \bibinfo {author} {\bibfnamefont {M.}~\bibnamefont {Johanning}}, \bibinfo
  {author} {\bibfnamefont {A.~F.}\ \bibnamefont {Var\'on}}, \bibinfo {author}
  {\bibfnamefont {M.~B.}\ \bibnamefont {Plenio}}, \bibinfo {author}
  {\bibfnamefont {A.}~\bibnamefont {Retzker}},\ and\ \bibinfo {author}
  {\bibfnamefont {C.}~\bibnamefont {Wunderlich}},\ }\bibfield  {title}
  {\bibinfo {title} {\textit{Quantum Gates and Memory Using Microwave-Dressed
  States}},\ }\href {https://doi.org/https://doi.org/10.1038/nature10319}
  {\bibfield  {journal} {\bibinfo  {journal} {Nature}\ }\textbf {\bibinfo
  {volume} {476}},\ \bibinfo {pages} {185} (\bibinfo {year}
  {2011})}\BibitemShut {NoStop}%
\bibitem [{\citenamefont {W\"olk}\ and\ \citenamefont
  {Wunderlich}(2017)}]{Wunderlich_dressed_qubit_NJP}%
  \BibitemOpen
  \bibfield  {author} {\bibinfo {author} {\bibfnamefont {S.}~\bibnamefont
  {W\"olk}}\ and\ \bibinfo {author} {\bibfnamefont {C.}~\bibnamefont
  {Wunderlich}},\ }\bibfield  {title} {\bibinfo {title} {\textit{Quantum
  Dynamics of Trapped Ions in a Dynamic Field Gradient Using Dressed States}},\
  }\href {https://doi.org/https://doi.org/10.1088/1367-2630/aa7b22} {\bibfield
  {journal} {\bibinfo  {journal} {New J. Phys.}\ }\textbf {\bibinfo {volume}
  {19}},\ \bibinfo {pages} {083021} (\bibinfo {year} {2017})}\BibitemShut
  {NoStop}%
\bibitem [{\citenamefont {Yang}\ \emph {et~al.}(2019)\citenamefont {Yang},
  \citenamefont {Coppersmith},\ and\ \citenamefont
  {Friesen}}]{Friesen_charge_qubit_Floquet}%
  \BibitemOpen
  \bibfield  {author} {\bibinfo {author} {\bibfnamefont {Y.-C.}\ \bibnamefont
  {Yang}}, \bibinfo {author} {\bibfnamefont {S.~N.}\ \bibnamefont
  {Coppersmith}},\ and\ \bibinfo {author} {\bibfnamefont {M.}~\bibnamefont
  {Friesen}},\ }\bibfield  {title} {\bibinfo {title} {\textit{Achieving
  High-Fidelity Single-Qubit Gates in a Strongly Driven Charge Qubit with $1/f$
  Charge Noise}},\ }\href {https://doi.org/10.1038/s41534-019-0127-1}
  {\bibfield  {journal} {\bibinfo  {journal} {npj Quantum Inf.}\ }\textbf
  {\bibinfo {volume} {5}},\ \bibinfo {pages} {12} (\bibinfo {year}
  {2019})}\BibitemShut {NoStop}%
\bibitem [{\citenamefont {Petta}\ \emph {et~al.}(2005)\citenamefont {Petta},
  \citenamefont {Johnson}, \citenamefont {Taylor}, \citenamefont {Laird},
  \citenamefont {Yacoby}, \citenamefont {Lukin}, \citenamefont {Marcus},
  \citenamefont {Hanson},\ and\ \citenamefont {Gossard}}]{Gossard_spin_qubit}%
  \BibitemOpen
  \bibfield  {author} {\bibinfo {author} {\bibfnamefont {J.~R.}\ \bibnamefont
  {Petta}}, \bibinfo {author} {\bibfnamefont {A.~C.}\ \bibnamefont {Johnson}},
  \bibinfo {author} {\bibfnamefont {J.~M.}\ \bibnamefont {Taylor}}, \bibinfo
  {author} {\bibfnamefont {E.~A.}\ \bibnamefont {Laird}}, \bibinfo {author}
  {\bibfnamefont {A.}~\bibnamefont {Yacoby}}, \bibinfo {author} {\bibfnamefont
  {M.~D.}\ \bibnamefont {Lukin}}, \bibinfo {author} {\bibfnamefont {C.~M.}\
  \bibnamefont {Marcus}}, \bibinfo {author} {\bibfnamefont {M.~P.}\
  \bibnamefont {Hanson}},\ and\ \bibinfo {author} {\bibfnamefont {A.~C.}\
  \bibnamefont {Gossard}},\ }\bibfield  {title} {\bibinfo {title}
  {\textit{Coherent Manipulation of Coupled Electron Spins in Semiconductor
  Quantum Dots}},\ }\href {https://doi.org/10.1126/science.1116955} {\bibfield
  {journal} {\bibinfo  {journal} {Science}\ }\textbf {\bibinfo {volume}
  {309}},\ \bibinfo {pages} {2180} (\bibinfo {year} {2005})}\BibitemShut
  {NoStop}%
\bibitem [{\citenamefont {Koppens}\ \emph {et~al.}(2008)\citenamefont
  {Koppens}, \citenamefont {Nowack},\ and\ \citenamefont
  {Vandersypen}}]{Vandersypen_spin_qubit_echo}%
  \BibitemOpen
  \bibfield  {author} {\bibinfo {author} {\bibfnamefont {F.~H.~L.}\
  \bibnamefont {Koppens}}, \bibinfo {author} {\bibfnamefont {K.~C.}\
  \bibnamefont {Nowack}},\ and\ \bibinfo {author} {\bibfnamefont {L.~M.~K.}\
  \bibnamefont {Vandersypen}},\ }\bibfield  {title} {\bibinfo {title}
  {\textit{Spin Echo of a Single Electron Spin in a Quantum Dot}},\ }\href
  {https://doi.org/10.1103/PhysRevLett.100.236802} {\bibfield  {journal}
  {\bibinfo  {journal} {Phys. Rev. Lett.}\ }\textbf {\bibinfo {volume} {100}},\
  \bibinfo {pages} {236802} (\bibinfo {year} {2008})}\BibitemShut {NoStop}%
\bibitem [{\citenamefont {Bluhm}\ \emph {et~al.}(2011)\citenamefont {Bluhm},
  \citenamefont {Foletti}, \citenamefont {Neder}, \citenamefont {Rudner},
  \citenamefont {Mahalu}, \citenamefont {Umansky},\ and\ \citenamefont
  {Yacoby}}]{Yacoby_spin_qubit_CPMG}%
  \BibitemOpen
  \bibfield  {author} {\bibinfo {author} {\bibfnamefont {H.}~\bibnamefont
  {Bluhm}}, \bibinfo {author} {\bibfnamefont {S.}~\bibnamefont {Foletti}},
  \bibinfo {author} {\bibfnamefont {I.}~\bibnamefont {Neder}}, \bibinfo
  {author} {\bibfnamefont {M.}~\bibnamefont {Rudner}}, \bibinfo {author}
  {\bibfnamefont {D.}~\bibnamefont {Mahalu}}, \bibinfo {author} {\bibfnamefont
  {V.}~\bibnamefont {Umansky}},\ and\ \bibinfo {author} {\bibfnamefont
  {A.}~\bibnamefont {Yacoby}},\ }\bibfield  {title} {\bibinfo {title}
  {\textit{Dephasing Time of GaAs Electron-Spin Qubits Coupled to a Nuclear
  Bath Exceeding 200 $\mu$s}},\ }\href
  {https://doi.org/10.1103/PhysRevLett.100.236802} {\bibfield  {journal}
  {\bibinfo  {journal} {Nat. Phys.}\ }\textbf {\bibinfo {volume} {2}},\
  \bibinfo {pages} {109} (\bibinfo {year} {2011})}\BibitemShut {NoStop}%
\bibitem [{\citenamefont {Safavi-Naini}\ \emph {et~al.}(2011)\citenamefont
  {Safavi-Naini}, \citenamefont {Rabl}, \citenamefont {Weck},\ and\
  \citenamefont {Sadeghpour}}]{Sadeghpour_trapped_ion_1/f_theory}%
  \BibitemOpen
  \bibfield  {author} {\bibinfo {author} {\bibfnamefont {A.}~\bibnamefont
  {Safavi-Naini}}, \bibinfo {author} {\bibfnamefont {P.}~\bibnamefont {Rabl}},
  \bibinfo {author} {\bibfnamefont {P.~F.}\ \bibnamefont {Weck}},\ and\
  \bibinfo {author} {\bibfnamefont {H.~R.}\ \bibnamefont {Sadeghpour}},\
  }\bibfield  {title} {\bibinfo {title} {\textit{Microscopic Model of
  Electric-Field-Noise Heating in Ion Traps}},\ }\href
  {https://doi.org/10.1103/PhysRevA.84.023412} {\bibfield  {journal} {\bibinfo
  {journal} {Phys. Rev. A}\ }\textbf {\bibinfo {volume} {84}},\ \bibinfo
  {pages} {023412} (\bibinfo {year} {2011})}\BibitemShut {NoStop}%
\bibitem [{\citenamefont {Daniilidis}\ \emph {et~al.}(2011)\citenamefont
  {Daniilidis}, \citenamefont {Narayanan}, \citenamefont {M\"oller},
  \citenamefont {Clark}, \citenamefont {Lee}, \citenamefont {Leek},
  \citenamefont {Wallraff}, \citenamefont {Schulz}, \citenamefont
  {Schmidt-Kaler},\ and\ \citenamefont
  {H\"affner}}]{Haffner_ion_trap_heat_rate}%
  \BibitemOpen
  \bibfield  {author} {\bibinfo {author} {\bibfnamefont {N.}~\bibnamefont
  {Daniilidis}}, \bibinfo {author} {\bibfnamefont {S.}~\bibnamefont
  {Narayanan}}, \bibinfo {author} {\bibfnamefont {S.~A.}\ \bibnamefont
  {M\"oller}}, \bibinfo {author} {\bibfnamefont {R.}~\bibnamefont {Clark}},
  \bibinfo {author} {\bibfnamefont {T.~E.}\ \bibnamefont {Lee}}, \bibinfo
  {author} {\bibfnamefont {P.~J.}\ \bibnamefont {Leek}}, \bibinfo {author}
  {\bibfnamefont {A.}~\bibnamefont {Wallraff}}, \bibinfo {author}
  {\bibfnamefont {S.}~\bibnamefont {Schulz}}, \bibinfo {author} {\bibfnamefont
  {F.}~\bibnamefont {Schmidt-Kaler}},\ and\ \bibinfo {author} {\bibfnamefont
  {H.}~\bibnamefont {H\"affner}},\ }\bibfield  {title} {\bibinfo {title}
  {\textit{Fabrication and Heating Rate Study of Microscopic Surface Electrode
  Ion Traps}},\ }\href {https://doi.org/10.1088/1367-2630/13/1/013032}
  {\bibfield  {journal} {\bibinfo  {journal} {New J. Phys.}\ }\textbf {\bibinfo
  {volume} {13}},\ \bibinfo {pages} {013032} (\bibinfo {year}
  {2011})}\BibitemShut {NoStop}%
\bibitem [{\citenamefont {Brownnutt}\ \emph {et~al.}(2015)\citenamefont
  {Brownnutt}, \citenamefont {Kumph}, \citenamefont {Rabl},\ and\ \citenamefont
  {Blatt}}]{Blatt_ion_trap_ef_noise}%
  \BibitemOpen
  \bibfield  {author} {\bibinfo {author} {\bibfnamefont {M.}~\bibnamefont
  {Brownnutt}}, \bibinfo {author} {\bibfnamefont {M.}~\bibnamefont {Kumph}},
  \bibinfo {author} {\bibfnamefont {P.}~\bibnamefont {Rabl}},\ and\ \bibinfo
  {author} {\bibfnamefont {R.}~\bibnamefont {Blatt}},\ }\bibfield  {title}
  {\bibinfo {title} {\textit{Ion-Trap Measurements of Electric-Field Noise near
  Surfaces}},\ }\href {https://doi.org/10.1103/RevModPhys.87.1419} {\bibfield
  {journal} {\bibinfo  {journal} {Rev. Mod. Phys.}\ }\textbf {\bibinfo {volume}
  {87}},\ \bibinfo {pages} {1419} (\bibinfo {year} {2015})}\BibitemShut
  {NoStop}%
\bibitem [{\citenamefont {Kalashnikov}\ \emph {et~al.}(2020)\citenamefont
  {Kalashnikov}, \citenamefont {Hsieh}, \citenamefont {Zhang}, \citenamefont
  {Lu}, \citenamefont {Kamenov}, \citenamefont {Di~Paolo}, \citenamefont
  {Blais}, \citenamefont {Gershenson},\ and\ \citenamefont
  {Bell}}]{Matthew_bifluxon}%
  \BibitemOpen
  \bibfield  {author} {\bibinfo {author} {\bibfnamefont {K.}~\bibnamefont
  {Kalashnikov}}, \bibinfo {author} {\bibfnamefont {W.~T.}\ \bibnamefont
  {Hsieh}}, \bibinfo {author} {\bibfnamefont {W.}~\bibnamefont {Zhang}},
  \bibinfo {author} {\bibfnamefont {W.-S.}\ \bibnamefont {Lu}}, \bibinfo
  {author} {\bibfnamefont {P.}~\bibnamefont {Kamenov}}, \bibinfo {author}
  {\bibfnamefont {A.}~\bibnamefont {Di~Paolo}}, \bibinfo {author}
  {\bibfnamefont {A.}~\bibnamefont {Blais}}, \bibinfo {author} {\bibfnamefont
  {M.~E.}\ \bibnamefont {Gershenson}},\ and\ \bibinfo {author} {\bibfnamefont
  {M.}~\bibnamefont {Bell}},\ }\bibfield  {title} {\bibinfo {title}
  {\textit{Bifluxon: Fluxon-Parity-Protected Superconducting Qubit}},\ }\href
  {https://doi.org/10.1103/PRXQuantum.1.010307} {\bibfield  {journal} {\bibinfo
   {journal} {PRX Quantum}\ }\textbf {\bibinfo {volume} {1}},\ \bibinfo {pages}
  {010307} (\bibinfo {year} {2020})}\BibitemShut {NoStop}%
\bibitem [{\citenamefont {Constantin}\ and\ \citenamefont
  {Yu}(2007)}]{Yu_critical_current}%
  \BibitemOpen
  \bibfield  {author} {\bibinfo {author} {\bibfnamefont {M.}~\bibnamefont
  {Constantin}}\ and\ \bibinfo {author} {\bibfnamefont {C.~C.}\ \bibnamefont
  {Yu}},\ }\bibfield  {title} {\bibinfo {title} {\textit{Microscopic Model of
  Critical Current Noise in Josephson Junctions}},\ }\href
  {https://doi.org/10.1103/PhysRevLett.99.207001} {\bibfield  {journal}
  {\bibinfo  {journal} {Phys. Rev. Lett.}\ }\textbf {\bibinfo {volume} {99}},\
  \bibinfo {pages} {207001} (\bibinfo {year} {2007})}\BibitemShut {NoStop}%
\bibitem [{\citenamefont {Mück}\ \emph {et~al.}(2005)\citenamefont {Mück},
  \citenamefont {Korn}, \citenamefont {Mugford}, \citenamefont {Kycia},\ and\
  \citenamefont {Clarke}}]{Clarke_critical_current}%
  \BibitemOpen
  \bibfield  {author} {\bibinfo {author} {\bibfnamefont {M.}~\bibnamefont
  {Mück}}, \bibinfo {author} {\bibfnamefont {M.}~\bibnamefont {Korn}},
  \bibinfo {author} {\bibfnamefont {C.~G.~A.}\ \bibnamefont {Mugford}},
  \bibinfo {author} {\bibfnamefont {J.~B.}\ \bibnamefont {Kycia}},\ and\
  \bibinfo {author} {\bibfnamefont {J.}~\bibnamefont {Clarke}},\ }\bibfield
  {title} {\bibinfo {title} {\textit{Measurements of $1/f$ Noise in Josephson
  Junctions at Zero Voltage: Implications for Decoherence in Superconducting
  Quantum Bits}},\ }\href {https://doi.org/10.1063/1.1846157} {\bibfield
  {journal} {\bibinfo  {journal} {Appl. Phys. Lett.}\ }\textbf {\bibinfo
  {volume} {86}},\ \bibinfo {pages} {012510} (\bibinfo {year} {2005})},\
  \Eprint {https://arxiv.org/abs/https://doi.org/10.1063/1.1846157}
  {https://doi.org/10.1063/1.1846157} \BibitemShut {NoStop}%
\bibitem [{\citenamefont {Khodjasteh}\ and\ \citenamefont
  {Lidar}(2005)}]{Lidar_DD_2005}%
  \BibitemOpen
  \bibfield  {author} {\bibinfo {author} {\bibfnamefont {K.}~\bibnamefont
  {Khodjasteh}}\ and\ \bibinfo {author} {\bibfnamefont {D.~A.}\ \bibnamefont
  {Lidar}},\ }\bibfield  {title} {\bibinfo {title} {\textit{Fault-Tolerant
  Quantum Dynamical Decoupling}},\ }\href
  {https://doi.org/10.1103/PhysRevLett.95.180501} {\bibfield  {journal}
  {\bibinfo  {journal} {Phys. Rev. Lett.}\ }\textbf {\bibinfo {volume} {95}},\
  \bibinfo {pages} {180501} (\bibinfo {year} {2005})}\BibitemShut {NoStop}%
\bibitem [{\citenamefont {Pokharel}\ \emph {et~al.}(2018)\citenamefont
  {Pokharel}, \citenamefont {Anand}, \citenamefont {Fortman},\ and\
  \citenamefont {Lidar}}]{Lidar_transmon_dd}%
  \BibitemOpen
  \bibfield  {author} {\bibinfo {author} {\bibfnamefont {B.}~\bibnamefont
  {Pokharel}}, \bibinfo {author} {\bibfnamefont {N.}~\bibnamefont {Anand}},
  \bibinfo {author} {\bibfnamefont {B.}~\bibnamefont {Fortman}},\ and\ \bibinfo
  {author} {\bibfnamefont {D.~A.}\ \bibnamefont {Lidar}},\ }\bibfield  {title}
  {\bibinfo {title} {\textit{Demonstration of Fidelity Improvement Using
  Dynamical Decoupling with Superconducting Qubits}},\ }\href
  {https://doi.org/10.1103/PhysRevLett.121.220502} {\bibfield  {journal}
  {\bibinfo  {journal} {Phys. Rev. Lett.}\ }\textbf {\bibinfo {volume} {121}},\
  \bibinfo {pages} {220502} (\bibinfo {year} {2018})}\BibitemShut {NoStop}%
\bibitem [{\citenamefont {Cywi\ifmmode~\acute{n}\else \'{n}\fi{}ski}\ \emph
  {et~al.}(2008)\citenamefont {Cywi\ifmmode~\acute{n}\else \'{n}\fi{}ski},
  \citenamefont {Lutchyn}, \citenamefont {Nave},\ and\ \citenamefont
  {Das~Sarma}}]{DasSarma_DD}%
  \BibitemOpen
  \bibfield  {author} {\bibinfo {author} {\bibfnamefont {L.}~\bibnamefont
  {Cywi\ifmmode~\acute{n}\else \'{n}\fi{}ski}}, \bibinfo {author}
  {\bibfnamefont {R.~M.}\ \bibnamefont {Lutchyn}}, \bibinfo {author}
  {\bibfnamefont {C.~P.}\ \bibnamefont {Nave}},\ and\ \bibinfo {author}
  {\bibfnamefont {S.}~\bibnamefont {Das~Sarma}},\ }\bibfield  {title} {\bibinfo
  {title} {\textit{How to Enhance Dephasing Time in Superconducting Qubits}},\
  }\href {https://doi.org/10.1103/PhysRevB.77.174509} {\bibfield  {journal}
  {\bibinfo  {journal} {Phys. Rev. B}\ }\textbf {\bibinfo {volume} {77}},\
  \bibinfo {pages} {174509} (\bibinfo {year} {2008})}\BibitemShut {NoStop}%
\bibitem [{\citenamefont {Viola}\ \emph {et~al.}(1999)\citenamefont {Viola},
  \citenamefont {Knill},\ and\ \citenamefont {Lloyd}}]{Lloyd_DD_theory}%
  \BibitemOpen
  \bibfield  {author} {\bibinfo {author} {\bibfnamefont {L.}~\bibnamefont
  {Viola}}, \bibinfo {author} {\bibfnamefont {E.}~\bibnamefont {Knill}},\ and\
  \bibinfo {author} {\bibfnamefont {S.}~\bibnamefont {Lloyd}},\ }\bibfield
  {title} {\bibinfo {title} {\textit{Dynamical Decoupling of Open Quantum
  Systems}},\ }\href {https://doi.org/10.1103/PhysRevLett.82.2417} {\bibfield
  {journal} {\bibinfo  {journal} {Phys. Rev. Lett.}\ }\textbf {\bibinfo
  {volume} {82}},\ \bibinfo {pages} {2417} (\bibinfo {year}
  {1999})}\BibitemShut {NoStop}%
\bibitem [{\citenamefont {Uhrig}(2007)}]{Uhrig_DD}%
  \BibitemOpen
  \bibfield  {author} {\bibinfo {author} {\bibfnamefont {G.~S.}\ \bibnamefont
  {Uhrig}},\ }\bibfield  {title} {\bibinfo {title} {\textit{Keeping a Quantum
  Bit Alive by Optimized $\ensuremath{\pi}$-Pulse Sequences}},\ }\href
  {https://doi.org/10.1103/PhysRevLett.98.100504} {\bibfield  {journal}
  {\bibinfo  {journal} {Phys. Rev. Lett.}\ }\textbf {\bibinfo {volume} {98}},\
  \bibinfo {pages} {100504} (\bibinfo {year} {2007})}\BibitemShut {NoStop}%
\bibitem [{\citenamefont {Hahn}(1950)}]{Hahn_echo}%
  \BibitemOpen
  \bibfield  {author} {\bibinfo {author} {\bibfnamefont {E.~L.}\ \bibnamefont
  {Hahn}},\ }\bibfield  {title} {\bibinfo {title} {\textit{Spin Echoes}},\
  }\href {https://doi.org/10.1103/PhysRev.80.580} {\bibfield  {journal}
  {\bibinfo  {journal} {Phys. Rev.}\ }\textbf {\bibinfo {volume} {80}},\
  \bibinfo {pages} {580} (\bibinfo {year} {1950})}\BibitemShut {NoStop}%
\bibitem [{\citenamefont {Carr}\ and\ \citenamefont {Purcell}(1954)}]{Carr_DD}%
  \BibitemOpen
  \bibfield  {author} {\bibinfo {author} {\bibfnamefont {H.~Y.}\ \bibnamefont
  {Carr}}\ and\ \bibinfo {author} {\bibfnamefont {E.~M.}\ \bibnamefont
  {Purcell}},\ }\bibfield  {title} {\bibinfo {title} {\textit{Effects of
  Diffusion on Free Precession in Nuclear Magnetic Resonance Experiments}},\
  }\href {https://doi.org/10.1103/PhysRev.94.630} {\bibfield  {journal}
  {\bibinfo  {journal} {Phys. Rev.}\ }\textbf {\bibinfo {volume} {94}},\
  \bibinfo {pages} {630} (\bibinfo {year} {1954})}\BibitemShut {NoStop}%
\bibitem [{\citenamefont {Meiboom}\ and\ \citenamefont
  {Gill}(1958)}]{Meiboom_DD}%
  \BibitemOpen
  \bibfield  {author} {\bibinfo {author} {\bibfnamefont {S.}~\bibnamefont
  {Meiboom}}\ and\ \bibinfo {author} {\bibfnamefont {D.}~\bibnamefont {Gill}},\
  }\bibfield  {title} {\bibinfo {title} {\textit{Modified Spin-Echo Method for
  Measuring Nuclear Relaxation Times}},\ }\href
  {https://doi.org/10.1063/1.1716296} {\bibfield  {journal} {\bibinfo
  {journal} {Rev. Sci. Instrum.}\ }\textbf {\bibinfo {volume} {29}},\ \bibinfo
  {pages} {688} (\bibinfo {year} {1958})}\BibitemShut {NoStop}%
\bibitem [{\citenamefont {Viola}\ and\ \citenamefont {Knill}(2003)}]{Knill_DD}%
  \BibitemOpen
  \bibfield  {author} {\bibinfo {author} {\bibfnamefont {L.}~\bibnamefont
  {Viola}}\ and\ \bibinfo {author} {\bibfnamefont {E.}~\bibnamefont {Knill}},\
  }\bibfield  {title} {\bibinfo {title} {\textit{Robust Dynamical Decoupling of
  Quantum Systems with Bounded Controls}},\ }\href
  {https://doi.org/10.1103/PhysRevLett.90.037901} {\bibfield  {journal}
  {\bibinfo  {journal} {Phys. Rev. Lett.}\ }\textbf {\bibinfo {volume} {90}},\
  \bibinfo {pages} {037901} (\bibinfo {year} {2003})}\BibitemShut {NoStop}%
\bibitem [{\citenamefont {Guo}\ \emph {et~al.}(2018)\citenamefont {Guo},
  \citenamefont {Zheng}, \citenamefont {Wang}, \citenamefont {Song},
  \citenamefont {Zhang}, \citenamefont {Li}, \citenamefont {Liu}, \citenamefont
  {Deng}, \citenamefont {Huang}, \citenamefont {Zheng}, \citenamefont {Zhu},
  \citenamefont {Wang}, \citenamefont {Lu},\ and\ \citenamefont
  {Pan}}]{Pan_transmon_dd}%
  \BibitemOpen
  \bibfield  {author} {\bibinfo {author} {\bibfnamefont {Q.}~\bibnamefont
  {Guo}}, \bibinfo {author} {\bibfnamefont {S.-B.}\ \bibnamefont {Zheng}},
  \bibinfo {author} {\bibfnamefont {J.}~\bibnamefont {Wang}}, \bibinfo {author}
  {\bibfnamefont {C.}~\bibnamefont {Song}}, \bibinfo {author} {\bibfnamefont
  {P.}~\bibnamefont {Zhang}}, \bibinfo {author} {\bibfnamefont
  {K.}~\bibnamefont {Li}}, \bibinfo {author} {\bibfnamefont {W.}~\bibnamefont
  {Liu}}, \bibinfo {author} {\bibfnamefont {H.}~\bibnamefont {Deng}}, \bibinfo
  {author} {\bibfnamefont {K.}~\bibnamefont {Huang}}, \bibinfo {author}
  {\bibfnamefont {D.}~\bibnamefont {Zheng}}, \bibinfo {author} {\bibfnamefont
  {X.}~\bibnamefont {Zhu}}, \bibinfo {author} {\bibfnamefont {H.}~\bibnamefont
  {Wang}}, \bibinfo {author} {\bibfnamefont {C.-Y.}\ \bibnamefont {Lu}},\ and\
  \bibinfo {author} {\bibfnamefont {J.-W.}\ \bibnamefont {Pan}},\ }\bibfield
  {title} {\bibinfo {title} {\textit{Dephasing-Insensitive Quantum Information
  Storage and Processing with Superconducting Qubits}},\ }\href
  {https://doi.org/10.1103/PhysRevLett.121.130501} {\bibfield  {journal}
  {\bibinfo  {journal} {Phys. Rev. Lett.}\ }\textbf {\bibinfo {volume} {121}},\
  \bibinfo {pages} {130501} (\bibinfo {year} {2018})}\BibitemShut {NoStop}%
\bibitem [{\citenamefont {Yan}\ \emph {et~al.}(2013)\citenamefont {Yan},
  \citenamefont {Gustavsson}, \citenamefont {Bylander}, \citenamefont {Jin},
  \citenamefont {Yoshihara}, \citenamefont {Cory}, \citenamefont {Nakamura},
  \citenamefont {Orlando},\ and\ \citenamefont
  {Oliver}}]{Oliver_rotating_frame_relaxation}%
  \BibitemOpen
  \bibfield  {author} {\bibinfo {author} {\bibfnamefont {F.}~\bibnamefont
  {Yan}}, \bibinfo {author} {\bibfnamefont {S.}~\bibnamefont {Gustavsson}},
  \bibinfo {author} {\bibfnamefont {J.}~\bibnamefont {Bylander}}, \bibinfo
  {author} {\bibfnamefont {X.}~\bibnamefont {Jin}}, \bibinfo {author}
  {\bibfnamefont {F.}~\bibnamefont {Yoshihara}}, \bibinfo {author}
  {\bibfnamefont {D.~G.}\ \bibnamefont {Cory}}, \bibinfo {author}
  {\bibfnamefont {Y.}~\bibnamefont {Nakamura}}, \bibinfo {author}
  {\bibfnamefont {T.~P.}\ \bibnamefont {Orlando}},\ and\ \bibinfo {author}
  {\bibfnamefont {W.~D.}\ \bibnamefont {Oliver}},\ }\bibfield  {title}
  {\bibinfo {title} {\textit{Rotating-Frame Relaxation as a Noise Spectrum
  Analyser of a Superconducting Qubit Undergoing Driven Evolution}},\ }\href
  {https://doi.org/10.1038/ncomms3337} {\bibfield  {journal} {\bibinfo
  {journal} {Nat. Commun.}\ }\textbf {\bibinfo {volume} {4}},\ \bibinfo {pages}
  {2337} (\bibinfo {year} {2013})}\BibitemShut {NoStop}%
\bibitem [{\citenamefont {Jing}\ \emph {et~al.}(2014)\citenamefont {Jing},
  \citenamefont {Huang},\ and\ \citenamefont {Hu}}]{Hu_drive_spin_qubit}%
  \BibitemOpen
  \bibfield  {author} {\bibinfo {author} {\bibfnamefont {J.}~\bibnamefont
  {Jing}}, \bibinfo {author} {\bibfnamefont {P.}~\bibnamefont {Huang}},\ and\
  \bibinfo {author} {\bibfnamefont {X.}~\bibnamefont {Hu}},\ }\bibfield
  {title} {\bibinfo {title} {\textit{Decoherence of an Electrically Driven Spin
  Qubit}},\ }\href {https://doi.org/10.1103/PhysRevA.90.022118} {\bibfield
  {journal} {\bibinfo  {journal} {Phys. Rev. A}\ }\textbf {\bibinfo {volume}
  {90}},\ \bibinfo {pages} {022118} (\bibinfo {year} {2014})}\BibitemShut
  {NoStop}%
\bibitem [{\citenamefont {Smirnov}(2003)}]{Smirnov_Raib_decoherence}%
  \BibitemOpen
  \bibfield  {author} {\bibinfo {author} {\bibfnamefont {A.~Y.}\ \bibnamefont
  {Smirnov}},\ }\bibfield  {title} {\bibinfo {title} {\textit{Decoherence and
  Relaxation of a Quantum Bit in the Presence of Rabi Oscillations}},\ }\href
  {https://doi.org/10.1103/PhysRevB.67.155104} {\bibfield  {journal} {\bibinfo
  {journal} {Phys. Rev. B}\ }\textbf {\bibinfo {volume} {67}},\ \bibinfo
  {pages} {155104} (\bibinfo {year} {2003})}\BibitemShut {NoStop}%
\bibitem [{sup()}]{supplementary}%
  \BibitemOpen
  \href@noop {} {}\bibinfo {note} {See Supplemental Material attached below for
  comparison between our theory and previously developed protection schemes
  \cite{Didier_dynamical_sweet_spot,Rigetti_ac_sweet_spot,Rigetti_ac_sweet_spot_exp,Pan_transmon_dd,Oliver_rotating_frame_relaxation,Coppersmith_two_qubit_sweet_spot,Hu_drive_spin_qubit,Rigetti_broad_band_noise}}\BibitemShut
  {NoStop}%
\bibitem [{\citenamefont {Mundada}\ \emph {et~al.}(2020)\citenamefont
  {Mundada}, \citenamefont {Gyenis}, \citenamefont {Huang}, \citenamefont
  {Koch},\ and\ \citenamefont {Houck}}]{Houck_Floquet_exp}%
  \BibitemOpen
  \bibfield  {author} {\bibinfo {author} {\bibfnamefont {P.~S.}\ \bibnamefont
  {Mundada}}, \bibinfo {author} {\bibfnamefont {A.}~\bibnamefont {Gyenis}},
  \bibinfo {author} {\bibfnamefont {Z.}~\bibnamefont {Huang}}, \bibinfo
  {author} {\bibfnamefont {J.}~\bibnamefont {Koch}},\ and\ \bibinfo {author}
  {\bibfnamefont {A.~A.}\ \bibnamefont {Houck}},\ }\bibfield  {title} {\bibinfo
  {title} {\textit{Floquet-Engineered Enhancement of Coherence Times in a
  Driven Fluxonium Qubit}},\ }\href
  {https://doi.org/10.1103/PhysRevApplied.14.054033} {\bibfield  {journal}
  {\bibinfo  {journal} {Phys. Rev. Applied}\ }\textbf {\bibinfo {volume}
  {14}},\ \bibinfo {pages} {054033} (\bibinfo {year} {2020})}\BibitemShut
  {NoStop}%
\bibitem [{\citenamefont {Manucharyan}\ \emph {et~al.}(2009)\citenamefont
  {Manucharyan}, \citenamefont {Koch}, \citenamefont {Glazman},\ and\
  \citenamefont {Devoret}}]{Fluxonium_Devoret}%
  \BibitemOpen
  \bibfield  {author} {\bibinfo {author} {\bibfnamefont {V.~E.}\ \bibnamefont
  {Manucharyan}}, \bibinfo {author} {\bibfnamefont {J.}~\bibnamefont {Koch}},
  \bibinfo {author} {\bibfnamefont {L.~I.}\ \bibnamefont {Glazman}},\ and\
  \bibinfo {author} {\bibfnamefont {M.~H.}\ \bibnamefont {Devoret}},\
  }\bibfield  {title} {\bibinfo {title} {\textit{Fluxonium: Single Cooper-Pair
  Circuit Free of Charge Offsets}},\ }\href
  {https://doi.org/10.1126/science.1175552} {\bibfield  {journal} {\bibinfo
  {journal} {Science}\ }\textbf {\bibinfo {volume} {326}},\ \bibinfo {pages}
  {113} (\bibinfo {year} {2009})}\BibitemShut {NoStop}%
\bibitem [{\citenamefont {Pop}\ \emph {et~al.}(2014)\citenamefont {Pop},
  \citenamefont {Geerlings}, \citenamefont {Catelani}, \citenamefont
  {Schoelkopf}, \citenamefont {Glazman},\ and\ \citenamefont
  {Devoret}}]{Devoret_fluxonium_quasiparticle}%
  \BibitemOpen
  \bibfield  {author} {\bibinfo {author} {\bibfnamefont {I.~M.}\ \bibnamefont
  {Pop}}, \bibinfo {author} {\bibfnamefont {K.}~\bibnamefont {Geerlings}},
  \bibinfo {author} {\bibfnamefont {G.}~\bibnamefont {Catelani}}, \bibinfo
  {author} {\bibfnamefont {R.~J.}\ \bibnamefont {Schoelkopf}}, \bibinfo
  {author} {\bibfnamefont {L.~I.}\ \bibnamefont {Glazman}},\ and\ \bibinfo
  {author} {\bibfnamefont {M.~H.}\ \bibnamefont {Devoret}},\ }\bibfield
  {title} {\bibinfo {title} {\textit{Coherent Suppression of Electromagnetic
  Dissipation due to Superconducting Quasiparticles}},\ }\href
  {https://doi.org/10.1038/nature13017} {\bibfield  {journal} {\bibinfo
  {journal} {Nature}\ }\textbf {\bibinfo {volume} {508}},\ \bibinfo {pages}
  {369} (\bibinfo {year} {2014})}\BibitemShut {NoStop}%
\bibitem [{\citenamefont {Gr{\"u}nhaupt}\ \emph {et~al.}(2019)\citenamefont
  {Gr{\"u}nhaupt}, \citenamefont {Spiecker}, \citenamefont {Gusenkova},
  \citenamefont {Maleeva}, \citenamefont {Skacel}, \citenamefont {Takmakov},
  \citenamefont {Valenti}, \citenamefont {Winkel}, \citenamefont {Rotzinger},
  \citenamefont {Wernsdorfer}, \citenamefont {Ustinov},\ and\ \citenamefont
  {Pop}}]{Pop_granular_fluxonium}%
  \BibitemOpen
  \bibfield  {author} {\bibinfo {author} {\bibfnamefont {L.}~\bibnamefont
  {Gr{\"u}nhaupt}}, \bibinfo {author} {\bibfnamefont {M.}~\bibnamefont
  {Spiecker}}, \bibinfo {author} {\bibfnamefont {D.}~\bibnamefont {Gusenkova}},
  \bibinfo {author} {\bibfnamefont {N.}~\bibnamefont {Maleeva}}, \bibinfo
  {author} {\bibfnamefont {S.~T.}\ \bibnamefont {Skacel}}, \bibinfo {author}
  {\bibfnamefont {I.}~\bibnamefont {Takmakov}}, \bibinfo {author}
  {\bibfnamefont {F.}~\bibnamefont {Valenti}}, \bibinfo {author} {\bibfnamefont
  {P.}~\bibnamefont {Winkel}}, \bibinfo {author} {\bibfnamefont
  {H.}~\bibnamefont {Rotzinger}}, \bibinfo {author} {\bibfnamefont
  {W.}~\bibnamefont {Wernsdorfer}}, \bibinfo {author} {\bibfnamefont {A.~V.}\
  \bibnamefont {Ustinov}},\ and\ \bibinfo {author} {\bibfnamefont {I.~M.}\
  \bibnamefont {Pop}},\ }\bibfield  {title} {\bibinfo {title} {\textit{Granular
  Aluminium as a Superconducting Material for High-Impedance Quantum
  Circuits}},\ }\href {https://doi.org/10.1038/s41563-019-0350-3} {\bibfield
  {journal} {\bibinfo  {journal} {Nature Materials}\ }\textbf {\bibinfo
  {volume} {18}},\ \bibinfo {pages} {816} (\bibinfo {year} {2019})}\BibitemShut
  {NoStop}%
\bibitem [{foo()}]{footnote_one_over_f}%
  \BibitemOpen
  \href@noop {} {}\bibinfo {note} {At low frequencies $\omega \ll k_BT$, the
  asymmetry in the spectrum is negligible and a symmetric 1/f noise spectrum
  may be used.}\BibitemShut {Stop}%
\bibitem [{\citenamefont {Groszkowski}\ \emph {et~al.}(2018)\citenamefont
  {Groszkowski}, \citenamefont {Di~Paolo}, \citenamefont {Grimsmo},
  \citenamefont {Blais}, \citenamefont {Schuster}, \citenamefont {Houck},\ and\
  \citenamefont {Koch}}]{Groszkowski_Zero_pi_theory}%
  \BibitemOpen
  \bibfield  {author} {\bibinfo {author} {\bibfnamefont {P.}~\bibnamefont
  {Groszkowski}}, \bibinfo {author} {\bibfnamefont {A.}~\bibnamefont
  {Di~Paolo}}, \bibinfo {author} {\bibfnamefont {A.~L.}\ \bibnamefont
  {Grimsmo}}, \bibinfo {author} {\bibfnamefont {A.}~\bibnamefont {Blais}},
  \bibinfo {author} {\bibfnamefont {D.~I.}\ \bibnamefont {Schuster}}, \bibinfo
  {author} {\bibfnamefont {A.~A.}\ \bibnamefont {Houck}},\ and\ \bibinfo
  {author} {\bibfnamefont {J.}~\bibnamefont {Koch}},\ }\bibfield  {title}
  {\bibinfo {title} {\textit{Coherence Properties of the 0-$\pi$ Qubit}},\
  }\href {https://doi.org/10.1088/1367-2630/aab7cd} {\bibfield  {journal}
  {\bibinfo  {journal} {New J. Phys.}\ }\textbf {\bibinfo {volume} {20}},\
  \bibinfo {pages} {043053} (\bibinfo {year} {2018})}\BibitemShut {NoStop}%
\bibitem [{\citenamefont {Smith}\ \emph {et~al.}(2020)\citenamefont {Smith},
  \citenamefont {Kou}, \citenamefont {Xiao}, \citenamefont {Vool},\ and\
  \citenamefont {Devoret}}]{Devoret_two_Cooper_pair_qubit}%
  \BibitemOpen
  \bibfield  {author} {\bibinfo {author} {\bibfnamefont {W.}~\bibnamefont
  {Smith}}, \bibinfo {author} {\bibfnamefont {A.}~\bibnamefont {Kou}}, \bibinfo
  {author} {\bibfnamefont {X.}~\bibnamefont {Xiao}}, \bibinfo {author}
  {\bibfnamefont {U.}~\bibnamefont {Vool}},\ and\ \bibinfo {author}
  {\bibfnamefont {M.}~\bibnamefont {Devoret}},\ }\bibfield  {title} {\bibinfo
  {title} {\textit{Superconducting Circuit Protected by Two-Cooper-Pair
  Tunneling}},\ }\href {https://doi.org/10.1038/s41534-019-0231-2} {\bibfield
  {journal} {\bibinfo  {journal} {npj Quantum Inf.}\ }\textbf {\bibinfo
  {volume} {6}},\ \bibinfo {pages} {8} (\bibinfo {year} {2020})}\BibitemShut
  {NoStop}%
\bibitem [{\citenamefont {Rigetti}\ \emph {et~al.}(2012)\citenamefont
  {Rigetti}, \citenamefont {Gambetta}, \citenamefont {Poletto}, \citenamefont
  {Plourde}, \citenamefont {Chow}, \citenamefont {C\'orcoles}, \citenamefont
  {Smolin}, \citenamefont {Merkel}, \citenamefont {Rozen}, \citenamefont
  {Keefe}, \citenamefont {Rothwell}, \citenamefont {Ketchen},\ and\
  \citenamefont {Steffen}}]{Rigetti_photon_shot}%
  \BibitemOpen
  \bibfield  {author} {\bibinfo {author} {\bibfnamefont {C.}~\bibnamefont
  {Rigetti}}, \bibinfo {author} {\bibfnamefont {J.~M.}\ \bibnamefont
  {Gambetta}}, \bibinfo {author} {\bibfnamefont {S.}~\bibnamefont {Poletto}},
  \bibinfo {author} {\bibfnamefont {B.~L.~T.}\ \bibnamefont {Plourde}},
  \bibinfo {author} {\bibfnamefont {J.~M.}\ \bibnamefont {Chow}}, \bibinfo
  {author} {\bibfnamefont {A.~D.}\ \bibnamefont {C\'orcoles}}, \bibinfo
  {author} {\bibfnamefont {J.~A.}\ \bibnamefont {Smolin}}, \bibinfo {author}
  {\bibfnamefont {S.~T.}\ \bibnamefont {Merkel}}, \bibinfo {author}
  {\bibfnamefont {J.~R.}\ \bibnamefont {Rozen}}, \bibinfo {author}
  {\bibfnamefont {G.~A.}\ \bibnamefont {Keefe}}, \bibinfo {author}
  {\bibfnamefont {M.~B.}\ \bibnamefont {Rothwell}}, \bibinfo {author}
  {\bibfnamefont {M.~B.}\ \bibnamefont {Ketchen}},\ and\ \bibinfo {author}
  {\bibfnamefont {M.}~\bibnamefont {Steffen}},\ }\bibfield  {title} {\bibinfo
  {title} {\textit{Superconducting Qubit in a Waveguide Cavity with a Coherence
  Time Approaching 0.1 ms}},\ }\href
  {https://doi.org/10.1103/PhysRevB.86.100506} {\bibfield  {journal} {\bibinfo
  {journal} {Phys. Rev. B}\ }\textbf {\bibinfo {volume} {86}},\ \bibinfo
  {pages} {100506} (\bibinfo {year} {2012})}\BibitemShut {NoStop}%
\bibitem [{\citenamefont {Sears}\ \emph {et~al.}(2012)\citenamefont {Sears},
  \citenamefont {Petrenko}, \citenamefont {Catelani}, \citenamefont {Sun},
  \citenamefont {Paik}, \citenamefont {Kirchmair}, \citenamefont {Frunzio},
  \citenamefont {Glazman}, \citenamefont {Girvin},\ and\ \citenamefont
  {Schoelkopf}}]{Schoelkopf_photon_shot_noise}%
  \BibitemOpen
  \bibfield  {author} {\bibinfo {author} {\bibfnamefont {A.~P.}\ \bibnamefont
  {Sears}}, \bibinfo {author} {\bibfnamefont {A.}~\bibnamefont {Petrenko}},
  \bibinfo {author} {\bibfnamefont {G.}~\bibnamefont {Catelani}}, \bibinfo
  {author} {\bibfnamefont {L.}~\bibnamefont {Sun}}, \bibinfo {author}
  {\bibfnamefont {H.}~\bibnamefont {Paik}}, \bibinfo {author} {\bibfnamefont
  {G.}~\bibnamefont {Kirchmair}}, \bibinfo {author} {\bibfnamefont
  {L.}~\bibnamefont {Frunzio}}, \bibinfo {author} {\bibfnamefont {L.~I.}\
  \bibnamefont {Glazman}}, \bibinfo {author} {\bibfnamefont {S.~M.}\
  \bibnamefont {Girvin}},\ and\ \bibinfo {author} {\bibfnamefont {R.~J.}\
  \bibnamefont {Schoelkopf}},\ }\bibfield  {title} {\bibinfo {title}
  {\textit{Photon Shot Noise Dephasing in the Strong-Dispersive Limit of
  Circuit QED}},\ }\href {https://doi.org/10.1103/PhysRevB.86.180504}
  {\bibfield  {journal} {\bibinfo  {journal} {Phys. Rev. B}\ }\textbf {\bibinfo
  {volume} {86}},\ \bibinfo {pages} {180504} (\bibinfo {year}
  {2012})}\BibitemShut {NoStop}%
\bibitem [{\citenamefont {Kohler}\ \emph {et~al.}(1997)\citenamefont {Kohler},
  \citenamefont {Dittrich},\ and\ \citenamefont
  {H\"anggi}}]{Hanggi_Floquet_Markovian}%
  \BibitemOpen
  \bibfield  {author} {\bibinfo {author} {\bibfnamefont {S.}~\bibnamefont
  {Kohler}}, \bibinfo {author} {\bibfnamefont {T.}~\bibnamefont {Dittrich}},\
  and\ \bibinfo {author} {\bibfnamefont {P.}~\bibnamefont {H\"anggi}},\
  }\bibfield  {title} {\bibinfo {title} {\textit{Floquet-Markovian Description
  of the Parametrically Driven, Dissipative Harmonic Quantum Oscillator}},\
  }\href {https://doi.org/10.1103/PhysRevE.55.300} {\bibfield  {journal}
  {\bibinfo  {journal} {Phys. Rev. E}\ }\textbf {\bibinfo {volume} {55}},\
  \bibinfo {pages} {300} (\bibinfo {year} {1997})}\BibitemShut {NoStop}%
\bibitem [{\citenamefont {Breuer}\ and\ \citenamefont
  {Petruccione}(2007)}]{Breuer_open_quantum_system}%
  \BibitemOpen
  \bibfield  {author} {\bibinfo {author} {\bibfnamefont {H.}~\bibnamefont
  {Breuer}}\ and\ \bibinfo {author} {\bibfnamefont {F.}~\bibnamefont
  {Petruccione}},\ }\href@noop {} {\emph {\bibinfo {title} {The Theory of Open
  Quantum Systems}}}\ (\bibinfo  {publisher} {Oxford University Press, New
  York},\ \bibinfo {year} {2007})\BibitemShut {NoStop}%
\bibitem [{\citenamefont {Son}\ \emph {et~al.}(2009)\citenamefont {Son},
  \citenamefont {Han},\ and\ \citenamefont {Chu}}]{Chu_flux_qubit_Floquet}%
  \BibitemOpen
  \bibfield  {author} {\bibinfo {author} {\bibfnamefont {S.-K.}\ \bibnamefont
  {Son}}, \bibinfo {author} {\bibfnamefont {S.}~\bibnamefont {Han}},\ and\
  \bibinfo {author} {\bibfnamefont {S.-I.}\ \bibnamefont {Chu}},\ }\bibfield
  {title} {\bibinfo {title} {\textit{Floquet Formulation for the Investigation
  of Multiphoton Quantum Interference in a Superconducting Qubit Driven by a
  Strong ac Field}},\ }\href {https://doi.org/10.1103/PhysRevA.79.032301}
  {\bibfield  {journal} {\bibinfo  {journal} {Phys. Rev. A}\ }\textbf {\bibinfo
  {volume} {79}},\ \bibinfo {pages} {032301} (\bibinfo {year}
  {2009})}\BibitemShut {NoStop}%
\bibitem [{\citenamefont {Deng}\ \emph {et~al.}(2015)\citenamefont {Deng},
  \citenamefont {Orgiazzi}, \citenamefont {Shen}, \citenamefont {Ashhab},\ and\
  \citenamefont {Lupascu}}]{Lupascu_flux_qubit_Floquet}%
  \BibitemOpen
  \bibfield  {author} {\bibinfo {author} {\bibfnamefont {C.}~\bibnamefont
  {Deng}}, \bibinfo {author} {\bibfnamefont {J.-L.}\ \bibnamefont {Orgiazzi}},
  \bibinfo {author} {\bibfnamefont {F.}~\bibnamefont {Shen}}, \bibinfo {author}
  {\bibfnamefont {S.}~\bibnamefont {Ashhab}},\ and\ \bibinfo {author}
  {\bibfnamefont {A.}~\bibnamefont {Lupascu}},\ }\bibfield  {title} {\bibinfo
  {title} {\textit{Observation of Floquet States in a Strongly Driven
  Artificial Atom}},\ }\href {https://doi.org/10.1103/PhysRevLett.115.133601}
  {\bibfield  {journal} {\bibinfo  {journal} {Phys. Rev. Lett.}\ }\textbf
  {\bibinfo {volume} {115}},\ \bibinfo {pages} {133601} (\bibinfo {year}
  {2015})}\BibitemShut {NoStop}%
\bibitem [{\citenamefont {Hausinger}\ and\ \citenamefont
  {Grifoni}(2010)}]{Grifoni_dissipative_Floquet}%
  \BibitemOpen
  \bibfield  {author} {\bibinfo {author} {\bibfnamefont {J.}~\bibnamefont
  {Hausinger}}\ and\ \bibinfo {author} {\bibfnamefont {M.}~\bibnamefont
  {Grifoni}},\ }\bibfield  {title} {\bibinfo {title} {\textit{Dissipative
  Two-Level System Under Strong ac Driving: A Combination of Floquet and Van
  Vleck Perturbation Theory}},\ }\href
  {https://doi.org/10.1103/PhysRevA.81.022117} {\bibfield  {journal} {\bibinfo
  {journal} {Phys. Rev. A}\ }\textbf {\bibinfo {volume} {81}},\ \bibinfo
  {pages} {022117} (\bibinfo {year} {2010})}\BibitemShut {NoStop}%
\bibitem [{sid()}]{sidenote_pure_dephasing}%
  \BibitemOpen
  \href@noop {} {}\bibinfo {note} {Due to the non-exponential decay for $1/f$
  noise, the relationship between rate and filter function has to be modified
  (see Appendix C).}\BibitemShut {Stop}%
\bibitem [{\citenamefont {Stehlik}\ \emph {et~al.}(2016)\citenamefont
  {Stehlik}, \citenamefont {Liu}, \citenamefont {Eichler}, \citenamefont
  {Hartke}, \citenamefont {Mi}, \citenamefont {Gullans}, \citenamefont
  {Taylor},\ and\ \citenamefont {Petta}}]{Petta_double_quantum_well_Floquet}%
  \BibitemOpen
  \bibfield  {author} {\bibinfo {author} {\bibfnamefont {J.}~\bibnamefont
  {Stehlik}}, \bibinfo {author} {\bibfnamefont {Y.-Y.}\ \bibnamefont {Liu}},
  \bibinfo {author} {\bibfnamefont {C.}~\bibnamefont {Eichler}}, \bibinfo
  {author} {\bibfnamefont {T.~R.}\ \bibnamefont {Hartke}}, \bibinfo {author}
  {\bibfnamefont {X.}~\bibnamefont {Mi}}, \bibinfo {author} {\bibfnamefont
  {M.~J.}\ \bibnamefont {Gullans}}, \bibinfo {author} {\bibfnamefont {J.~M.}\
  \bibnamefont {Taylor}},\ and\ \bibinfo {author} {\bibfnamefont {J.~R.}\
  \bibnamefont {Petta}},\ }\bibfield  {title} {\bibinfo {title} {\textit{Double
  Quantum Dot Floquet Gain Medium}},\ }\href
  {https://doi.org/10.1103/PhysRevX.6.041027} {\bibfield  {journal} {\bibinfo
  {journal} {Phys. Rev. X}\ }\textbf {\bibinfo {volume} {6}},\ \bibinfo {pages}
  {041027} (\bibinfo {year} {2016})}\BibitemShut {NoStop}%
\bibitem [{\citenamefont {Ashhab}\ \emph {et~al.}(2007)\citenamefont {Ashhab},
  \citenamefont {Johansson}, \citenamefont {Zagoskin},\ and\ \citenamefont
  {Nori}}]{Nori_strong_drive_Floquet}%
  \BibitemOpen
  \bibfield  {author} {\bibinfo {author} {\bibfnamefont {S.}~\bibnamefont
  {Ashhab}}, \bibinfo {author} {\bibfnamefont {J.~R.}\ \bibnamefont
  {Johansson}}, \bibinfo {author} {\bibfnamefont {A.~M.}\ \bibnamefont
  {Zagoskin}},\ and\ \bibinfo {author} {\bibfnamefont {F.}~\bibnamefont
  {Nori}},\ }\bibfield  {title} {\bibinfo {title} {\textit{Two-Level Systems
  Driven by Large-Amplitude Fields}},\ }\href
  {https://doi.org/10.1103/PhysRevA.75.063414} {\bibfield  {journal} {\bibinfo
  {journal} {Phys. Rev. A}\ }\textbf {\bibinfo {volume} {75}},\ \bibinfo
  {pages} {063414} (\bibinfo {year} {2007})}\BibitemShut {NoStop}%
\bibitem [{\citenamefont {Silveri}\ \emph {et~al.}(2017)\citenamefont
  {Silveri}, \citenamefont {Tuorila}, \citenamefont {Thuneberg},\ and\
  \citenamefont {Paraoanu}}]{Paraoanu_frequency_modulation_review}%
  \BibitemOpen
  \bibfield  {author} {\bibinfo {author} {\bibfnamefont {M.~P.}\ \bibnamefont
  {Silveri}}, \bibinfo {author} {\bibfnamefont {J.~A.}\ \bibnamefont
  {Tuorila}}, \bibinfo {author} {\bibfnamefont {E.~V.}\ \bibnamefont
  {Thuneberg}},\ and\ \bibinfo {author} {\bibfnamefont {G.~S.}\ \bibnamefont
  {Paraoanu}},\ }\bibfield  {title} {\bibinfo {title} {\textit{Quantum Systems
  Under Frequency Modulation}},\ }\href
  {https://doi.org/https://doi.org/10.1088/1361-6633/aa5170} {\bibfield
  {journal} {\bibinfo  {journal} {Rep. Prog. Phys.}\ }\textbf {\bibinfo
  {volume} {80}},\ \bibinfo {pages} {056002} (\bibinfo {year}
  {2017})}\BibitemShut {NoStop}%
\bibitem [{swe()}]{sweet_spot_stability}%
  \BibitemOpen
  \href@noop {} {}\bibinfo {note} {The stability of the doubly sweet spots with
  respect to the dc flux bias can be estimated by the second derivative of the
  quasi-energy difference. We have verified numerically that our protection
  scheme does not significantly deteriorate the sweet-spot
  stability.}\BibitemShut {Stop}%
\bibitem [{\citenamefont {Li}\ \emph {et~al.}(2013)\citenamefont {Li},
  \citenamefont {Silveri}, \citenamefont {Kumar}, \citenamefont {Pirkkalainen},
  \citenamefont {Veps{\"a}l{\"a}inen}, \citenamefont {Chien}, \citenamefont
  {Tuorila}, \citenamefont {Sillanp{\"a}{\"a}}, \citenamefont {Hakonen},
  \citenamefont {Thuneberg},\ and\ \citenamefont
  {Paraoanu}}]{Paraoanu_dressed_qubit}%
  \BibitemOpen
  \bibfield  {author} {\bibinfo {author} {\bibfnamefont {J.}~\bibnamefont
  {Li}}, \bibinfo {author} {\bibfnamefont {M.~P.}\ \bibnamefont {Silveri}},
  \bibinfo {author} {\bibfnamefont {K.~S.}\ \bibnamefont {Kumar}}, \bibinfo
  {author} {\bibfnamefont {J.~M.}\ \bibnamefont {Pirkkalainen}}, \bibinfo
  {author} {\bibfnamefont {A.}~\bibnamefont {Veps{\"a}l{\"a}inen}}, \bibinfo
  {author} {\bibfnamefont {W.~C.}\ \bibnamefont {Chien}}, \bibinfo {author}
  {\bibfnamefont {J.}~\bibnamefont {Tuorila}}, \bibinfo {author} {\bibfnamefont
  {M.~A.}\ \bibnamefont {Sillanp{\"a}{\"a}}}, \bibinfo {author} {\bibfnamefont
  {P.~J.}\ \bibnamefont {Hakonen}}, \bibinfo {author} {\bibfnamefont {E.~V.}\
  \bibnamefont {Thuneberg}},\ and\ \bibinfo {author} {\bibfnamefont {G.~S.}\
  \bibnamefont {Paraoanu}},\ }\bibfield  {title} {\bibinfo {title}
  {\textit{Motional Averaging in A Superconducting Qubit}},\ }\href
  {https://doi.org/10.1038/ncomms2383} {\bibfield  {journal} {\bibinfo
  {journal} {Nat. Commun.}\ }\textbf {\bibinfo {volume} {4}},\ \bibinfo {pages}
  {1420} (\bibinfo {year} {2013})}\BibitemShut {NoStop}%
\bibitem [{\citenamefont {Tan}\ \emph {et~al.}(2013)\citenamefont {Tan},
  \citenamefont {Gaebler}, \citenamefont {Bowler}, \citenamefont {Lin},
  \citenamefont {Jost}, \citenamefont {Leibfried},\ and\ \citenamefont
  {Wineland}}]{Wineland_dressed_qubit}%
  \BibitemOpen
  \bibfield  {author} {\bibinfo {author} {\bibfnamefont {T.~R.}\ \bibnamefont
  {Tan}}, \bibinfo {author} {\bibfnamefont {J.~P.}\ \bibnamefont {Gaebler}},
  \bibinfo {author} {\bibfnamefont {R.}~\bibnamefont {Bowler}}, \bibinfo
  {author} {\bibfnamefont {Y.}~\bibnamefont {Lin}}, \bibinfo {author}
  {\bibfnamefont {J.~D.}\ \bibnamefont {Jost}}, \bibinfo {author}
  {\bibfnamefont {D.}~\bibnamefont {Leibfried}},\ and\ \bibinfo {author}
  {\bibfnamefont {D.~J.}\ \bibnamefont {Wineland}},\ }\bibfield  {title}
  {\bibinfo {title} {\textit{Demonstration of a Dressed-State Phase Gate for
  Trapped Ions}},\ }\href {https://doi.org/10.1103/PhysRevLett.110.263002}
  {\bibfield  {journal} {\bibinfo  {journal} {Phys. Rev. Lett.}\ }\textbf
  {\bibinfo {volume} {110}},\ \bibinfo {pages} {263002} (\bibinfo {year}
  {2013})}\BibitemShut {NoStop}%
\bibitem [{\citenamefont {Mikelsons}\ \emph {et~al.}(2015)\citenamefont
  {Mikelsons}, \citenamefont {Cohen}, \citenamefont {Retzker},\ and\
  \citenamefont {Plenio}}]{Plenio_dressed_qubit_gates}%
  \BibitemOpen
  \bibfield  {author} {\bibinfo {author} {\bibfnamefont {G.}~\bibnamefont
  {Mikelsons}}, \bibinfo {author} {\bibfnamefont {I.}~\bibnamefont {Cohen}},
  \bibinfo {author} {\bibfnamefont {A.}~\bibnamefont {Retzker}},\ and\ \bibinfo
  {author} {\bibfnamefont {M.~B.}\ \bibnamefont {Plenio}},\ }\bibfield  {title}
  {\bibinfo {title} {\textit{Universal Set of Gates for Microwave Dressed-State
  Quantum Computing}},\ }\href
  {https://doi.org/https://doi.org/10.1088/1367-2630/17/5/053032} {\bibfield
  {journal} {\bibinfo  {journal} {New J. Phys.}\ }\textbf {\bibinfo {volume}
  {17}},\ \bibinfo {pages} {053032} (\bibinfo {year} {2015})}\BibitemShut
  {NoStop}%
\bibitem [{\citenamefont {Gu\'erin}(1997)}]{GuerinFloquet}%
  \BibitemOpen
  \bibfield  {author} {\bibinfo {author} {\bibfnamefont {S.}~\bibnamefont
  {Gu\'erin}},\ }\bibfield  {title} {\bibinfo {title} {\textit{Complete
  Dissociation by Chirped Laser Pulses Designed by Adiabatic Floquet
  Analysis}},\ }\href {https://doi.org/10.1103/PhysRevA.56.1458} {\bibfield
  {journal} {\bibinfo  {journal} {Phys. Rev. A}\ }\textbf {\bibinfo {volume}
  {56}},\ \bibinfo {pages} {1458} (\bibinfo {year} {1997})}\BibitemShut
  {NoStop}%
\bibitem [{\citenamefont {Yang}\ \emph {et~al.}(2020)\citenamefont {Yang},
  \citenamefont {Coppersmith},\ and\ \citenamefont
  {Friesen}}]{Friesen_all_sweet_two_q_gate}%
  \BibitemOpen
  \bibfield  {author} {\bibinfo {author} {\bibfnamefont {Y.-C.}\ \bibnamefont
  {Yang}}, \bibinfo {author} {\bibfnamefont {S.~N.}\ \bibnamefont
  {Coppersmith}},\ and\ \bibinfo {author} {\bibfnamefont {M.}~\bibnamefont
  {Friesen}},\ }\bibfield  {title} {\bibinfo {title} {\textit{High-Fidelity
  Entangling Gates for Quantum-Dot Hybrid Qubits Based on Exchange
  Interactions}},\ }\href {https://doi.org/10.1103/PhysRevA.101.012338}
  {\bibfield  {journal} {\bibinfo  {journal} {Phys. Rev. A}\ }\textbf {\bibinfo
  {volume} {101}},\ \bibinfo {pages} {012338} (\bibinfo {year}
  {2020})}\BibitemShut {NoStop}%
\bibitem [{\citenamefont {Yan}\ \emph {et~al.}(2018)\citenamefont {Yan},
  \citenamefont {Krantz}, \citenamefont {Sung}, \citenamefont {Kjaergaard},
  \citenamefont {Campbell}, \citenamefont {Orlando}, \citenamefont
  {Gustavsson},\ and\ \citenamefont {Oliver}}]{Oliver_tunable_coupling}%
  \BibitemOpen
  \bibfield  {author} {\bibinfo {author} {\bibfnamefont {F.}~\bibnamefont
  {Yan}}, \bibinfo {author} {\bibfnamefont {P.}~\bibnamefont {Krantz}},
  \bibinfo {author} {\bibfnamefont {Y.}~\bibnamefont {Sung}}, \bibinfo {author}
  {\bibfnamefont {M.}~\bibnamefont {Kjaergaard}}, \bibinfo {author}
  {\bibfnamefont {D.~L.}\ \bibnamefont {Campbell}}, \bibinfo {author}
  {\bibfnamefont {T.~P.}\ \bibnamefont {Orlando}}, \bibinfo {author}
  {\bibfnamefont {S.}~\bibnamefont {Gustavsson}},\ and\ \bibinfo {author}
  {\bibfnamefont {W.~D.}\ \bibnamefont {Oliver}},\ }\bibfield  {title}
  {\bibinfo {title} {\textit{Tunable Coupling Scheme for Implementing
  High-Fidelity Two-Qubit Gates}},\ }\href
  {https://doi.org/10.1103/PhysRevApplied.10.054062} {\bibfield  {journal}
  {\bibinfo  {journal} {Phys. Rev. Applied}\ }\textbf {\bibinfo {volume}
  {10}},\ \bibinfo {pages} {054062} (\bibinfo {year} {2018})}\BibitemShut
  {NoStop}%
\bibitem [{\citenamefont {Chen}\ \emph {et~al.}(2014)\citenamefont {Chen},
  \citenamefont {Neill}, \citenamefont {Roushan}, \citenamefont {Leung},
  \citenamefont {Fang}, \citenamefont {Barends}, \citenamefont {Kelly},
  \citenamefont {Campbell}, \citenamefont {Chen}, \citenamefont {Chiaro},
  \citenamefont {Dunsworth}, \citenamefont {Jeffrey}, \citenamefont {Megrant},
  \citenamefont {Mutus}, \citenamefont {O'Malley}, \citenamefont {Quintana},
  \citenamefont {Sank}, \citenamefont {Vainsencher}, \citenamefont {Wenner},
  \citenamefont {White}, \citenamefont {Geller}, \citenamefont {Cleland},\ and\
  \citenamefont {Martinis}}]{Martinis_tunable_coupling}%
  \BibitemOpen
  \bibfield  {author} {\bibinfo {author} {\bibfnamefont {Y.}~\bibnamefont
  {Chen}}, \bibinfo {author} {\bibfnamefont {C.}~\bibnamefont {Neill}},
  \bibinfo {author} {\bibfnamefont {P.}~\bibnamefont {Roushan}}, \bibinfo
  {author} {\bibfnamefont {N.}~\bibnamefont {Leung}}, \bibinfo {author}
  {\bibfnamefont {M.}~\bibnamefont {Fang}}, \bibinfo {author} {\bibfnamefont
  {R.}~\bibnamefont {Barends}}, \bibinfo {author} {\bibfnamefont
  {J.}~\bibnamefont {Kelly}}, \bibinfo {author} {\bibfnamefont
  {B.}~\bibnamefont {Campbell}}, \bibinfo {author} {\bibfnamefont
  {Z.}~\bibnamefont {Chen}}, \bibinfo {author} {\bibfnamefont {B.}~\bibnamefont
  {Chiaro}}, \bibinfo {author} {\bibfnamefont {A.}~\bibnamefont {Dunsworth}},
  \bibinfo {author} {\bibfnamefont {E.}~\bibnamefont {Jeffrey}}, \bibinfo
  {author} {\bibfnamefont {A.}~\bibnamefont {Megrant}}, \bibinfo {author}
  {\bibfnamefont {J.~Y.}\ \bibnamefont {Mutus}}, \bibinfo {author}
  {\bibfnamefont {P.~J.~J.}\ \bibnamefont {O'Malley}}, \bibinfo {author}
  {\bibfnamefont {C.~M.}\ \bibnamefont {Quintana}}, \bibinfo {author}
  {\bibfnamefont {D.}~\bibnamefont {Sank}}, \bibinfo {author} {\bibfnamefont
  {A.}~\bibnamefont {Vainsencher}}, \bibinfo {author} {\bibfnamefont
  {J.}~\bibnamefont {Wenner}}, \bibinfo {author} {\bibfnamefont {T.~C.}\
  \bibnamefont {White}}, \bibinfo {author} {\bibfnamefont {M.~R.}\ \bibnamefont
  {Geller}}, \bibinfo {author} {\bibfnamefont {A.~N.}\ \bibnamefont
  {Cleland}},\ and\ \bibinfo {author} {\bibfnamefont {J.~M.}\ \bibnamefont
  {Martinis}},\ }\bibfield  {title} {\bibinfo {title} {\textit{Qubit
  Architecture with High Coherence and Fast Tunable Coupling}},\ }\href
  {https://doi.org/10.1103/PhysRevLett.113.220502} {\bibfield  {journal}
  {\bibinfo  {journal} {Phys. Rev. Lett.}\ }\textbf {\bibinfo {volume} {113}},\
  \bibinfo {pages} {220502} (\bibinfo {year} {2014})}\BibitemShut {NoStop}%
\bibitem [{\citenamefont {Xu}\ \emph {et~al.}()\citenamefont {Xu},
  \citenamefont {Chu}, \citenamefont {Yuan}, \citenamefont {Qiu}, \citenamefont
  {Zhou}, \citenamefont {Zhang}, \citenamefont {Tan}, \citenamefont {Yu},
  \citenamefont {Liu}, \citenamefont {Li}, \citenamefont {Yan},\ and\
  \citenamefont {Yu}}]{Yu_two_qubit_gates}%
  \BibitemOpen
  \bibfield  {author} {\bibinfo {author} {\bibfnamefont {Y.}~\bibnamefont
  {Xu}}, \bibinfo {author} {\bibfnamefont {J.}~\bibnamefont {Chu}}, \bibinfo
  {author} {\bibfnamefont {J.}~\bibnamefont {Yuan}}, \bibinfo {author}
  {\bibfnamefont {J.}~\bibnamefont {Qiu}}, \bibinfo {author} {\bibfnamefont
  {Y.}~\bibnamefont {Zhou}}, \bibinfo {author} {\bibfnamefont {L.}~\bibnamefont
  {Zhang}}, \bibinfo {author} {\bibfnamefont {X.}~\bibnamefont {Tan}}, \bibinfo
  {author} {\bibfnamefont {Y.}~\bibnamefont {Yu}}, \bibinfo {author}
  {\bibfnamefont {S.}~\bibnamefont {Liu}}, \bibinfo {author} {\bibfnamefont
  {J.}~\bibnamefont {Li}}, \bibinfo {author} {\bibfnamefont {F.}~\bibnamefont
  {Yan}},\ and\ \bibinfo {author} {\bibfnamefont {D.}~\bibnamefont {Yu}},\
  }\bibfield  {title} {\bibinfo {title} {\textit{High-Fidelity,
  High-Scalability Two-Qubit Gate Scheme for Superconducting Qubits}},\
  }\Eprint {https://arxiv.org/abs/arXiv:2006.11860} {arXiv:2006.11860}
  \BibitemShut {NoStop}%
\bibitem [{\citenamefont {Mundada}\ \emph {et~al.}(2019)\citenamefont
  {Mundada}, \citenamefont {Zhang}, \citenamefont {Hazard},\ and\ \citenamefont
  {Houck}}]{Houck_two_qubit_gates}%
  \BibitemOpen
  \bibfield  {author} {\bibinfo {author} {\bibfnamefont {P.}~\bibnamefont
  {Mundada}}, \bibinfo {author} {\bibfnamefont {G.}~\bibnamefont {Zhang}},
  \bibinfo {author} {\bibfnamefont {T.}~\bibnamefont {Hazard}},\ and\ \bibinfo
  {author} {\bibfnamefont {A.}~\bibnamefont {Houck}},\ }\bibfield  {title}
  {\bibinfo {title} {\textit{Suppression of Qubit Crosstalk in a Tunable
  Coupling Superconducting Circuit}},\ }\href
  {https://doi.org/10.1103/PhysRevApplied.12.054023} {\bibfield  {journal}
  {\bibinfo  {journal} {Phys. Rev. Applied}\ }\textbf {\bibinfo {volume}
  {12}},\ \bibinfo {pages} {054023} (\bibinfo {year} {2019})}\BibitemShut
  {NoStop}%
\bibitem [{\citenamefont {Srinivasan}\ \emph {et~al.}(2011)\citenamefont
  {Srinivasan}, \citenamefont {Hoffman}, \citenamefont {Gambetta},\ and\
  \citenamefont {Houck}}]{Houck_tunable_coupling}%
  \BibitemOpen
  \bibfield  {author} {\bibinfo {author} {\bibfnamefont {S.~J.}\ \bibnamefont
  {Srinivasan}}, \bibinfo {author} {\bibfnamefont {A.~J.}\ \bibnamefont
  {Hoffman}}, \bibinfo {author} {\bibfnamefont {J.~M.}\ \bibnamefont
  {Gambetta}},\ and\ \bibinfo {author} {\bibfnamefont {A.~A.}\ \bibnamefont
  {Houck}},\ }\bibfield  {title} {\bibinfo {title} {\textit{Tunable Coupling in
  Circuit Quantum Electrodynamics Using a Superconducting Charge Qubit with a
  $V$-Shaped Energy Level Diagram}},\ }\href
  {https://doi.org/10.1103/PhysRevLett.106.083601} {\bibfield  {journal}
  {\bibinfo  {journal} {Phys. Rev. Lett.}\ }\textbf {\bibinfo {volume} {106}},\
  \bibinfo {pages} {083601} (\bibinfo {year} {2011})}\BibitemShut {NoStop}%
\bibitem [{\citenamefont {You}\ \emph {et~al.}(2019)\citenamefont {You},
  \citenamefont {Sauls},\ and\ \citenamefont {Koch}}]{You_flux_allocation}%
  \BibitemOpen
  \bibfield  {author} {\bibinfo {author} {\bibfnamefont {X.}~\bibnamefont
  {You}}, \bibinfo {author} {\bibfnamefont {J.~A.}\ \bibnamefont {Sauls}},\
  and\ \bibinfo {author} {\bibfnamefont {J.}~\bibnamefont {Koch}},\ }\bibfield
  {title} {\bibinfo {title} {\textit{Circuit Quantization in the Presence of
  Time-Dependent External Flux}},\ }\href
  {https://doi.org/10.1103/PhysRevB.99.174512} {\bibfield  {journal} {\bibinfo
  {journal} {Phys. Rev. B}\ }\textbf {\bibinfo {volume} {99}},\ \bibinfo
  {pages} {174512} (\bibinfo {year} {2019})}\BibitemShut {NoStop}%
\end{thebibliography}%


\begin{thebibliography}{10}%
\makeatletter
\providecommand \@ifxundefined [1]{%
 \@ifx{#1\undefined}
}%
\providecommand \@ifnum [1]{%
 \ifnum #1\expandafter \@firstoftwo
 \else \expandafter \@secondoftwo
 \fi
}%
\providecommand \@ifx [1]{%
 \ifx #1\expandafter \@firstoftwo
 \else \expandafter \@secondoftwo
 \fi
}%
\providecommand \natexlab [1]{#1}%
\providecommand \enquote  [1]{``#1''}%
\providecommand \bibnamefont  [1]{#1}%
\providecommand \bibfnamefont [1]{#1}%
\providecommand \citenamefont [1]{#1}%
\providecommand \href@noop [0]{\@secondoftwo}%
\providecommand \href [0]{\begingroup \@sanitize@url \@href}%
\providecommand \@href[1]{\@@startlink{#1}\@@href}%
\providecommand \@@href[1]{\endgroup#1\@@endlink}%
\providecommand \@sanitize@url [0]{\catcode `\\12\catcode `\$12\catcode
  `\&12\catcode `\#12\catcode `\^12\catcode `\_12\catcode `\%12\relax}%
\providecommand \@@startlink[1]{}%
\providecommand \@@endlink[0]{}%
\providecommand \url  [0]{\begingroup\@sanitize@url \@url }%
\providecommand \@url [1]{\endgroup\@href {#1}{\urlprefix }}%
\providecommand \urlprefix  [0]{URL }%
\providecommand \Eprint [0]{\href }%
\providecommand \doibase [0]{https://doi.org/}%
\providecommand \selectlanguage [0]{\@gobble}%
\providecommand \bibinfo  [0]{\@secondoftwo}%
\providecommand \bibfield  [0]{\@secondoftwo}%
\providecommand \translation [1]{[#1]}%
\providecommand \BibitemOpen [0]{}%
\providecommand \bibitemStop [0]{}%
\providecommand \bibitemNoStop [0]{.\EOS\space}%
\providecommand \EOS [0]{\spacefactor3000\relax}%
\providecommand \BibitemShut  [1]{\csname bibitem#1\endcsname}%
\let\auto@bib@innerbib\@empty
\bibitem [{\citenamefont {Frees}\ \emph {et~al.}(2019)\citenamefont {Frees},
  \citenamefont {Mehl}, \citenamefont {Gamble}, \citenamefont {Friesen},\ and\
  \citenamefont {Coppersmith}}]{Coppersmith_two_qubit_sweet_spot}%
  \BibitemOpen
  \bibfield  {author} {\bibinfo {author} {\bibfnamefont {A.}~\bibnamefont
  {Frees}}, \bibinfo {author} {\bibfnamefont {S.}~\bibnamefont {Mehl}},
  \bibinfo {author} {\bibfnamefont {J.~K.}\ \bibnamefont {Gamble}}, \bibinfo
  {author} {\bibfnamefont {M.}~\bibnamefont {Friesen}},\ and\ \bibinfo {author}
  {\bibfnamefont {S.~N.}\ \bibnamefont {Coppersmith}},\ }\bibfield  {title}
  {\bibinfo {title} {\textit{Adiabatic Two-Qubit Gates in Capacitively Coupled
  Quantum Dot Hybrid Qubits}},\ }\href
  {https://doi.org/10.1038/s41534-019-0190-7} {\bibfield  {journal} {\bibinfo
  {journal} {npj Quantum Inform.}\ }\textbf {\bibinfo {volume} {5}},\ \bibinfo
  {pages} {73} (\bibinfo {year} {2019})}\BibitemShut {NoStop}%
\bibitem [{\citenamefont {Didier}()}]{Didier_dynamical_sweet_spot}%
  \BibitemOpen
  \bibfield  {author} {\bibinfo {author} {\bibfnamefont {N.}~\bibnamefont
  {Didier}},\ }\bibfield  {title} {\bibinfo {title} {\textit{Flux Control of
  Superconducting Qubits at Dynamical Sweet Spots}},\ }\Eprint
  {https://arxiv.org/abs/arXiv:1912.09416} {arXiv:1912.09416} \BibitemShut
  {NoStop}%
\bibitem [{\citenamefont {Didier}\ \emph {et~al.}(2019)\citenamefont {Didier},
  \citenamefont {Sete}, \citenamefont {Combes},\ and\ \citenamefont
  {da~Silva}}]{Rigetti_ac_sweet_spot}%
  \BibitemOpen
  \bibfield  {author} {\bibinfo {author} {\bibfnamefont {N.}~\bibnamefont
  {Didier}}, \bibinfo {author} {\bibfnamefont {E.~A.}\ \bibnamefont {Sete}},
  \bibinfo {author} {\bibfnamefont {J.}~\bibnamefont {Combes}},\ and\ \bibinfo
  {author} {\bibfnamefont {M.~P.}\ \bibnamefont {da~Silva}},\ }\bibfield
  {title} {\bibinfo {title} {\textit{ac Flux Sweet Spots in Parametrically
  Modulated Superconducting Qubits}},\ }\href
  {https://doi.org/10.1103/PhysRevApplied.12.054015} {\bibfield  {journal}
  {\bibinfo  {journal} {Phys. Rev. Applied}\ }\textbf {\bibinfo {volume}
  {12}},\ \bibinfo {pages} {054015} (\bibinfo {year} {2019})}\BibitemShut
  {NoStop}%
\bibitem [{\citenamefont {Hong}\ \emph {et~al.}(2020)\citenamefont {Hong},
  \citenamefont {Papageorge}, \citenamefont {Sivarajah}, \citenamefont
  {Crossman}, \citenamefont {Didier}, \citenamefont {Polloreno}, \citenamefont
  {Sete}, \citenamefont {Turkowski}, \citenamefont {da~Silva},\ and\
  \citenamefont {Johnson}}]{Rigetti_ac_sweet_spot_exp}%
  \BibitemOpen
  \bibfield  {author} {\bibinfo {author} {\bibfnamefont {S.~S.}\ \bibnamefont
  {Hong}}, \bibinfo {author} {\bibfnamefont {A.~T.}\ \bibnamefont
  {Papageorge}}, \bibinfo {author} {\bibfnamefont {P.}~\bibnamefont
  {Sivarajah}}, \bibinfo {author} {\bibfnamefont {G.}~\bibnamefont {Crossman}},
  \bibinfo {author} {\bibfnamefont {N.}~\bibnamefont {Didier}}, \bibinfo
  {author} {\bibfnamefont {A.~M.}\ \bibnamefont {Polloreno}}, \bibinfo {author}
  {\bibfnamefont {E.~A.}\ \bibnamefont {Sete}}, \bibinfo {author}
  {\bibfnamefont {S.~W.}\ \bibnamefont {Turkowski}}, \bibinfo {author}
  {\bibfnamefont {M.~P.}\ \bibnamefont {da~Silva}},\ and\ \bibinfo {author}
  {\bibfnamefont {B.~R.}\ \bibnamefont {Johnson}},\ }\bibfield  {title}
  {\bibinfo {title} {\textit{Demonstration of a Parametrically Activated
  Entangling Gate Protected from Flux Noise}},\ }\href
  {https://doi.org/10.1103/PhysRevA.101.012302} {\bibfield  {journal} {\bibinfo
   {journal} {Phys. Rev. A}\ }\textbf {\bibinfo {volume} {101}},\ \bibinfo
  {pages} {012302} (\bibinfo {year} {2020})}\BibitemShut {NoStop}%
\bibitem [{\citenamefont {Fried}\ \emph {et~al.}()\citenamefont {Fried},
  \citenamefont {Sivarajah}, \citenamefont {Didier}, \citenamefont {Sete},
  \citenamefont {da~Silva}, \citenamefont {Johnson},\ and\ \citenamefont
  {Ryan}}]{Rigetti_broad_band_noise}%
  \BibitemOpen
  \bibfield  {author} {\bibinfo {author} {\bibfnamefont {E.~S.}\ \bibnamefont
  {Fried}}, \bibinfo {author} {\bibfnamefont {P.}~\bibnamefont {Sivarajah}},
  \bibinfo {author} {\bibfnamefont {N.}~\bibnamefont {Didier}}, \bibinfo
  {author} {\bibfnamefont {E.~A.}\ \bibnamefont {Sete}}, \bibinfo {author}
  {\bibfnamefont {M.~P.}\ \bibnamefont {da~Silva}}, \bibinfo {author}
  {\bibfnamefont {B.~R.}\ \bibnamefont {Johnson}},\ and\ \bibinfo {author}
  {\bibfnamefont {C.~A.}\ \bibnamefont {Ryan}},\ }\bibfield  {title} {\bibinfo
  {title} {\textit{Assessing the Influence of Broadband Instrumentation Noise
  on Parametrically Modulated Superconducting Qubits}},\ }\Eprint
  {https://arxiv.org/abs/arXiv:1908.11370} {arXiv:1908.11370} \BibitemShut
  {NoStop}%
\bibitem [{\citenamefont {Guo}\ \emph {et~al.}(2018)\citenamefont {Guo},
  \citenamefont {Zheng}, \citenamefont {Wang}, \citenamefont {Song},
  \citenamefont {Zhang}, \citenamefont {Li}, \citenamefont {Liu}, \citenamefont
  {Deng}, \citenamefont {Huang}, \citenamefont {Zheng}, \citenamefont {Zhu},
  \citenamefont {Wang}, \citenamefont {Lu},\ and\ \citenamefont
  {Pan}}]{Pan_transmon_dd}%
  \BibitemOpen
  \bibfield  {author} {\bibinfo {author} {\bibfnamefont {Q.}~\bibnamefont
  {Guo}}, \bibinfo {author} {\bibfnamefont {S.-B.}\ \bibnamefont {Zheng}},
  \bibinfo {author} {\bibfnamefont {J.}~\bibnamefont {Wang}}, \bibinfo {author}
  {\bibfnamefont {C.}~\bibnamefont {Song}}, \bibinfo {author} {\bibfnamefont
  {P.}~\bibnamefont {Zhang}}, \bibinfo {author} {\bibfnamefont
  {K.}~\bibnamefont {Li}}, \bibinfo {author} {\bibfnamefont {W.}~\bibnamefont
  {Liu}}, \bibinfo {author} {\bibfnamefont {H.}~\bibnamefont {Deng}}, \bibinfo
  {author} {\bibfnamefont {K.}~\bibnamefont {Huang}}, \bibinfo {author}
  {\bibfnamefont {D.}~\bibnamefont {Zheng}}, \bibinfo {author} {\bibfnamefont
  {X.}~\bibnamefont {Zhu}}, \bibinfo {author} {\bibfnamefont {H.}~\bibnamefont
  {Wang}}, \bibinfo {author} {\bibfnamefont {C.-Y.}\ \bibnamefont {Lu}},\ and\
  \bibinfo {author} {\bibfnamefont {J.-W.}\ \bibnamefont {Pan}},\ }\bibfield
  {title} {\bibinfo {title} {\textit{Dephasing-Insensitive Quantum Information
  Storage and Processing with Superconducting Qubits}},\ }\href
  {https://doi.org/10.1103/PhysRevLett.121.130501} {\bibfield  {journal}
  {\bibinfo  {journal} {Phys. Rev. Lett.}\ }\textbf {\bibinfo {volume} {121}},\
  \bibinfo {pages} {130501} (\bibinfo {year} {2018})}\BibitemShut {NoStop}%
\bibitem [{\citenamefont {Yan}\ \emph {et~al.}(2013)\citenamefont {Yan},
  \citenamefont {Gustavsson}, \citenamefont {Bylander}, \citenamefont {Jin},
  \citenamefont {Yoshihara}, \citenamefont {Cory}, \citenamefont {Nakamura},
  \citenamefont {Orlando},\ and\ \citenamefont
  {Oliver}}]{Oliver_rotating_frame_relaxation}%
  \BibitemOpen
  \bibfield  {author} {\bibinfo {author} {\bibfnamefont {F.}~\bibnamefont
  {Yan}}, \bibinfo {author} {\bibfnamefont {S.}~\bibnamefont {Gustavsson}},
  \bibinfo {author} {\bibfnamefont {J.}~\bibnamefont {Bylander}}, \bibinfo
  {author} {\bibfnamefont {X.}~\bibnamefont {Jin}}, \bibinfo {author}
  {\bibfnamefont {F.}~\bibnamefont {Yoshihara}}, \bibinfo {author}
  {\bibfnamefont {D.~G.}\ \bibnamefont {Cory}}, \bibinfo {author}
  {\bibfnamefont {Y.}~\bibnamefont {Nakamura}}, \bibinfo {author}
  {\bibfnamefont {T.~P.}\ \bibnamefont {Orlando}},\ and\ \bibinfo {author}
  {\bibfnamefont {W.~D.}\ \bibnamefont {Oliver}},\ }\bibfield  {title}
  {\bibinfo {title} {\textit{Rotating-Frame Relaxation as a Noise Spectrum
  Analyser of a Superconducting Qubit Undergoing Driven Evolution}},\ }\href
  {https://doi.org/10.1038/ncomms3337} {\bibfield  {journal} {\bibinfo
  {journal} {Nat. Commun.}\ }\textbf {\bibinfo {volume} {4}},\ \bibinfo {pages}
  {2337} (\bibinfo {year} {2013})}\BibitemShut {NoStop}%
\bibitem [{\citenamefont {Jing}\ \emph {et~al.}(2014)\citenamefont {Jing},
  \citenamefont {Huang},\ and\ \citenamefont {Hu}}]{Hu_drive_spin_qubit}%
  \BibitemOpen
  \bibfield  {author} {\bibinfo {author} {\bibfnamefont {J.}~\bibnamefont
  {Jing}}, \bibinfo {author} {\bibfnamefont {P.}~\bibnamefont {Huang}},\ and\
  \bibinfo {author} {\bibfnamefont {X.}~\bibnamefont {Hu}},\ }\bibfield
  {title} {\bibinfo {title} {\textit{Decoherence of an Electrically Driven Spin
  Qubit}},\ }\href {https://doi.org/10.1103/PhysRevA.90.022118} {\bibfield
  {journal} {\bibinfo  {journal} {Phys. Rev. A}\ }\textbf {\bibinfo {volume}
  {90}},\ \bibinfo {pages} {022118} (\bibinfo {year} {2014})}\BibitemShut
  {NoStop}%
\bibitem [{\citenamefont {Smirnov}(2003)}]{Smirnov_Raib_decoherence}%
  \BibitemOpen
  \bibfield  {author} {\bibinfo {author} {\bibfnamefont {A.~Y.}\ \bibnamefont
  {Smirnov}},\ }\bibfield  {title} {\bibinfo {title} {\textit{Decoherence and
  Relaxation of a Quantum Bit in the Presence of Rabi Oscillations}},\ }\href
  {https://doi.org/10.1103/PhysRevB.67.155104} {\bibfield  {journal} {\bibinfo
  {journal} {Phys. Rev. B}\ }\textbf {\bibinfo {volume} {67}},\ \bibinfo
  {pages} {155104} (\bibinfo {year} {2003})}\BibitemShut {NoStop}%
\bibitem [{\citenamefont {Ithier}\ \emph {et~al.}(2005)\citenamefont {Ithier},
  \citenamefont {Collin}, \citenamefont {Joyez}, \citenamefont {Meeson},
  \citenamefont {Vion}, \citenamefont {Esteve}, \citenamefont {Chiarello},
  \citenamefont {Shnirman}, \citenamefont {Makhlin}, \citenamefont {Schriefl},\
  and\ \citenamefont {Sch\"on}}]{Ithier_decoherence_analysis}%
  \BibitemOpen
  \bibfield  {author} {\bibinfo {author} {\bibfnamefont {G.}~\bibnamefont
  {Ithier}}, \bibinfo {author} {\bibfnamefont {E.}~\bibnamefont {Collin}},
  \bibinfo {author} {\bibfnamefont {P.}~\bibnamefont {Joyez}}, \bibinfo
  {author} {\bibfnamefont {P.~J.}\ \bibnamefont {Meeson}}, \bibinfo {author}
  {\bibfnamefont {D.}~\bibnamefont {Vion}}, \bibinfo {author} {\bibfnamefont
  {D.}~\bibnamefont {Esteve}}, \bibinfo {author} {\bibfnamefont
  {F.}~\bibnamefont {Chiarello}}, \bibinfo {author} {\bibfnamefont
  {A.}~\bibnamefont {Shnirman}}, \bibinfo {author} {\bibfnamefont
  {Y.}~\bibnamefont {Makhlin}}, \bibinfo {author} {\bibfnamefont
  {J.}~\bibnamefont {Schriefl}},\ and\ \bibinfo {author} {\bibfnamefont
  {G.}~\bibnamefont {Sch\"on}},\ }\bibfield  {title} {\bibinfo {title}
  {\textit{Decoherence in a Superconducting Quantum Bit Circuit}},\ }\href
  {https://doi.org/10.1103/PhysRevB.72.134519} {\bibfield  {journal} {\bibinfo
  {journal} {Phys. Rev. B}\ }\textbf {\bibinfo {volume} {72}},\ \bibinfo
  {pages} {134519} (\bibinfo {year} {2005})}\BibitemShut {NoStop}%
\end{thebibliography}%
\end{document}


\author{Ziwen Huang}
\address{Department of Physics and Astronomy, Northwestern University, Evanston, Illinois 60208, USA}
\author{Pranav S. Mundada}
\address{Department of Electrical Engineering, Princeton University, Princeton, New Jersey 08544, USA}
\author{Andr\'as Gyenis}
\address{Department of Electrical Engineering, Princeton University, Princeton, New Jersey 08544, USA}
\author{David I. Schuster}
\address{ The James Franck Institute and Department of Physics, University of Chicago,  Chicago, Illinois 60637, USA}
\author{Andrew A. Houck}
\address{Department of Electrical Engineering, Princeton University, Princeton, New Jersey 08544, USA}
\author{Jens Koch}
\address{Department of Physics and Astronomy, Northwestern University, Evanston, Illinois 60208, USA}
\address{Northwestern–Fermilab Center for Applied Physics and Superconducting Technologies, Northwestern University, Evanston, Illinois 60208, USA}
\title{Supplemental Material for ``Engineering Dynamical Sweet Spots to Protect Qubits from 1/$f$ Noise"}
\maketitle

The discussion in the main text focuses on the connection between dynamical sweet spots and extrema of quasi-energy differences for the example of a driven fluxonium qubit. In this note, we first extend this connection to a general periodically driven qubit system, using a procedure similar to the one discussed in Appendix B. Subsequently, we employ the Floquet framework to rephrase several previously developed protection schemes \cite{Coppersmith_two_qubit_sweet_spot,Didier_dynamical_sweet_spot,Rigetti_ac_sweet_spot,Rigetti_ac_sweet_spot_exp,Rigetti_broad_band_noise,Pan_transmon_dd,Oliver_rotating_frame_relaxation,Hu_drive_spin_qubit,Smirnov_Raib_decoherence,Ithier_decoherence_analysis} as special limits of our theory.

\section{General discussion: dynamical sweet spots}
We consider a periodically driven qubit described by an abstract Hamiltonian $\hat{H}(\lambda, t)$. Here, $\lambda$ is a control parameter, and the Hamiltonian is time-periodic with period $T_\mathrm{d}=2\pi/\omega_\mathrm{d}$, i.e., $\hat{H}(\lambda,t+T_\mathrm{d}) = \hat{H}(\lambda,t)$.  (As in the main text, $\omega_\mathrm{d}$ denotes the drive frequency.) Due to environmental noise, $\lambda$ is subjected to low-frequency fluctuations, $\lambda(t) = \lambda_0+\delta\lambda(t)$. Here, $\delta\lambda(t)$ captures the random fluctuations in $\lambda$. (Note that in this simple model there is only a single noise channel.)


If the amplitude of the classical noise is sufficiently weak, then the original Hamiltonian may be expanded up to leading order in $\delta\lambda(t)$ which yields the unperturbed qubit Hamiltonian and the time-dependent perturbation
\begin{align}
    \hat{H}_\mathrm{q}(t) = \hat{H}(\lambda_0,t),\quad\hat{H}_\mathrm{int}(t) = \frac{\partial \hat{H}(\lambda, t)}{\partial \lambda}\Big\vert_{\lambda=\lambda_0}\delta\lambda(t),
    \label{general_hamiltonian}
\end{align}
respectively. For convenience, we define $\hat{\sigma}(t)\equiv [{\partial \hat{H}(\lambda, t)}/{\partial \lambda}]\vert_{\lambda=\lambda_0}$. The time-periodicity of the Hamiltonian renders $\hat{\sigma}(t)$ time-periodic as well. While the example discussed in the main text leads to a constant $\hat{\sigma}(t)$, time dependence is present in other cases such as discussed in  Refs.\  \cite{Rigetti_ac_sweet_spot,Didier_dynamical_sweet_spot}. With $\hat{H}_\mathrm{q}$ and $\hat{H}_\mathrm{int}$ specified, we are ready to employ the Floquet framework developed in Appendix B, and calculate the pure-dephasing rate. The result is given by
\begin{align}
    \gamma_\phi = \sum_{k\in\mathbb{Z}}2|g^\lambda_{k\phi}|^2 S_\lambda (k\omega_\mathrm{d}),
    \label{gamma_phi_appE}
\end{align}
where 
\begin{align}
    g^\lambda_{k\phi} = \frac{1}{2T_\mathrm{d}}\int_0^{T_\mathrm{d}}\mathrm{d}t\,e^{i\omega_\mathrm{d}t}\tr_\mathrm{q}\left[ \hat{\sigma}(t)\hat{c}_\phi(t)\right],
\end{align}
and $S_\lambda(\omega) =\int_{-\infty}^{\infty}\mathrm{d}t\,e^{i\omega t} \langle \delta\lambda(t)\delta\lambda(0)\rangle$ is the noise spectrum. As a regularized variant of 1/$f$ noise it is appropriate to consider an $S_\lambda(\omega)$ that is strongly peaked at $\omega=0$. For such a spectrum the pure-dephasing rate \eqref{gamma_phi_appE} of the qubit is generically dominated by the term $2|g^\lambda_{0,\phi}|^2S_\lambda(0)$. However, the pure dephasing rate can be decreased significantly by choosing a working point where $g^\lambda_{0,\phi}=0$, thus eliminating the dominant contribution. In this case, weaker contributions of terms sampling the noise spectral density at non-zero frequencies will become relevant. These $g^\lambda_{0,\phi}=0$ working points are the \textit{dynamical sweet spots}.

Based on a similar argument as in Appendix D.1, we can prove that the condition $g^\lambda_{0,\phi}=0$ is closely related to the extrema of the quasi-energy difference, according to the relation
\begin{align}
     g^\lambda_{0\phi}=\frac{1}{2}\frac{\partial \epsilon_{01}}{\partial \lambda}.
    \label{sweet_spot_general}
\end{align}
Setting both sides of Eq.~\eqref{sweet_spot_general} to zero establishes the connection between the dynamical sweet spots and the quasi-energy extrema. 

In the following, we show how the theoretical framework outlined above can be used to understand  noise protection schemes based on qubit frequency modulation and Rabi drives, as presented in Refs.\ \cite{Coppersmith_two_qubit_sweet_spot,Didier_dynamical_sweet_spot,Rigetti_ac_sweet_spot,Rigetti_ac_sweet_spot_exp,Rigetti_broad_band_noise,Pan_transmon_dd,Oliver_rotating_frame_relaxation,Hu_drive_spin_qubit,Smirnov_Raib_decoherence,Ithier_decoherence_analysis}.

\section{Dynamical sweet spots realized through qubit-frequency modulation}
In Refs.~\cite{Didier_dynamical_sweet_spot,Rigetti_ac_sweet_spot_exp,Coppersmith_two_qubit_sweet_spot,Rigetti_ac_sweet_spot,Rigetti_broad_band_noise}, it is pointed out that qubit-frequency modulation can be harnessed for protecting a qubit from low-frequency noise. Such protection can be established by choosing modulation parameters for which the averaged, instantaneous transition frequency exhibits an extremum with respect to the noise parameter. Here, we confirm that the Floquet framework presented above indeed reproduces this condition in a certain limit.

We specifically consider the case of a frequency-modulated qubit using a purely longitudinal drive as discussed in Ref.\ \cite{Rigetti_ac_sweet_spot}. The model Hamiltonian in this case is given by $\hat{H}(\lambda,t) ={\Omega_{ge}(\lambda,t)}\hat{\sigma}_z/2$, where $\Omega_{ge}(\lambda,t)$ is the time-dependent instantaneous eigenenergy splitting, and $\lambda$ is an external control parameter that determines the splitting. The unperturbed Hamiltonian and perturbation operator from Eq.~\eqref{general_hamiltonian} now take on the concrete form  $\hat{H}_\mathrm{q} = \Omega_{ge}(\lambda_0,t)\hat{\sigma}_z/2$ and $\hat{H}_\mathrm{int}=[{\partial \Omega_{ge}(\lambda,t)}/{\partial \lambda}]\vert_{\lambda = \lambda_0}\delta\lambda(t)\hat{\sigma}_z/2$. (We assume that $H_\mathrm{int}$ is nonzero.)
To phrase the calculation of decoherence rates in our previous Floquet language, we obtain the unperturbed Floquet states and corresponding quasi-energies of the frequency-modulated qubit (in the absence of fluctuations $\delta\lambda(t)$). The Floquet states read
\begin{align}
    \vert w_{0(1)}(t)\rangle = \vert g(e)\rangle \exp\left[\varmp \frac{i}{2}\int_0^t\mathrm{d}t'(\Omega_{ge}(t')-\bar{\Omega}_{ge})\right],
\end{align}
and the corresponding quasi-energies are given by $\epsilon_{0(1)} = \varmp\bar{\Omega}_{ge}/2$. Different from notation in the main text,  $\vert g(e)\rangle$ here denote the eigenstates of  $\sigma_z$, and we have defined the averaged transition frequency $\bar{\Omega}_{ge} = \int_0^{T_\mathrm{m}}\mathrm{d}t\,\Omega_{ge}(\lambda,t)/T_\mathrm{m}$. Within this model, the quasi-energy difference is evidently given by the averaged qubit transition frequency, i.e., $\epsilon_{01}=\bar{\Omega}_{ge}$. According to Eq.~\eqref{sweet_spot_general}, dynamical sweet spots now manifest whenever the time-averaged transition frequency vanishes, $\partial \bar{\Omega}_{ge}/\partial\lambda = 0$.  This is in full agreement with the sweet-spot condition as formulated in Refs.~\cite{Didier_dynamical_sweet_spot,Rigetti_ac_sweet_spot_exp,Coppersmith_two_qubit_sweet_spot,Rigetti_ac_sweet_spot,Rigetti_broad_band_noise}.

The full expression of the pure-dephasing rate is calculated using Eq.~\eqref{gamma_phi_appE}, where the coefficients are given by
\begin{align}
    g^\lambda_{k\phi}
    = \frac{1}{2T_\mathrm{d}}\int_0^{T_\mathrm{d}}\mathrm{d}t\,e^{ik\omega_\mathrm{d}t}\frac{\partial \Omega_{ge}(\lambda,t)}{\partial \lambda}\Big\vert_{\lambda=\lambda_0}\!=\frac{1}{2} \frac{\partial\Omega_{ge,k}}{\partial\lambda}\Big\vert_{\lambda=\lambda_0}.
    \label{gkzphi}
\end{align}
Here, $\Omega_{ge,k}$ denotes the $k$th Fourier coefficient of the time-periodic transition frequency. We find that this result reproduces the one reported in Ref.~\cite{Rigetti_ac_sweet_spot}.

\section{Dynamical sweet spots induced by on-resonance Rabi driving}
It has been demonstrated that an on-resonance Rabi drive can dynamically decouple a qubit from low-frequency noise affecting its transition frequency \cite{Hu_drive_spin_qubit,Pan_transmon_dd,Oliver_rotating_frame_relaxation,Smirnov_Raib_decoherence,Ithier_decoherence_analysis}. Such decoupling is sometimes also referred to as \textit{spin-locking} \cite{Oliver_rotating_frame_relaxation, Ithier_decoherence_analysis}. In the following, we confirm that this protection scheme can also be understood as an instance of the dynamical sweet-spot operation discussed in this paper.

Consider the Hamiltonian of a transversely driven qubit within RWA,
\begin{align}
    H(\lambda,t) = \frac{\Omega_{ge}(\lambda)}{2}\hat{\sigma}_z + d(\sigma^+e^{-i\omega_\mathrm{d}t}+\textrm{h.c.}),
\end{align}
where $d$ denotes the drive strength. As before, $\lambda=\lambda_0 + \delta\lambda(t)$ is an external control parameter subjected to random fluctuations $\delta\lambda$. Note that, in contrast to the previous case, there is no separate modulation (AC component) of $\Omega_{ge}$ here. Employing series expansion in $\delta\lambda$, the perturbation describing the effect of noise to leading order is $\hat{H}_\mathrm{int}={\delta\lambda(t)}[\partial\Omega_{ge}(\lambda)/\partial\lambda]_{\lambda=\lambda_0}\hat{\sigma}_z/2$. (We again assume operation away from static sweet spots so that $\hat{H}_\mathrm{int}\not=0$.) To evaluate the decoherence rates, we need to invoke the  Floquet states of the noise-free qubit, which are given by
\begin{align}
    \vert w_{0}(t)\rangle =&\, \cos\frac{\theta}{2}\vert g\rangle-\exp\left(-{i\omega_\mathrm{d} t}\right) \sin\frac{\theta}{2} \vert e\rangle ,\nonumber\\
    \vert w_{1}(t)\rangle =&\, \sin\frac{\theta}{2}\vert g\rangle + \exp\left(-i\omega_\mathrm{d} t\right) \cos\frac{\theta}{2} \vert e\rangle,
\end{align}
where $\theta = \tan^{-1}(d/\delta\Omega_{ge})$ and $\delta\Omega_{ge}=\Omega_{ge}(\lambda)- \omega_\mathrm{d}$. The corresponding quasi-energies are given by $\epsilon_{0(1)} = \varmp \Omega_\mathrm{R}/2$, where $\Omega_\mathrm{R}=\sqrt{\delta\Omega_{ge}^2+d^2}$ is the Rabi frequency. As a result, the quasi-energy difference for this model is given by the Rabi frequency, i.e., $\epsilon_{01}=\Omega_\mathrm{R}$. The dynamical sweet spots therefore obey the condition $\partial \epsilon_{01}/\partial \lambda = \partial \Omega_\mathrm{R}/\partial\lambda = 0$, which is implies $\delta\Omega_{ge} = 0$.

We further compare the explicit expressions of the depolarization and pure-dephasing rates derived within our framework with the results previously reported in the literature. We find for these rates
\begin{align}
    \gamma_\phi =&\, \frac{1}{2} \left[\frac{\partial\Omega_{ge}(\lambda)}{\partial \lambda}\Big\vert_{\lambda=\lambda_0}\cos\theta\right]^2 S_\lambda(0),\nonumber\\
    \gamma_{\mp} =&\, \frac{1}{4} \left[\frac{\partial\Omega_{ge}(\lambda)}{\partial \lambda}\Big\vert_{\lambda=\lambda_0}\sin\theta\right]^2S_\lambda(\pm\Omega_\mathrm{R}),
\end{align}
which reproduce the results presented in  Ref.~\cite{Ithier_decoherence_analysis}.

\bibliography{mybib_1}